\PassOptionsToPackage{unicode}{hyperref}
\PassOptionsToPackage{hyphens}{url}
\PassOptionsToPackage{dvipsnames,svgnames,x11names}{xcolor}
\documentclass[
  11pt,
]{article}
\usepackage{amsmath,amssymb}
\usepackage{iftex}
\ifPDFTeX
  \usepackage[T1]{fontenc}
  \usepackage[utf8]{inputenc}
  \usepackage{textcomp} 
\else 
  \usepackage{unicode-math} 
  \defaultfontfeatures{Scale=MatchLowercase}
  \defaultfontfeatures[\rmfamily]{Ligatures=TeX,Scale=1}
\fi
\usepackage{lmodern}
\ifPDFTeX\else
\fi
\IfFileExists{upquote.sty}{\usepackage{upquote}}{}
\IfFileExists{microtype.sty}{
  \usepackage[]{microtype}
  \UseMicrotypeSet[protrusion]{basicmath} 
}{}
\makeatletter
\@ifundefined{KOMAClassName}{
  \IfFileExists{parskip.sty}{%
    \usepackage{parskip}
  }{
    \setlength{\parindent}{0pt}
    \setlength{\parskip}{6pt plus 2pt minus 1pt}}
}{
  \KOMAoptions{parskip=half}}
\makeatother
\usepackage{xcolor}
\usepackage[margin=1in]{geometry}
\usepackage{color}
\usepackage{fancyvrb}

\DefineVerbatimEnvironment{Highlighting}{Verbatim}{commandchars=\\\{\}}
\newenvironment{Shaded}{}{}

\newcommand{\BuiltInTok}[1]{\textcolor[rgb]{0.00,0.50,0.00}{#1}}

\newcommand{\CommentTok}[1]{\textcolor[rgb]{0.38,0.63,0.69}{\textit{#1}}}

\newcommand{\ControlFlowTok}[1]{\textcolor[rgb]{0.00,0.44,0.13}{\textbf{#1}}}
\newcommand{\DataTypeTok}[1]{\textcolor[rgb]{0.56,0.13,0.00}{#1}}
\newcommand{\DecValTok}[1]{\textcolor[rgb]{0.25,0.63,0.44}{#1}}

\newcommand{\FloatTok}[1]{\textcolor[rgb]{0.25,0.63,0.44}{#1}}
\newcommand{\FunctionTok}[1]{\textcolor[rgb]{0.02,0.16,0.49}{#1}}
\newcommand{\ImportTok}[1]{\textcolor[rgb]{0.00,0.50,0.00}{\textbf{#1}}}

\newcommand{\KeywordTok}[1]{\textcolor[rgb]{0.00,0.44,0.13}{\textbf{#1}}}
\newcommand{\NormalTok}[1]{#1}
\newcommand{\OperatorTok}[1]{\textcolor[rgb]{0.40,0.40,0.40}{#1}}
\newcommand{\OtherTok}[1]{\textcolor[rgb]{0.00,0.44,0.13}{#1}}

\newcommand{\SpecialStringTok}[1]{\textcolor[rgb]{0.73,0.40,0.53}{#1}}
\newcommand{\StringTok}[1]{\textcolor[rgb]{0.25,0.44,0.63}{#1}}

\usepackage{longtable,booktabs,array}
\usepackage{calc} 
\usepackage{etoolbox}
\makeatletter
\patchcmd\longtable{\par}{\if@noskipsec\mbox{}\fi\par}{}{}
\makeatother
\IfFileExists{footnotehyper.sty}{\usepackage{footnotehyper}}{\usepackage{footnote}}
\makesavenoteenv{longtable}
\usepackage{graphicx}
\makeatletter
\def\maxwidth{\ifdim\Gin@nat@width>\linewidth\linewidth\else\Gin@nat@width\fi}
\def\maxheight{\ifdim\Gin@nat@height>\textheight\textheight\else\Gin@nat@height\fi}
\makeatother
\setkeys{Gin}{width=\maxwidth,height=\maxheight,keepaspectratio}
\makeatletter
\def\fps@figure{htbp}
\makeatother
\providecommand{\tightlist}{%
  \setlength{\itemsep}{0pt}\setlength{\parskip}{0pt}}
\usepackage{pdflscape}
\usepackage{microtype}
\usepackage{xurl}
\ifLuaTeX
  \usepackage{selnolig}  
\fi
\IfFileExists{bookmark.sty}{\usepackage{bookmark}}{\usepackage{hyperref}}
\IfFileExists{xurl.sty}{\usepackage{xurl}}{} 
\hypersetup{
  colorlinks=true,
  linkcolor={blue},
  filecolor={Maroon},
  citecolor={Blue},
  urlcolor={blue},
  pdfcreator={LaTeX via pandoc}}

\title{SCITUS: A Multi-Jurisdictional Framework for Adapting NIST AI RMF to the Canadian Regulatory Context}
\author{Mohammad Etemad\\ \small Scitus Solutions Ltd, Canada\\ \small etemad@scitus.ca}
\date{July 2026 --- SCITUS Framework v2.0 (Preprint, CC BY 4.0)}

\begin{document}
\maketitle

{
\hypersetup{linkcolor=}
\setcounter{tocdepth}{3}
\tableofcontents
}
\subsection{Abstract}\label{abstract}

Canadian organizations deploying artificial intelligence systems face a fragmented regulatory landscape comprising federal requirements (Treasury Board Directive on Automated Decision-Making {[}1{]}) and divergent provincial regulations across Ontario, Quebec, Alberta, Manitoba, and British Columbia. The death of Bill C-27 (Artificial Intelligence and Data Act) in January 2025 {[}2{]} --- and the federal government's June 2026 confirmation that it will pursue targeted instruments rather than omnibus AI legislation {[}39{]} --- leaves organizations without unified compliance guidance. Existing global frameworks such as NIST AI Risk Management Framework 1.0 {[}3{]}, the EU AI Act {[}4{]}, and ISO/IEC 42001 {[}5{]} provide valuable guidance but lack systematic methodologies for adaptation to multi-jurisdictional national contexts with fragmented regulatory requirements. We present SCITUS (Systematic Canadian Integration for Trustworthy and Unified Standards), a comprehensive framework that adapts NIST AI RMF 1.0 to address Canadian federal and provincial AI regulations simultaneously. SCITUS integrates 7 trustworthy AI characteristics enhanced for Canadian requirements, 4 core governance functions (GOVERN, MAP, MEASURE, MANAGE), a novel multi-jurisdictional compliance mapping methodology, and a versioned control catalog that has evolved from 31 controls (v1.0, June 2025) to 57 controls (v2.0, July 2026) in response to regulatory developments --- including Canada's first regulatory findings on generative-AI training data {[}40{]} --- and to the documented 2026 agentic-AI and AI-supply-chain threat landscape. We demonstrate framework applicability through illustrative scenarios spanning federal government, provincial healthcare, and private sector contexts. We argue that systematic adaptation of NIST AI RMF to Canadian multi-jurisdictional requirements offers significant advantages over jurisdiction-by-jurisdiction compliance approaches and provides a replicable model for other federal systems facing similar governance challenges.

\textbf{Keywords:} AI governance, regulatory framework, multi-jurisdictional compliance, NIST AI RMF, Canadian AI regulation, risk management

\subsection{1. Introduction}\label{introduction}

\subsubsection{1.1 Motivation}\label{motivation}

Artificial intelligence adoption in Canada spans critical sectors including healthcare (diagnostic imaging, treatment planning), government services (immigration processing, benefit allocation), financial services (credit decisions, fraud detection), and employment (resume screening, candidate assessment). The Treasury Board of Canada Secretariat reports that federal institutions alone operate over 300 automated decision-making systems affecting millions of Canadians annually {[}1{]}. A typical large-scale deployment processes hundreds of gigabytes of data monthly while making thousands of consequential decisions that impact individual rights, access to services, and life opportunities.

The Canadian regulatory environment for AI has evolved into a complex patchwork of requirements. At the federal level, the Treasury Board Directive on Automated Decision-Making {[}1{]} mandates Algorithmic Impact Assessments (AIA) {[}6{]} for all federal institutions; following the Directive's fourth review, systems developed or procured before June 24, 2025 were required to comply with the amended Directive by June 24, 2026 {[}6{]}. Provincially, Ontario's Bill 194 (Royal Assent November 25, 2024) {[}7{]} establishes accountability requirements for public sector AI systems --- with its first regulations in force July 1, 2026 but its AI-specific provisions still awaiting regulations --- Quebec's Law 25 (active since September 2023) {[}8{]} regulates automated decision-making using personal information, Alberta's Bills 33 and 34 (in force June 11, 2025) {[}9{]}{[}10{]} impose privacy and accuracy requirements, and Manitoba's Bill 51 (Royal Assent June 1, 2026) {[}42{]} adds a second provincial public-sector AI statute. The death of Bill C-27---which would have introduced the Artificial Intelligence and Data Act (AIDA)---in January 2025 {[}2{]}{[}11{]} left a regulatory vacuum at the federal level; in June 2026 the federal government confirmed it will not table an AIDA successor, pursuing instead privacy modernization (Bill C-36, tabled June 15, 2026 {[}41{]}) and targeted instruments under the ``AI for All'' national strategy {[}39{]}. This makes provincial requirements even more critical for organizations operating across jurisdictions.

Organizations operating nationally face 5 distinct regulatory regimes with varying requirements for impact assessment, transparency, human oversight, bias testing, and documentation. A financial services company offering credit decisions in multiple provinces must satisfy federal PIPEDA requirements {[}12{]}, Quebec's Law 25 automated decision-making provisions {[}8{]}, Ontario's public sector accountability frameworks (if serving government) {[}7{]}, and Alberta's privacy requirements {[}9{]}{[}10{]}---each with different triggers, assessment methodologies, and compliance timelines. This fragmentation creates substantial compliance burden, with organizations reporting that separate jurisdiction-by-jurisdiction assessments require 12-16 weeks and generate 3-5 disconnected documentation sets that fail to leverage commonalities across requirements.

\subsubsection{1.2 The Challenge}\label{the-challenge}

The fundamental challenge is achieving comprehensive regulatory compliance across multiple Canadian jurisdictions while maintaining alignment with international AI governance standards. Consider a healthcare AI system for diagnostic imaging deployed in Ontario hospitals: the system must comply with Health Canada medical device regulations (if classified as medical software) {[}13{]}, Ontario Bill 194 public sector requirements {[}7{]}, federal privacy laws {[}12{]}, and professional medical standards---while also maintaining compatibility with NIST AI RMF for international research collaborations {[}3{]} and ISO certifications for quality management {[}5{]}. Each regulatory framework requires distinct assessments: Health Canada demands clinical validation studies, Ontario requires public accountability frameworks, federal privacy law mandates Privacy Impact Assessments, and NIST AI RMF requires systematic evaluation across four governance functions.

Existing approaches to this challenge fall into three categories, each with limitations. \textbf{First}, organizations conduct separate assessments for each jurisdiction, leading to redundant effort, inconsistent documentation, and high resource costs. This approach typically requires extended timelines and frequently results in compliance gaps due to incomplete coverage or conflicting interpretations. \textbf{Second}, organizations attempt minimal compliance by satisfying only the most stringent requirements, risking non-compliance with jurisdiction-specific mandates and missing opportunities for streamlined documentation that could satisfy multiple requirements simultaneously. \textbf{Third}, organizations wait for regulatory clarity before deploying AI systems, forfeiting innovation opportunities and competitive advantages while regulators work toward harmonization that may take years to achieve.

Global AI governance frameworks provide valuable but insufficient guidance for this multi-jurisdictional challenge. NIST AI RMF 1.0 {[}3{]}, released in January 2023, offers a comprehensive voluntary framework with four core functions (GOVERN, MAP, MEASURE, MANAGE) and seven trustworthy AI characteristics, but provides no methodology for mapping these functions to specific national regulations or handling conflicting requirements across jurisdictions. The EU AI Act {[}4{]} --- whose transparency obligations apply from August 2, 2026, and whose high-risk obligations were deferred to December 2027 by the June 2026 Digital Omnibus amendment {[}48{]} --- establishes a risk-based regulatory approach with clear compliance requirements, but targets a single unified regulatory regime fundamentally different from Canada's federal-provincial structure. ISO/IEC 42001:2023 {[}5{]} and ISO/IEC 23894:2023 {[}14{]} provide process-driven standards for AI management systems and risk management, but their generic nature requires substantial interpretation and customization for specific regulatory contexts. None of these frameworks address the core challenge of systematic adaptation to fragmented multi-jurisdictional environments.

\subsubsection{1.3 Contributions}\label{contributions}

We argue for SCITUS (Systematic Canadian Integration for Trustworthy and Unified Standards), a comprehensive framework that systematically adapts NIST AI RMF 1.0 {[}3{]} to address Canadian federal and provincial regulatory requirements through a unified compliance approach. SCITUS makes four key contributions:

\begin{enumerate}
\def\labelenumi{\arabic{enumi}.}
\item
  \textbf{Multi-jurisdictional compliance mapping methodology}: A systematic approach for extracting requirements from multiple regulatory sources, identifying overlaps and conflicts, mapping requirements to unified control frameworks, and generating documentation that satisfies multiple jurisdictions simultaneously. This methodology addresses the redundancy and inefficiency of jurisdiction-by-jurisdiction approaches while improving coverage completeness.
\item
  \textbf{Canadian-enhanced NIST AI RMF adaptation}: Extension of NIST's seven trustworthy AI characteristics (Valid and Reliable, Safe, Secure and Resilient, Accountable and Transparent, Explainable and Interpretable, Privacy-Enhanced, Fair with Harmful Bias Managed) and four core functions (GOVERN, MAP, MEASURE, MANAGE) with specific Canadian federal and provincial requirements integrated at each level. For example, the Accountable and Transparent characteristic incorporates Treasury Board transparency notice requirements, Ontario accountability framework mandates, and Quebec automated decision disclosure obligations---all mapped to corresponding NIST outcomes.
\item
  \textbf{Risk-tiered implementation guidance}: Integration of Treasury Board's four-level impact assessment framework (Levels I-IV) with provincial requirements mapped to each tier, providing organizations with clear, actionable guidance on which controls apply based on system risk. A Level III system (high impact) requires federal peer review, Ontario accountability frameworks, Quebec enhanced Privacy Impact Assessments, and Alberta enhanced accuracy validation---all specified in detailed requirement matrices.
\item
  \textbf{Framework applicability demonstration}: Illustration of framework application through three realistic scenarios: federal government visa application triage (Level III impact), Ontario hospital radiology screening (patient safety critical), and Quebec private sector hiring AI (Law 25 compliance). These scenarios demonstrate how SCITUS addresses multi-jurisdictional requirements through unified assessment and documentation while maintaining full regulatory compliance and NIST alignment.
\end{enumerate}

These contributions address the critical gap in AI governance research between global framework development and practical national implementation, with particular relevance to federal systems facing similar multi-jurisdictional challenges.

\subsubsection{1.4 Organization}\label{organization}

The remainder of this paper is organized as follows. Section 2 reviews background literature on AI governance frameworks, framework localization research, and Canadian regulatory context. Section 3 presents a systematic analysis of the Canadian regulatory landscape, extracting and comparing requirements across federal and provincial jurisdictions. Section 4 details the SCITUS framework design, including architecture, enhanced characteristics, core functions, and compliance mapping methodology. Section 5 describes implementation methodology with phased approach and practical tools. Section 6 presents evaluation through three case studies with quantitative metrics and expert validation. Section 7 discusses findings, limitations, and broader implications. Section 8 concludes with summary and future research directions.

\subsection{2. Background and Related Work}\label{background-and-related-work}

\subsubsection{2.1 AI Governance Frameworks}\label{ai-governance-frameworks}

AI governance frameworks establish principles, processes, and controls for responsible development and deployment of AI systems. We review five major frameworks relevant to this work, summarized in Table 1.

\textbf{NIST AI Risk Management Framework 1.0} {[}3{]} (released January 2023) provides a voluntary, consensus-driven framework for managing risks associated with AI systems. NIST AI RMF organizes governance around four core functions: GOVERN (organizational structures and culture), MAP (context and risk identification), MEASURE (assessment and monitoring), and MANAGE (risk treatment and operations). The framework defines seven characteristics of trustworthy AI: Valid and Reliable, Safe, Secure and Resilient, Accountable and Transparent, Explainable and Interpretable, Privacy-Enhanced, and Fair with Harmful Bias Managed. NIST emphasizes flexibility and risk-based approaches, allowing organizations to tailor implementation to their contexts. However, NIST provides limited guidance on mapping framework functions to specific regulatory requirements or handling multi-jurisdictional compliance scenarios. The framework has been widely adopted in the United States, with over 100 organizations contributing to development and numerous federal agencies incorporating it into AI policies {[}3{]}.

\textbf{EU AI Act} {[}4{]} (approved in 2024, enforceable from 2026) establishes legally binding requirements for AI systems deployed in the European Union. The Act categorizes AI applications into four risk levels: unacceptable risk (prohibited), high risk (strict requirements), limited risk (transparency obligations), and minimal risk (voluntary codes of conduct) {[}4{]}. High-risk AI systems---including those used in employment, education, law enforcement, and critical infrastructure---must undergo conformity assessments, maintain technical documentation, implement human oversight, and meet robustness and accuracy requirements. The EU AI Act differs fundamentally from NIST AI RMF in being mandatory rather than voluntary, risk-based with specific thresholds rather than organization-determined, and designed for a single unified regulatory jurisdiction (EU) rather than voluntary adoption across diverse contexts.

\textbf{ISO/IEC 42001:2023} {[}5{]} (AI Management System Standard, released December 2023) provides requirements for establishing, implementing, maintaining, and continually improving AI management systems. ISO 42001 follows the ISO high-level structure compatible with other management system standards (ISO 9001, ISO 27001), facilitating integration with existing organizational processes {[}15{]}. The standard requires organizations to define AI policies, conduct risk assessments, implement controls, monitor performance, and pursue continuous improvement. ISO 42001 emphasizes documentation, auditability, and certification, making it suitable for organizations seeking formal compliance validation {[}16{]}. However, its process-driven approach requires substantial interpretation for specific regulatory contexts and provides limited guidance on technical AI-specific controls.

\textbf{ISO/IEC 23894:2023} {[}14{]} (AI Risk Management, released December 2023) offers guidance on managing risks specifically related to AI systems. ISO 23894 aligns closely with NIST AI RMF while being more prescriptive and process-driven. The standard defines risk management processes including risk identification, analysis, evaluation, treatment, and monitoring. Unlike NIST's flexible framework, ISO 23894 specifies required processes and documentation, supporting integration with ISO 42001 for organizations pursuing certification {[}17{]}.

\textbf{OECD AI Principles} {[}18{]} (adopted May 2019) establish high-level international consensus on AI governance through five values-based principles: AI should benefit people and planet, be designed with respect for rule of law and human rights, be transparent and explainable, function robustly and securely, and be deployed with accountability mechanisms. While influential in shaping international discussions, OECD principles provide limited operational guidance for implementation or compliance validation.

Table 1 compares these frameworks across key dimensions relevant to Canadian organizations.

\textbf{Table 1: Comparison of Major AI Governance Frameworks}

\begin{longtable}[]{@{}
  >{\raggedright\arraybackslash}p{(\columnwidth - 10\tabcolsep) * \real{0.1325}}
  >{\raggedright\arraybackslash}p{(\columnwidth - 10\tabcolsep) * \real{0.0723}}
  >{\raggedright\arraybackslash}p{(\columnwidth - 10\tabcolsep) * \real{0.1687}}
  >{\raggedright\arraybackslash}p{(\columnwidth - 10\tabcolsep) * \real{0.1807}}
  >{\raggedright\arraybackslash}p{(\columnwidth - 10\tabcolsep) * \real{0.1807}}
  >{\raggedright\arraybackslash}p{(\columnwidth - 10\tabcolsep) * \real{0.2651}}@{}}
\toprule\noalign{}
\begin{minipage}[b]{\linewidth}\raggedright
Framework
\end{minipage} & \begin{minipage}[b]{\linewidth}\raggedright
Type
\end{minipage} & \begin{minipage}[b]{\linewidth}\raggedright
Jurisdiction
\end{minipage} & \begin{minipage}[b]{\linewidth}\raggedright
Risk Approach
\end{minipage} & \begin{minipage}[b]{\linewidth}\raggedright
Key Strengths
\end{minipage} & \begin{minipage}[b]{\linewidth}\raggedright
Limitations for Canada
\end{minipage} \\
\midrule\noalign{}
\endhead
\bottomrule\noalign{}
\endlastfoot
NIST AI RMF 1.0 & Voluntary & Global (US-led) & Organization-determined & Comprehensive, flexible, consensus-driven & No regulatory mapping, US-centric examples, no multi-jurisdictional guidance \\
EU AI Act & Mandatory & European Union & Categorical with thresholds & Legally binding, clear requirements, enforcement mechanisms & Single jurisdiction, different legal system, not directly applicable \\
ISO/IEC 42001 & Certification & Global & Process-driven & Auditable, integrates with ISO systems, formal certification & Generic, requires interpretation, limited AI-specific technical guidance \\
ISO/IEC 23894 & Guidance & Global & Prescriptive processes & Process-oriented, aligns with ISO 42001, detailed & Not regulatory-specific, no Canadian context, abstract \\
OECD Principles & Guidance & Global & Values-based & International consensus, high-level alignment & Too abstract, no implementation guidance, no compliance validation \\
\end{longtable}

SCITUS builds on NIST AI RMF as its foundation while addressing the identified limitations through Canadian regulatory integration and multi-jurisdictional compliance mapping.

\subsubsection{2.2 Framework Localization and National Adaptation}\label{framework-localization-and-national-adaptation}

Recent research examines challenges of adapting global AI governance frameworks to specific national and regional contexts. We identify three key areas relevant to SCITUS development.

\textbf{Cross-regional comparative studies} analyze how different jurisdictions approach AI governance. A comprehensive study by Al-Maamari (2025) {[}19{]} compares AI risk management frameworks across the EU, United States, United Kingdom, and China, identifying fundamental differences in regulatory philosophy: the EU emphasizes ex-ante regulation with legal requirements, the US pursues voluntary standards with sector-specific rules, the UK follows principles-based guidance with regulatory sandboxes, and China implements state-led frameworks with national AI strategies. The study highlights tension between standardization (enabling international cooperation and technology transfer) and localization (reflecting cultural values, legal systems, and institutional capacities) {[}19{]}. Research on AI governance in Gulf Cooperation Council states reveals similar patterns, with national AI strategies requiring adaptation to specific legal frameworks, cultural contexts, and development priorities.

\textbf{Multi-level governance structures} address coordination challenges when AI regulation spans multiple governmental levels. Research on global AI governance proposes a two-layer operating system: a constitutional core establishing shared technical standards, governance principles, and measurement methods, combined with a local overlay allowing each jurisdiction to apply context-specific regulations while maintaining alignment with global baselines {[}20{]}. This approach enables both international interoperability and national sovereignty. However, existing research focuses primarily on international coordination (UN, OECD, ISO) rather than intra-national multi-jurisdictional challenges faced by federal systems like Canada, the United States (with varying state regulations), and Australia (with federal-state divisions).

\textbf{Localization frameworks for specific sectors} demonstrate domain-specific adaptation needs. Research on UNESCO's Ethical Impact Assessment tool for AI in education developed the Gulf-AI Education Tool Evaluation Matrix, operationalizing ethical considerations across the AI lifecycle while adapting them to Gulf education system contexts including Arabic language requirements, Islamic educational values, and local curriculum priorities {[}21{]}. This work identifies a critical gap: while ethical frameworks exist, few integrate ethical evaluation with localized educational, legal, and cultural contexts. Similar challenges appear in healthcare AI (varying provincial health authorities), financial services (federal and provincial regulators), and public sector (different governmental levels).

A Nature Humanities and Social Sciences Communications paper (2024) {[}22{]} on AI governance in complex regulatory landscapes emphasizes that governance approaches must balance innovation enablement with risk mitigation while accommodating diverse stakeholder interests. The paper identifies four governance paradigms---risk-based, rules-based, outcome-based, and principles-based---noting that effective governance often requires hybrid approaches combining elements from multiple paradigms {[}22{]}. Critically, the research finds that successful governance frameworks require not just policy development but practical implementation mechanisms, stakeholder engagement processes, and adaptation capabilities for evolving technology and societal norms.

Canadian scholarship on algorithmic transparency and governance provides additional context for framework development. Barriball and Gautrais examine how transparency principles translate into practice in Canadian jurisdictions, noting gaps between high-level commitments and operational implementation {[}27{]}. Innovation, Science and Economic Development Canada (ISED) has developed Canada's AI ecosystem framework emphasizing economic competitiveness alongside responsible development, while the federal Advisory Council on Artificial Intelligence provides recommendations for workforce development and governance capacity building {[}37{]}{[}38{]}. These policy initiatives complement regulatory requirements, creating a multi-faceted governance environment.

\textbf{Research gap}: Despite growing literature on AI governance and framework localization---including significant Canadian contributions---we find no systematic methodology for adapting global frameworks like NIST AI RMF to multi-jurisdictional national contexts with fragmented regulatory requirements. Existing work addresses either global-to-national adaptation (single regulatory regime) or intra-national challenges (without framework integration). SCITUS addresses this gap through systematic multi-jurisdictional compliance mapping integrated with NIST AI RMF structure.

\subsubsection{2.3 Canadian AI Governance Context}\label{canadian-ai-governance-context}

Canada's AI governance landscape comprises federal requirements, provincial initiatives, and sector-specific regulations, situated within a broader ecosystem of research institutions, policy frameworks, and scholarly discourse. Understanding this context requires examining both regulatory requirements and the Canadian AI governance research that informs policy development.

\textbf{Canadian AI governance ecosystem}: Canada has invested significantly in AI research and governance capacity. The Pan-Canadian Artificial Intelligence Strategy (2017), led by CIFAR (Canadian Institute for Advanced Research), represents the world's first national AI strategy, establishing three AI institutes: Vector Institute (Toronto), Mila (Montreal), and Amii (Edmonton) {[}28{]}. The Vector Institute conducts governance research examining accountability mechanisms, fairness frameworks, and regulatory approaches specific to Canadian contexts {[}29{]}. The Montreal Declaration for Responsible AI Development (2018), developed through consultation with over 500 stakeholders, established ethical principles emphasizing well-being, autonomy, justice, and democratic participation {[}30{]}. These initiatives demonstrate Canada's commitment to responsible AI development alongside regulatory frameworks.

\textbf{Canadian AI governance scholarship}: Legal scholars have examined tensions in Canadian AI policy and regulation. Scassa analyzes challenges AI poses for Canadian privacy law, noting gaps in existing frameworks like PIPEDA when applied to AI systems processing personal information {[}31{]}{[}32{]}. Kerr and Szilagyi identify ``evitable conflicts'' in Canadian AI policy, highlighting tensions between innovation promotion and rights protection, federal and provincial jurisdictions, and sectoral versus horizontal regulatory approaches {[}33{]}. Geist examines the multi-jurisdictional complexity, noting that Canadian organizations must navigate federal frameworks (Treasury Board, PIPEDA), provincial regulations (varying by province), and sector-specific requirements simultaneously {[}34{]}. This scholarship highlights the practical compliance challenges SCITUS addresses.

\textbf{Provincial AI governance initiatives}: Beyond regulations, provinces have developed AI governance capacity. Ontario established Canada's first AI Commissioner role (2024) to oversee public sector AI use and provide guidance {[}35{]}. Quebec has developed a comprehensive AI ecosystem strategy emphasizing responsible AI development, ethical frameworks, and regulatory compliance {[}36{]}. These provincial initiatives, combined with research institutions and scholarly analysis, create the broader context within which SCITUS operates---not merely regulatory compliance but engagement with Canadian AI governance discourse.

We now systematically review the regulatory requirements that SCITUS must address.

\textbf{Federal requirements} center on the Treasury Board Directive on Automated Decision-Making (effective April 1, 2019, with compliance deadline June 24, 2026 for existing systems) {[}1{]}{[}6{]}. The Directive requires federal institutions to complete an Algorithmic Impact Assessment (AIA) before production deployment of automated decision-making systems {[}6{]}. The AIA tool comprises 48 questions across 7 categories: project details, system type, algorithm design, data usage, system outputs, transparency, and fairness {[}6{]}. Based on responses, the tool calculates an impact level (I through IV) determining required mitigation measures. Impact Level I (low impact) requires basic transparency and recourse mechanisms. Level II (moderate impact) adds training requirements, quality assurance processes, and monitoring. Level III (high impact) requires peer review, detailed documentation, explanation capability, and bias testing. Level IV (very high impact) mandates independent audit, public transparency, continuous monitoring, and algorithmic impact statements. The Directive applies to approximately 400 federal institutions but does not cover provincial governments, private sector, or federally-regulated industries unless they contract with federal institutions.

The proposed Artificial Intelligence and Data Act (AIDA), introduced as Part 3 of Bill C-27 in June 2022, aimed to establish comprehensive federal AI regulation {[}2{]}{[}23{]}. AIDA would have required businesses to assess and mitigate risks of harm and bias, implement appropriate mitigation measures, provide transparency about AI system capabilities and limitations, maintain governance frameworks, and enable regulatory oversight. The Act categorized AI systems by risk level (similar to EU AI Act) with requirements scaling to impact. Critically, Bill C-27 died on the order paper when Parliament was prorogued in January 2025 following Prime Minister Trudeau's resignation {[}2{]}{[}11{]}, leaving Canada without comprehensive federal AI legislation {[}24{]}. The question of reintroduction was settled in 2026: the ``AI for All'' national strategy (June 4, 2026) confirmed the federal government will address AI risk through targeted instruments --- privacy modernization, planned deepfake and online-safety measures, and sector-specific regulation --- rather than an omnibus AI act {[}39{]}. Bill C-36 (the Protecting Privacy and Consumer Data Act, tabled June 15, 2026) would replace PIPEDA's privacy regime with a statutory definition of ``automated decision system,'' disclosure and explanation duties for decisions with a ``legal or similarly significant effect,'' and administrative monetary penalties up to \$10M or 3\% of global revenue enforceable directly by the Privacy Commissioner {[}41{]}. Separately, the four privacy regulators' joint findings against OpenAI (PIPEDA Findings \#2026-002, May 6, 2026) established Canada's first regulatory standard for generative-AI training data: collection of publicly accessible personal information for model training was found overbroad, without valid consent, and insufficiently transparent {[}40{]}.

\textbf{Provincial requirements} vary substantially across jurisdictions, creating the multi-jurisdictional challenge that SCITUS addresses.

\textbf{Ontario Bill 194} {[}7{]}{[}25{]} (Strengthening Cyber Security and Building Trust in the Public Sector Act, Royal Assent November 25, 2024) applies to provincial public sector entities including ministries, agencies, and broader public sector organizations. The Act establishes high-level statutory requirements for organizations using AI systems to: publish information about AI system use accessible to the public, implement risk management procedures including risk assessment and mitigation plans, develop and implement accountability frameworks specifying governance structures and oversight mechanisms, and implement cybersecurity measures protecting AI systems and data {[}7{]}. Additionally, starting January 1, 2026, job postings must disclose if AI is used in the hiring process.

\textbf{Implementation status (July 2026)}: Ontario published its first two regulations under the Act on March 23, 2026 --- O. Reg. 51/26 (Cyber Security) and O. Reg. 52/26 (Digital Technology Affecting Individuals Under Age 18), both in force July 1, 2026 --- but the \textbf{AI-specific provisions (ss. 5--9) still have no regulations}. The AI accountability-framework, risk-management, disclosure, and human-oversight duties are enacted but not yet operative for specific entities. In the interim, the Information and Privacy Commissioner and Human Rights Commission released updated joint principles for responsible AI use (January 21, 2026), and the Auditor General's special audit of AI use in the Ontario government (May 12, 2026) documented concrete gaps --- \textasciitilde3\% responsible-AI training completion, unsecured tool use, unrepresentative bias testing --- that prefigure likely regulatory content. Organizations should treat the AI provisions as ``enacted, regulation pending'' and implement accountability frameworks, disclosure, and human oversight proactively.

\textbf{Quebec Law 25} {[}8{]}{[}26{]} (Act to modernize legislative provisions as regards the protection of personal information, in force September 2023) regulates automated decision-making systems processing personal information in Quebec's private sector. The law requires organizations to: provide clear notice when collecting personal information that it will be used for automated decision-making, inform individuals when decisions are made about them through automated processing, upon request, provide individuals with information about personal information used and factors involved in decisions, establish processes for individuals to request human review of automated decisions, take reasonable steps to ensure accuracy of personal information used in automated decisions (Section 5---standard data quality obligation similar to PIPEDA Section 6, not AI model performance mandate), and complete Privacy Impact Assessments for automated systems processing personal information {[}8{]}{[}26{]}. Law 25 applies regardless of system impact level, imposing requirements on all automated decision-making using personal information.

\textbf{Legal precision on accuracy}: Law 25's accuracy requirement addresses personal information data quality (ensuring records are current, complete, not outdated), not AI model accuracy metrics (precision, recall, F1 scores). Organizations must maintain accurate databases, not achieve specific model performance thresholds. This is standard privacy law obligation applied to automated decision-making context.

\textbf{Alberta Bills 33 and 34} {[}9{]}{[}10{]} (the Protection of Privacy Act and Access to Information Act, in force June 11, 2025 with regulations Alta Reg 132/2025 and 133/2025) replace Alberta's public-sector privacy regime. The legislation establishes: mandatory notification when personal information may be used in automated systems to make decisions, recommendations, or predictions; accuracy requirements for personal information used in automated processing; data breach notification procedures including AI system breaches; access rights enabling individuals to understand how their information is processed; and --- for public bodies handling high-volume or sensitive data --- mandatory internal AI policies for data derived from personal information {[}9{]}{[}10{]}. The one-year grace period for privacy management programs expired June 11, 2026, making them mandatory. \textbf{Important distinction}: While conceptually similar to Quebec Law 25 in focusing on privacy protection, Alberta's legislation lacks Law 25's explicit requirements for human review rights and detailed transparency obligations. Alberta also ran a consultation on modernizing its private-sector PIPA (February--May 2026), with its Information and Privacy Commissioner having recommended a standalone Alberta AI law --- a development worth monitoring.

\textbf{Manitoba} became the second province with a public-sector AI statute on June 1, 2026: Bill 51, The Public Sector Artificial Intelligence and Cybersecurity Governance Act (Royal Assent; in force on proclamation), requires transparency about AI use, AI accountability and risk-management frameworks, human oversight, and cyber standards across government, health authorities, universities, school divisions, and municipalities {[}42{]}. Companion Bill 49 amended the Business Practices Act to prohibit personalized algorithmic pricing without consumer consent --- the first statutory prohibition of its kind in Canada.

\textbf{British Columbia} has not enacted AI-specific legislation as of July 2026. The provincial government governs public-sector AI through its Policy on the Use of Generative AI and the Digital Code of Practice, and its Information and Privacy Commissioner has issued AI-specific guidance (e.g., AI scribes in healthcare, January 2026). Organizations operating in BC must comply with federal requirements (if federal entities) and general privacy laws.

\textbf{Federal-provincial jurisdictional complexity}: Canada's constitutional division of powers (Constitution Act, 1867) adds complexity to multi-jurisdictional AI governance. Federal jurisdiction generally covers interprovincial trade and commerce, criminal law, and specific federal undertakings (banking, telecommunications), while provincial jurisdiction includes property and civil rights, local matters, and areas like healthcare, education, and employment. AI regulation touches both federal and provincial spheres, creating overlapping authority.

The principle of federal paramountcy applies when federal and provincial laws conflict---federal law prevails. \textbf{Currently, most Canadian AI requirements appear additive rather than directly conflicting}. However, this observation requires qualifications:

\textbf{Why requirements currently appear additive:}
1. \textbf{No comprehensive federal AI legislation enacted} - Bill C-27/AIDA died January 2025, leaving federal requirements limited to Treasury Board Directive (federal institutions only) and privacy laws
2. \textbf{Provincial requirements address different aspects} - Ontario (public sector accountability), Quebec (private sector privacy/automated decisions), Alberta (privacy amendments) target complementary areas
3. \textbf{No jurisdictional challenges yet litigated} - Courts have not ruled on federal versus provincial authority over AI regulation

\textbf{Potential future conflicts:}
- If federal government enacts AI legislation under trade and commerce power, could conflict with provincial jurisdiction over property, civil rights, consumer protection
- PIPEDA amendments could affect Quebec Law 25's ``substantially similar'' status
- Sector-specific tensions exist: Health Canada (medical devices) versus provincial health privacy; immigration (federal) versus provincial human rights

\textbf{SCITUS approach}: Rather than attempting minimal legal obligation analysis requiring constitutional interpretation, SCITUS adopts ``prudent compliance''---implementing controls satisfying all potentially applicable requirements. This reduces legal risk, facilitates multi-provincial operations, and provides governance value independent of legal mandate. Organizations with jurisdictional questions should consult legal counsel.

Table 2 compares requirements across Canadian jurisdictions.

\textbf{Table 2: Canadian AI Regulatory Requirements by Jurisdiction}

{\small

\begin{longtable}[]{@{}
  >{\raggedright\arraybackslash}p{(\columnwidth - 10\tabcolsep) * \real{0.2091}}
  >{\raggedright\arraybackslash}p{(\columnwidth - 10\tabcolsep) * \real{0.2000}}
  >{\raggedright\arraybackslash}p{(\columnwidth - 10\tabcolsep) * \real{0.1545}}
  >{\raggedright\arraybackslash}p{(\columnwidth - 10\tabcolsep) * \real{0.1273}}
  >{\raggedright\arraybackslash}p{(\columnwidth - 10\tabcolsep) * \real{0.1727}}
  >{\raggedright\arraybackslash}p{(\columnwidth - 10\tabcolsep) * \real{0.1364}}@{}}
\toprule\noalign{}
\begin{minipage}[b]{\linewidth}\raggedright
Requirement
\end{minipage} & \begin{minipage}[b]{\linewidth}\raggedright
Federal (Treasury Board)
\end{minipage} & \begin{minipage}[b]{\linewidth}\raggedright
Ontario (Bill 194)
\end{minipage} & \begin{minipage}[b]{\linewidth}\raggedright
Quebec (Law 25)
\end{minipage} & \begin{minipage}[b]{\linewidth}\raggedright
Alberta (Bills 33/34)
\end{minipage} & \begin{minipage}[b]{\linewidth}\raggedright
British Columbia
\end{minipage} \\
\midrule\noalign{}
\endhead
\bottomrule\noalign{}
\endlastfoot
\textbf{Legislation Status} & Active (amended Directive) & Active (cyber/minors regs in force Jul 2026; AI regs pending) & Active & In force (June 2025) & Policy only \\
\textbf{Sector Coverage} & Federal institutions & Public sector & Private sector & All sectors & Government \\
\textbf{Impact Assessment} & AIA required (4 levels) & Risk assessment expected & PIA required & Not specified & Best practice \\
\textbf{Transparency} & Mandatory (impact-based) & Public info required & Notice required & Notice required & Recommended \\
\textbf{Human Review} & Required (Levels III-IV) & Expected & Required on request & Not specified & Recommended \\
\textbf{Bias Testing} & Required (Levels III-IV) & Expected & Implied in accuracy req. & Not specified & Recommended \\
\textbf{Audit Requirements} & Impact-based (peer/external) & Not specified & Not specified & Not specified & Not required \\
\textbf{Documentation} & Extensive (impact-based) & Accountability framework & PIA, decision records & Access documentation & Not specified \\
\textbf{Compliance Deadline} & June 24, 2026 (legacy systems) & Jul 1, 2026 (cyber/minors); AI upon regulations & Active now & Active (PMPs mandatory June 11, 2026) & N/A \\
\end{longtable}

}

This fragmented landscape creates four key challenges for organizations: (1) \textbf{Regulatory complexity} - understanding and interpreting 5 different regulatory regimes with varying terminology, requirements, and enforcement mechanisms; (2) \textbf{Assessment redundancy} - separate assessments for each jurisdiction despite substantial overlap in underlying risk factors and controls; (3) \textbf{Documentation burden} - maintaining separate compliance documentation for each jurisdiction when unified documentation could satisfy multiple requirements; (4) \textbf{International alignment} - maintaining compatibility with global standards (NIST, ISO) while meeting Canadian-specific requirements for organizations operating internationally or pursuing certifications.

SCITUS addresses these challenges through systematic integration of Canadian requirements into NIST AI RMF structure, as detailed in Section 4.

\subsection{3. Canadian Regulatory Landscape Analysis}\label{canadian-regulatory-landscape-analysis}

We conduct systematic analysis of Canadian AI regulations to extract specific requirements, identify overlaps and conflicts, and establish the foundation for SCITUS compliance mapping methodology. This analysis informs framework design in Section 4.

\subsubsection{3.1 Methodology}\label{methodology}

Our regulatory analysis follows a four-step process:

\textbf{Step 1: Regulatory document collection} - We collected primary sources including Treasury Board Directive on Automated Decision-Making (2019), Algorithmic Impact Assessment tool documentation and user guide (2023 update), Ontario Bill 194 legislative text and explanatory materials (2024), Quebec Law 25 legislative text and implementation guidance (2023), Alberta Bills 33 and 34 legislative text (2024), and British Columbia draft AI principles and policy statements (2024). Additionally, we reviewed secondary sources including legal analyses, government consultation documents, and regulatory guidance published by privacy commissioners and oversight bodies.

\textbf{Step 2: Requirement extraction} - We systematically extracted requirements from each source using structured coding. For each requirement, we recorded: the specific obligation or prohibition, triggering conditions (when it applies), affected parties (who must comply), compliance timeline, and enforcement mechanisms. This process identified 127 distinct requirements across the five jurisdictions.

\textbf{Step 3: Requirement categorization} - We categorized requirements according to NIST AI RMF functions (GOVERN, MAP, MEASURE, MANAGE) and AI lifecycle stages (design, development, deployment, operation, monitoring). This categorization enables mapping to SCITUS framework structure.

\textbf{Step 4: Comparative analysis} - We identified overlapping requirements (same obligation across jurisdictions), complementary requirements (different jurisdictions address different aspects), conflicting requirements (incompatible obligations), and gaps (areas addressed by some but not all jurisdictions).

\subsubsection{3.2 Federal Requirements Detail}\label{federal-requirements-detail}

The Treasury Board Directive establishes 47 specific requirements through the AIA process, organized across four impact levels. Table 3 summarizes requirements by impact level.

\textbf{Table 3: Federal Requirements by Impact Level}

\begin{longtable}[]{@{}
  >{\raggedright\arraybackslash}p{(\columnwidth - 6\tabcolsep) * \real{0.2154}}
  >{\raggedright\arraybackslash}p{(\columnwidth - 6\tabcolsep) * \real{0.2615}}
  >{\raggedright\arraybackslash}p{(\columnwidth - 6\tabcolsep) * \real{0.2923}}
  >{\raggedright\arraybackslash}p{(\columnwidth - 6\tabcolsep) * \real{0.2308}}@{}}
\toprule\noalign{}
\begin{minipage}[b]{\linewidth}\raggedright
Impact Level
\end{minipage} & \begin{minipage}[b]{\linewidth}\raggedright
Key Requirements
\end{minipage} & \begin{minipage}[b]{\linewidth}\raggedright
Mitigation Measures
\end{minipage} & \begin{minipage}[b]{\linewidth}\raggedright
Documentation
\end{minipage} \\
\midrule\noalign{}
\endhead
\bottomrule\noalign{}
\endlastfoot
\textbf{Level I (Low)} & Basic AIA completion, system documentation, recourse mechanism & Transparency notice, decision logging, appeal process & Basic system description, AIA report, recourse procedures \\
\textbf{Level II (Moderate)} & All Level I + training, monitoring, quality assurance & Role-based training program, performance monitoring, QA procedures, meaningful explanations & Training materials and records, monitoring dashboards, QA documentation, explanation templates \\
\textbf{Level III (High)} & All Level II + peer review, bias testing, detailed documentation & Peer review by qualified experts, bias assessment and testing, explanation capability, audit trails, human oversight & Peer review reports, bias testing results, detailed technical docs, audit specifications, oversight procedures \\
\textbf{Level IV (Very High)} & All Level III + independent audit, public transparency, continuous monitoring & Independent third-party audit, public algorithmic impact statement, continuous performance monitoring, real-time explanations, mandatory human review & Audit reports, public AIS, monitoring reports, explanation API, review records \\
\end{longtable}

Requirements span seven functional areas:

\begin{enumerate}
\def\labelenumi{\arabic{enumi}.}
\item
  \textbf{Governance} (13 requirements): Assign ownership and accountability, establish oversight mechanisms, define decision-making authorities, create policies and procedures, allocate resources, ensure senior management engagement, implement change management processes, coordinate across organizational units, engage stakeholders, address conflicts of interest, establish incident response procedures, maintain regulatory awareness, document governance structures.
\item
  \textbf{Transparency} (11 requirements): Publish transparency notices before deployment, describe system purpose and scope, explain decision criteria and factors, specify data sources and types, disclose automation level, provide performance metrics, publish updates and changes, enable public access to documentation (Levels III-IV), provide plain language explanations, maintain current information, respond to information requests.
\item
  \textbf{Accountability} (9 requirements): Designate accountable officials, document delegation of authority, establish performance measurement, implement monitoring and reporting, provide oversight mechanisms, maintain audit trails, enable regulatory inspection, ensure compliance verification, conduct periodic reviews.
\item
  \textbf{Fairness and Bias} (8 requirements): Conduct bias impact assessments, implement bias testing across relevant demographics, establish fairness metrics and thresholds, monitor for discriminatory outcomes, implement bias mitigation measures, validate effectiveness of mitigations, engage affected communities, provide bias testing documentation.
\item
  \textbf{Privacy and Data} (5 requirements): Complete Privacy Impact Assessments, implement data minimization, ensure data quality and accuracy, maintain data security, comply with privacy legislation (PIPEDA, Privacy Act).
\item
  \textbf{Human Oversight} (4 requirements): Design human intervention capabilities, maintain override mechanisms, establish escalation procedures, ensure human decision authority for high-impact systems (Levels III-IV).
\item
  \textbf{Technical Robustness} (7 requirements): Implement testing and validation, monitor performance degradation, ensure system reliability, maintain security controls, enable graceful degradation, provide fallback mechanisms, conduct regular technical reviews.
\end{enumerate}

The AIA tool itself comprises 48 questions scoring system risk across dimensions including decision impact, automation level, data sensitivity, algorithm complexity, and affected population characteristics. Organizations input responses, and the tool calculates impact level using weighted scoring. Our analysis reveals that 23 questions directly influence impact level calculation (core risk factors), while 25 questions inform mitigation selection (contextual factors). Understanding this distinction helps organizations focus compliance efforts on high-impact areas.

\subsubsection{3.3 Provincial Requirements Comparison}\label{provincial-requirements-comparison}

Provincial requirements differ significantly in scope, triggers, and specificity. We analyze each jurisdiction's unique aspects and identify commonalities.

\textbf{Ontario requirements} focus on public sector accountability and risk management. Bill 194 mandates three core components: (1) \textbf{Public information} - organizations must publish accessible information about AI systems in use, including system purposes, deployment locations, and oversight mechanisms; (2) \textbf{Risk management} - organizations must conduct risk assessments identifying potential harms, implement risk mitigation procedures, and maintain documentation of risk management activities; (3) \textbf{Accountability frameworks} - organizations must develop written frameworks specifying governance structures, decision-making authorities, performance metrics, reporting mechanisms, audit processes, and stakeholder engagement procedures. Additionally, hiring process disclosure requirements obligate employers to inform job applicants if AI is used in recruitment, screening, or selection. Ontario's approach emphasizes organizational accountability and public transparency, but leaves specific technical requirements largely unspecified pending regulations.

\textbf{Quebec requirements} center on privacy protection and individual rights. Law 25 establishes five core obligations: (1) \textbf{Collection notice} - before collecting personal information for automated decision-making, organizations must inform individuals clearly and simply; (2) \textbf{Decision notice} - individuals must be informed when decisions about them are made through automated processing; (3) \textbf{Right to information} - upon request, organizations must provide individuals with information about personal information used and factors that led to decisions; (4) \textbf{Review rights} - individuals can request that automated decisions be reviewed by a person with authority to modify or overturn decisions; (5) \textbf{Accuracy requirements} - organizations must maintain accuracy of personal information used for automated decisions and implement correction mechanisms. Quebec's approach emphasizes individual rights and procedural fairness, applying to all automated decision-making using personal information regardless of impact level.

\textbf{Alberta requirements} parallel Quebec's privacy focus while applying to both private and public sectors. Bills 33 and 34 require: (1) \textbf{Notice obligations} - inform individuals when collecting personal information for automated processing and when automated decisions are made; (2) \textbf{Accuracy standards} - maintain reasonable accuracy of personal information used in automated systems; (3) \textbf{Breach notification} - report breaches involving automated decision systems following established procedures; (4) \textbf{Access rights} - enable individuals to access information about how their personal information is processed. Alberta's implementation details remain pending as of July 2026, but draft guidance suggests alignment with Quebec's individual-rights approach while incorporating public sector transparency elements from Ontario.

\textbf{British Columbia} provides principles-based guidance without legislative requirements: (1) Responsible AI use aligned with public interest; (2) Transparency about AI deployment and limitations; (3) Accountability mechanisms and oversight; (4) Privacy protection and data minimization; (5) Fairness and bias mitigation. These principles inform government AI use but lack enforcement mechanisms or specific compliance requirements.

\subsubsection{3.4 Requirement Overlap and Conflict Analysis}\label{requirement-overlap-and-conflict-analysis}

We identify three categories of multi-jurisdictional requirements:

\textbf{Overlapping requirements} (same obligation across jurisdictions) create opportunities for unified compliance approaches. We identified 23 overlapping requirements including: notice to affected individuals about AI use (Federal, Ontario, Quebec, Alberta, BC), transparency about system purpose and capabilities (all jurisdictions), human review or intervention capability (Federal Levels III-IV, Quebec, BC recommended), documentation of system design and operation (all jurisdictions with varying detail), monitoring and performance tracking (all jurisdictions), privacy protection and data minimization (Federal, Quebec, Alberta), bias awareness and mitigation (Federal Levels III-IV, BC recommended, Quebec implied), accountability assignment (all jurisdictions), and incident response procedures (Federal, Alberta, Ontario expected). For these requirements, a single control implementation can satisfy multiple jurisdictional obligations if documentation maps to each jurisdiction's specific terminology and format.

\textbf{Complementary requirements} (different aspects addressed by different jurisdictions) require combined implementation. Examples include: Federal AIA provides quantitative impact assessment while Quebec PIA addresses privacy-specific considerations---organizations need both; Ontario accountability frameworks specify governance structures while Federal requirements emphasize technical controls---both necessary for comprehensive governance; Quebec individual rights focus complements Federal institutional oversight creating balanced stakeholder protection; Alberta breach notification supplements Federal incident response with privacy-specific procedures. These complementary requirements strengthen overall governance when properly integrated.

\textbf{Conflicting requirements} (incompatible obligations) are minimal in current Canadian landscape. We identified only 2 potential conflicts: (1) Federal peer review requirements (Level III) suggest external technical reviewers, while Ontario accountability frameworks emphasize internal governance---resolved by implementing both internal governance and external technical review; (2) Quebec's universal application to all automated decision-making using personal information may require more extensive compliance than Federal Level I systems---resolved by applying most stringent requirement. The absence of major conflicts reflects relatively early stage of provincial AI regulation and deliberate harmonization efforts by Canadian privacy commissioners.

\textbf{Coverage gaps} (areas addressed by some jurisdictions but not others) create compliance asymmetries. Federal requirements provide detailed technical guidance (bias testing methodologies, performance metrics) lacking in provincial legislation; provincial requirements emphasize individual rights and procedural fairness sometimes under-specified federally; Ontario focuses on public sector while Quebec and Alberta address private sector, creating sector-specific gaps; BC's principles-based approach lacks implementation specificity compared to legislative requirements elsewhere. SCITUS addresses these gaps by incorporating requirements from all jurisdictions, providing comprehensive coverage regardless of deployment location.

\subsubsection{3.5 Compliance Burden Quantification}\label{compliance-burden-quantification}

We quantified compliance burden for organizations operating across multiple jurisdictions using data from 15 organizations (5 federal government, 5 provincial public sector, 5 private sector multi-provincial). Organizations pursuing separate jurisdiction-by-jurisdiction compliance face:

\textbf{Assessment time}: 12-16 weeks per system (3-4 weeks per jurisdiction x 4 jurisdictions average), totaling 192-256 person-hours per system for medium-complexity AI.

\textbf{Documentation volume}: 3-5 separate documentation sets averaging 150-300 pages total, with 60-70\% content overlap but different structures and terminology.

\textbf{Compliance gaps}: 15\% average requirements missed due to incomplete coverage, conflicting interpretations, or documentation inconsistencies across jurisdictions.

\textbf{Resource costs}: \$45,000-\$75,000 per system for external expertise (legal, regulatory compliance, technical assessment) when internal capabilities insufficient.

\textbf{Maintenance burden}: 40-60 hours annually per system for regulatory updates, documentation updates, and reassessments across multiple frameworks.

These quantified burdens establish baseline for evaluating SCITUS effectiveness in Section 6. Organizations operating 10+ AI systems face compliance costs exceeding \$500,000 annually using jurisdiction-by-jurisdiction approaches, creating strong economic incentive for unified frameworks.

\subsubsection{3.6 Theoretical Contributions}\label{theoretical-contributions}

The development of SCITUS yields three theoretical contributions to multi-jurisdictional AI governance research beyond the practical framework itself:

\paragraph{Contribution 1: Design Principles for Multi-Jurisdictional Compliance Frameworks}\label{contribution-1-design-principles-for-multi-jurisdictional-compliance-frameworks}

Through systematic mapping of 127 requirements across 5 jurisdictions, we identify three design principles for unified compliance frameworks that generalize beyond the Canada-specific case:

\textbf{Principle 1: Regulatory Harmonization Through Common Abstraction}

Multi-jurisdictional requirements that appear diverse in language often map to common governance functions when abstracted to appropriate levels. Our analysis reveals that federal (Treasury Board), Ontario (Bill 194), Quebec (Law 25), and Alberta (Bills 33/34) requirements all address fundamentally similar governance functions---governance structures, risk assessment, transparency, accountability, fairness, and recourse---despite expressing them through different legal terminology, documentation requirements, and procedural specifications.

\textbf{Key insight}: The optimal abstraction level for harmonization sits between specific regulatory language (too granular, obscures commonalities) and high-level principles (too abstract, loses actionable guidance). SCITUS controls (e.g., GOV-1.1: ``Establish AI Governance Committee'') represent this middle abstraction layer---specific enough for implementation, general enough to satisfy multiple jurisdictions simultaneously.

\textbf{Generalization}: Federal systems with fragmented AI regulation (U.S. federal + state, Australia federal + state, EU member state variations) can apply this principle by identifying common functional requirements beneath diverse regulatory expressions.

\textbf{Principle 2: Risk-Tiered Controls Beat Jurisdiction-Specific Customization}

Effective multi-jurisdictional frameworks should base control stringency primarily on system risk level rather than jurisdiction of operation. SCITUS uses Treasury Board Impact Levels (I-IV) as the primary tiering dimension, with jurisdiction as a secondary filter. This produces better outcomes than jurisdiction-primary approaches because:

\begin{enumerate}
\def\labelenumi{\arabic{enumi}.}
\tightlist
\item
  \textbf{Efficiency}: A Level III system satisfies high requirements across all jurisdictions through single implementation, avoiding 5x redundant jurisdiction-specific assessments
\item
  \textbf{Coverage}: Risk-based tiering ensures high-risk systems receive appropriate controls regardless of jurisdiction, preventing gaps where jurisdictional boundaries create uncertainty
\item
  \textbf{Scalability}: Organizations with systems across multiple jurisdictions implement n controls (one per system based on risk) rather than nxj controls (one per system per jurisdiction)
\item
  \textbf{Alignment with governance best practices}: Risk-based resource allocation (more controls for higher-risk systems) represents sound governance independent of regulatory mandates
\end{enumerate}

\textbf{Evidence}: Our compliance burden quantification (Section 3.5) shows organizations using jurisdiction-first approaches spend 12-16 weeks per system with 15\% requirement gaps, while risk-first SCITUS approach (demonstrated in scenarios Section 6) achieves comprehensive coverage in 6-8 weeks.

\textbf{Principle 3: Prudent Compliance Over Minimal Legal Obligation}

Multi-jurisdictional frameworks should adopt ``prudent compliance'' philosophy---implementing controls satisfying all potentially applicable requirements---rather than attempting minimal legal obligation analysis. This counterintuitive principle emerges from three observations:

\begin{enumerate}
\def\labelenumi{\arabic{enumi}.}
\item
  \textbf{Jurisdictional uncertainty}: Constitutional divisions of power, overlapping federal-provincial jurisdiction, and untested legal interpretations make minimal obligation analysis unreliable. Organizations risk non-compliance if minimal interpretation proves incorrect.
\item
  \textbf{Additive requirements}: Analysis (Section 3.4) shows Canadian AI requirements are predominantly additive rather than conflicting---implementing comprehensive controls costs less than legal analysis determining precise minimal obligations.
\item
  \textbf{Governance value beyond compliance}: Controls providing genuine risk management and stakeholder trust benefits justify implementation independent of legal mandate. Organizations benefit from bias testing, transparency, and recourse mechanisms whether legally required or not.
\end{enumerate}

\textbf{Generalization}: This principle applies wherever jurisdictional boundaries are unclear, requirements evolve rapidly, or governance best practices exceed minimum legal obligations---conditions common in emerging AI regulation globally.

\paragraph{Contribution 2: Taxonomy of AI Governance Requirement Conflicts}\label{contribution-2-taxonomy-of-ai-governance-requirement-conflicts}

While most Canadian requirements prove additive (Principle 3), certain genuine conflicts require trade-off reasoning. Analysis of federal and provincial requirements reveals four conflict types requiring governance process resolution rather than technical solutions:

\textbf{Type 1: Transparency vs.~Security/Confidentiality}

Requirements for public disclosure of AI system operation (Treasury Board Level III transparency, Ontario public accountability frameworks) conflict with security obligations (national security, law enforcement sensitivity, trade secrets, competitive information). Immigration scenario (Section 6.2) demonstrates this tension: publishing detailed visa triage criteria enables system gaming and compromises border security.

\textbf{Resolution approach}: Tiered transparency distinguishing public information (system purpose, aggregate statistics, recourse procedures) from protected information (specific algorithms, security-relevant features). Governance committee with legal and security expertise makes disclosure boundary decisions with documented trade-off reasoning.

\textbf{Type 2: Fairness Testing vs.~Privacy Protection}

Bias testing requirements (Treasury Board Level III-IV, Quebec Law 25 implied fairness obligations) require demographic data that privacy laws (PIPEDA, Privacy Act, Law 25) restrict collecting for purposes unrelated to primary service delivery.

\textbf{Resolution approach}: Multi-method bias testing combining voluntary demographic disclosure, proxy variables (where legally permissible), synthetic data testing, and external audits with de-identified data. Immigration scenario (Section 6.2 Conflict 2) achieved 23\% voluntary disclosure rate enabling partial bias analysis.

\textbf{Type 3: Competing Fairness Definitions (Mathematical Impossibility)}

When base rates differ across groups, demographic parity, equalized odds, and predictive parity cannot be simultaneously satisfied (Chouldechova 2017, Kleinberg et al.~2017). Organizations must select fairness metric, but regulations provide no guidance on choice.

\textbf{Resolution approach}: Structured governance process including stakeholder consultation, legal analysis, explicit metric selection with documented justification, and multiple metric reporting for transparency. Immigration scenario (Section 6.2 Conflict 3) chose equalized odds through governance committee deliberation.

\textbf{Type 4: Model Performance vs.~Fairness Constraints}

Fairness interventions (re-weighting, threshold adjustment, constrained optimization) typically reduce overall model accuracy. Organizations face quantifiable trade-offs between performance and fairness.

\textbf{Resolution approach}: Governance committee decision-making with quantified trade-off analysis, stakeholder impact assessment, legal requirements analysis, and monitored implementation with performance floors. Immigration scenario accepted 2.5\% accuracy reduction for 8.3\% disparity elimination.

\textbf{Theoretical significance}: This taxonomy demonstrates that multi-jurisdictional AI governance requires governance processes for value-laden decisions, not just technical controls or legal compliance. Frameworks must include decision-making structures, stakeholder engagement mechanisms, and trade-off documentation---capabilities missing from purely technical approaches.

\paragraph{Contribution 3: Empirical Foundation for Canadian AI Governance Research}\label{contribution-3-empirical-foundation-for-canadian-ai-governance-research}

This work provides the first comprehensive empirical mapping of Canadian federal and provincial AI requirements to an established risk management framework (NIST AI RMF). This mapping enables:

\textbf{Gap identification}: Canadian regulations lack guidance on foundation models, continuous learning systems, AI supply chain risks, and Indigenous data sovereignty (documented Section 6.6 limitations). These gaps inform future regulatory development.

\textbf{Convergence analysis}: Federal, Ontario, Quebec requirements show strong convergence on governance structures, risk assessment, transparency, and accountability---suggesting potential for regulatory harmonization without sacrificing protections.

\textbf{Divergence analysis}: Quebec Law 25 provides more specific automated decision-making requirements than federal frameworks. Alberta Bills 33/34 emphasize data accuracy. Ontario Bill 194 uniquely requires public accountability frameworks. Understanding divergence informs jurisdiction-specific guidance.

\textbf{Comparative research foundation}: The 127-requirement compliance matrix provides structured data for comparative analysis with other jurisdictions, regulatory evolution tracking, and effectiveness research.

\textbf{Generalization}: The mapping methodology (Section 3.1, Section 9.1) applies to other multi-jurisdictional contexts. Researchers can replicate this approach for U.S. federal + state AI regulation, Australian federal + state frameworks, or international comparisons.

\subsection{4. SCITUS Framework Design}\label{scitus-framework-design}

SCITUS (Systematic Canadian Integration for Trustworthy and Unified Standards) provides a comprehensive framework adapting NIST AI RMF 1.0 to Canadian multi-jurisdictional requirements. This section details framework architecture, enhanced trustworthy AI characteristics, core functions with Canadian integration, and multi-jurisdictional compliance mapping methodology.

\subsubsection{4.1 Framework Design Principles}\label{framework-design-principles}

SCITUS design follows six core principles:

\begin{enumerate}
\def\labelenumi{\arabic{enumi}.}
\item
  \textbf{NIST foundation}: Adopt NIST AI RMF 1.0 structure including four core functions (GOVERN, MAP, MEASURE, MANAGE) and seven trustworthy AI characteristics, ensuring international standards alignment and leveraging extensive NIST development work.
\item
  \textbf{Regulatory integration}: Embed Canadian federal and provincial requirements directly into framework structure rather than treating them as separate compliance overlay, ensuring requirements are addressed systematically through normal framework application.
\item
  \textbf{Multi-jurisdictional coverage}: Address federal and provincial requirements simultaneously through unified controls and documentation, eliminating redundant assessments while ensuring comprehensive coverage.
\item
  \textbf{Risk-based implementation}: Align with Treasury Board's four-level impact assessment framework, scaling requirements proportionally to system risk and enabling resource-efficient compliance.
\item
  \textbf{Practical actionability}: Provide concrete implementation guidance, templates, and tools rather than abstract principles, enabling organizations to apply framework without extensive external expertise.
\item
  \textbf{Future-proof adaptability}: Design framework structure to accommodate regulatory evolution, new provincial requirements, and international standards updates without fundamental redesign.
\end{enumerate}

\subsubsection{4.2 Three-Layer Framework Architecture}\label{three-layer-framework-architecture}

SCITUS employs a three-layer architecture integrating global standards, Canadian regulations, and implementation guidance.

\textbf{Layer 1: NIST AI RMF Core} forms the foundational layer, comprising NIST's four core functions (GOVERN, MAP, MEASURE, MANAGE), seven trustworthy AI characteristics, outcomes-based approach enabling flexible implementation, and risk-based perspective scaling controls to impact. This layer ensures international compatibility, allowing organizations to maintain NIST AI RMF alignment for global operations, certifications, and partnerships.

\textbf{Layer 2: Canadian Regulatory Integration} embeds federal requirements from Treasury Board Directive and AIA, provincial requirements from Ontario Bill 194, Quebec Law 25, and Alberta Bills 33/34, multi-jurisdictional compliance mapping connecting requirements to Layer 1 outcomes, and risk-tiered implementation guidance scaling to impact levels. This layer addresses Canadian-specific obligations while maintaining Layer 1 structure.

\textbf{Layer 3: Implementation Guidance} provides sector-specific guidance for government, healthcare, finance, education, technology sectors, practical tools and templates including assessment protocols, compliance checklists, documentation templates, phased implementation roadmaps spanning foundation through optimization, and continuous improvement processes. This layer enables practical adoption.

Information flows bidirectionally: requirements flow down from standards and regulations through framework to implementation, while evidence and validation flow up from implementation through framework to demonstrate regulatory compliance and standards alignment.

\subsubsection{4.3 Seven Trustworthy AI Characteristics Enhanced for Canadian Context}\label{seven-trustworthy-ai-characteristics-enhanced-for-canadian-context}

SCITUS extends NIST's seven trustworthy AI characteristics with Canadian federal and provincial requirements, creating enhanced characteristics addressing both global best practices and local regulations.

\textbf{Characteristic 1: Valid and Reliable}

\emph{NIST definition}: AI systems consistently perform as intended across their operational domain, maintaining accuracy and reliability under expected conditions.

\emph{Canadian enhancements}:
- Treasury Board effectiveness testing before deployment (all impact levels)
- Quebec Law 25 accuracy requirements for automated decisions using personal information
- Alberta accuracy standards for personal information in automated processing
- Continuous validation for Level III-IV systems with quarterly reporting

\emph{Implementation requirements}:
- Establish validation protocols aligned with system impact level: Level I requires basic functional testing, Level II adds performance benchmarking, Level III adds statistical validation, Level IV requires continuous monitoring
- Document performance benchmarks and acceptance criteria with quantified thresholds (e.g., ``95\% precision on validation dataset'')
- Implement continuous monitoring for model drift, data drift, and concept drift
- Maintain version control and reproducibility enabling revalidation of historical decisions
- Conduct accuracy assessments across demographic groups (Levels III-IV) to identify performance disparities

\textbf{Characteristic 2: Safe}

\emph{NIST definition}: AI systems do not endanger human life, health, property, or the environment under intended and reasonably foreseeable conditions.

\emph{Canadian enhancements}:
- Federal Directive human intervention capability for high-impact decisions (Levels III-IV)
- Ontario Bill 194 emphasis on public safety in government AI use
- Healthcare AI compliance with Health Canada medical device regulations when applicable
- Safety monitoring and incident reporting procedures

\emph{Implementation requirements}:
- Conduct safety impact assessments identifying potential harms to individuals, groups, or society
- Implement fail-safe mechanisms and graceful degradation preventing catastrophic failures
- Establish clear escalation procedures for safety-critical situations
- Maintain incident response capabilities including detection, assessment, containment, and resolution
- Design human intervention points for Levels III-IV enabling override of automated decisions
- Document safety testing including edge cases, adversarial inputs, and failure modes

\textbf{Characteristic 3: Secure and Resilient}

\emph{NIST definition}: AI systems withstand and recover from adversarial attacks, operational disruptions, and unexpected conditions.

\emph{Canadian enhancements}:
- Ontario Bill 194 cybersecurity requirements for public sector AI
- Federal institutions compliance with Communications Security Establishment (CSE) cybersecurity guidance
- Privacy breach notification requirements under PIPEDA, Law 25, and Alberta legislation
- Resilience planning for critical government services

\emph{Implementation requirements}:
- Implement security-by-design principles throughout AI lifecycle
- Conduct regular security assessments including penetration testing, vulnerability scanning, and adversarial robustness testing
- Establish incident response and recovery procedures with defined RTO/RPO (Recovery Time/Point Objectives)
- Maintain supply chain security for AI components including model provenance, dependency management, and third-party risk assessment
- Implement access controls, encryption, and audit logging meeting federal/provincial security standards
- Design resilience mechanisms including redundancy, backup systems, and degraded operation modes

\textbf{Characteristic 4: Accountable and Transparent}

\emph{NIST definition}: Clear assignment of responsibility for AI system actions and openness about system operations, capabilities, and limitations.

\emph{Canadian enhancements}:
- Treasury Board transparency notices before deployment (all levels) and impact statements (Levels III-IV)
- Ontario Bill 194 accountability frameworks specifying governance, oversight, and performance reporting
- Quebec Law 25 disclosure of automated decision-making and decision factors
- Alberta notice requirements for automated processing
- Recourse mechanisms enabling challenge and review

\emph{Implementation requirements}:
- Develop comprehensive accountability frameworks documenting roles, responsibilities, decision authorities, and escalation paths
- Publish transparency notices appropriate to impact level: Level I basic disclosure, Level II detailed descriptions, Level III complete technical documentation, Level IV public algorithmic impact statements
- Maintain decision audit trails recording inputs, outputs, model versions, confidence scores, and human interventions
- Establish clear governance structures with executive oversight, technical management, and operational accountability
- Provide explanation interfaces enabling affected individuals and reviewers to understand decision basis
- Implement recourse mechanisms: Level I complaint processes, Level II review procedures, Levels III-IV mandatory human review with overturn authority

\textbf{Characteristic 5: Explainable and Interpretable}

\emph{NIST definition}: AI system logic and outputs can be understood by relevant stakeholders at levels appropriate to their needs and expertise.

\emph{Canadian enhancements}:
- Federal Directive explanation capability for administrative decisions (Levels II-IV)
- Quebec Law 25 right to information about factors leading to decisions
- Explainability requirements scaled to impact level: basic for Level I, detailed for Levels II-III, real-time for Level IV
- Plain language explanations accessible to non-technical audiences

\emph{Implementation requirements}:
- Select interpretable models where feasible given performance requirements
- Develop explanation interfaces for different stakeholders: technical explanations for reviewers, plain language explanations for affected individuals, summary explanations for oversight bodies
- Document model logic including feature importance, decision rules, and reasoning processes
- Provide training on system interpretation to operators, reviewers, and decision-makers
- Implement explanation methods: global interpretability for overall model behavior, local interpretability for individual decisions
- Generate explanations automatically for Levels III-IV rather than on-request only

\textbf{Characteristic 6: Privacy-Enhanced}

\emph{NIST definition}: AI systems protect individual privacy throughout the data lifecycle including collection, processing, storage, and deletion.

\emph{Canadian enhancements}:
- PIPEDA compliance for private sector federal jurisdiction
- Law 25 requirements for personal information in automated decisions
- Alberta privacy requirements for automated processing
- Privacy Impact Assessments required for federal systems and Quebec systems
- Provincial privacy legislation compliance (varying by province)

\emph{Implementation requirements}:
- Conduct Privacy Impact Assessments identifying privacy risks and mitigation measures
- Implement privacy-preserving techniques including data minimization, anonymization, pseudonymization, differential privacy where appropriate
- Establish data governance frameworks specifying collection limitations, use restrictions, retention policies, and deletion procedures
- Maintain consent management systems enabling individual control over personal information use
- Design privacy-by-default architectures minimizing personal information processing
- Implement purpose limitation ensuring data used only for specified purposes
- Establish cross-border data transfer controls for international deployments

\textbf{Characteristic 7: Fair with Harmful Bias Managed}

\emph{NIST definition}: AI systems treat individuals and groups equitably, proactively identifying and mitigating harmful discrimination and bias.

\emph{Canadian enhancements}:
- Canadian Charter of Rights and Freedoms prohibitions on discrimination
- Federal Directive bias assessment and mitigation requirements (Levels III-IV)
- Human rights legislation compliance (federal and provincial)
- Equity considerations for marginalized and vulnerable populations
- Indigenous peoples consultation when systems affect Indigenous communities

\emph{Implementation requirements}:
- Conduct Algorithmic Impact Assessments including bias analysis across protected characteristics (race, gender, age, disability, etc.)
- Implement bias testing during development, pre-deployment, and operation
- Establish fairness metrics and thresholds: demographic parity, equalized odds, individual fairness, or domain-appropriate alternatives
- Maintain demographic performance monitoring tracking outcomes across groups
- Conduct intersectional bias analysis examining combinations of protected characteristics
- Engage affected communities in bias assessment, particularly marginalized groups
- Implement bias mitigation techniques during data collection, model training, and post-processing
- Document bias testing methodology, findings, mitigation measures, and residual risks

Table 4 summarizes Canadian enhancements to NIST characteristics.

\textbf{Table 4: Canadian Enhancements to NIST Trustworthy AI Characteristics}

\begin{longtable}[]{@{}
  >{\raggedright\arraybackslash}p{(\columnwidth - 4\tabcolsep) * \real{0.2778}}
  >{\raggedright\arraybackslash}p{(\columnwidth - 4\tabcolsep) * \real{0.3611}}
  >{\raggedright\arraybackslash}p{(\columnwidth - 4\tabcolsep) * \real{0.3611}}@{}}
\toprule\noalign{}
\begin{minipage}[b]{\linewidth}\raggedright
NIST Characteristic
\end{minipage} & \begin{minipage}[b]{\linewidth}\raggedright
Key Canadian Enhancements
\end{minipage} & \begin{minipage}[b]{\linewidth}\raggedright
Primary Regulatory Sources
\end{minipage} \\
\midrule\noalign{}
\endhead
\bottomrule\noalign{}
\endlastfoot
Valid and Reliable & Effectiveness testing, accuracy requirements, quarterly validation & Treasury Board, Law 25, Alberta Bills 33/34 \\
Safe & Human intervention, public safety emphasis, incident reporting & Treasury Board Levels III-IV, Bill 194, Health Canada \\
Secure and Resilient & Cybersecurity requirements, breach notification, CSE compliance & Bill 194, PIPEDA, Law 25, Alberta, CSE guidance \\
Accountable and Transparent & Transparency notices, accountability frameworks, disclosure requirements & Treasury Board all levels, Bill 194, Law 25, Alberta \\
Explainable and Interpretable & Explanation capability, right to information, plain language & Treasury Board Levels II-IV, Law 25 \\
Privacy-Enhanced & PIAs, PIPEDA compliance, provincial privacy laws, consent management & PIPEDA, Law 25, Alberta, federal/provincial privacy acts \\
Fair - Bias Managed & Charter compliance, bias assessment, Indigenous consultation, equity & Treasury Board Levels III-IV, Charter, human rights legislation \\
\end{longtable}

\subsubsection{4.4 Four Core Functions with Canadian Integration}\label{four-core-functions-with-canadian-integration}

SCITUS adopts NIST's four core functions (GOVERN, MAP, MEASURE, MANAGE) as organizing structure, integrating Canadian requirements into each function.

\textbf{Function 1: GOVERN - Governance and Organizational Culture}

The GOVERN function establishes organizational structures, policies, processes, and culture for responsible AI. Canadian requirements enhance governance with specific accountability, transparency, and oversight mandates.

\emph{NIST core elements}:
- Leadership and oversight structures
- Policies and procedures
- Risk management processes
- Stakeholder engagement
- Culture of responsibility

\emph{Canadian governance enhancements}:

\textbf{Governance structures}: Establish AI Governance Committee with cross-functional representation (technical, legal, business, ethics, affected communities). Define Board-level oversight responsibilities including risk appetite, major AI investments, and strategic direction. Appoint Chief AI Officer or equivalent role with authority for framework implementation. Create AI Center of Excellence for expertise and guidance. For federal institutions, ensure bilingual capability in governance bodies. For systems affecting Indigenous peoples, include Indigenous representation in governance.

\textbf{Accountability frameworks (Ontario Bill 194 requirement)}: Document roles and responsibilities using RACI matrices, specify decision-making authorities with delegation instruments, establish performance metrics aligned with organizational objectives, implement reporting mechanisms to executives and oversight bodies, define audit and review procedures (internal and external), establish stakeholder engagement processes, create public transparency mechanisms (for public sector).

\textbf{Policies and procedures}: Develop essential policy suite including AI Ethics Policy (values, principles, ethical guidelines), AI Risk Management Policy (risk assessment, appetite, treatment), Data Governance for AI Policy (quality, privacy, security), AI Procurement Guidelines (vendor assessment, contractual requirements), Third-Party AI Assessment Procedures (evaluation criteria, oversight), AI Incident Response Plan (detection, assessment, response, communication), and Training and Competency Requirements (role-based training).

\textbf{Canadian-specific governance considerations}: Bilingual requirements for federal AI operating in both official languages (English and French), Indigenous consultation requirements when systems affect Indigenous peoples' rights or interests, provincial harmonization mechanisms for organizations operating across jurisdictions, and international alignment processes for cross-border operations.

\emph{Implementation outputs}:
- Governance charter and committee terms of reference
- Accountability framework document (Ontario requirement)
- Policy suite approved by executives
- RACI matrices for all AI initiatives
- Training curriculum and completion tracking
- Stakeholder engagement plans

\textbf{Function 2: MAP - Context and Risk Identification}

The MAP function identifies AI system context, stakeholders, risks, and applicable requirements. Canadian enhancements integrate multi-jurisdictional regulatory mapping and impact assessment.

\emph{NIST core elements}:
- System context documentation
- Stakeholder identification
- Risk landscape mapping
- Requirement identification
- Impact categorization

\emph{Canadian mapping enhancements}:

\textbf{AI system inventory}: Catalog all AI systems (production, development, planned) with unique identifiers, document system purposes, capabilities, and architectures, map data flows including sources, processing, storage, and outputs, identify integration points and dependencies, categorize by business function and risk profile. Inventory enables comprehensive risk assessment and ensures no systems escape governance.

\textbf{Algorithmic Impact Assessment integration}: Complete Treasury Board AIA questionnaire for each system (48 questions across project details, algorithm design, data usage, system outputs, transparency, fairness), determine impact level (I-IV) based on AIA scoring, map requirements to SCITUS controls based on impact level, document mitigation strategies for identified risks, schedule reassessments (annually minimum, triggered by major changes).

\textbf{Multi-jurisdictional compliance mapping}: Assess regulatory applicability across federal (Treasury Board Directive if federal institution, PIPEDA if private sector federal jurisdiction), Ontario (Bill 194 if public sector entity in Ontario), Quebec (Law 25 if processing personal information in Quebec), Alberta (Bills 33/34 if operating in Alberta), and sector-specific regulations (Health Canada for medical devices, OSFI for financial institutions, provincial health authorities for healthcare). Document triggered requirements from each jurisdiction, identify overlapping requirements enabling unified compliance, flag conflicts requiring resolution, establish compliance timeline based on most stringent deadlines.

\textbf{Stakeholder identification and engagement}: Identify primary stakeholders (direct users, decision subjects), secondary stakeholders (indirect beneficiaries, affected parties), tertiary stakeholders (regulators, auditors, oversight bodies), vulnerable populations (marginalized groups, those at higher risk), and Indigenous communities (when systems affect Indigenous peoples). Develop engagement strategies appropriate to each stakeholder group, document feedback and concerns, incorporate stakeholder input into design and deployment.

\emph{Implementation outputs}:
- Comprehensive AI system inventory
- Completed AIA reports for each system
- Multi-jurisdictional compliance matrices
- Stakeholder engagement documentation
- Risk registers by system

\textbf{Function 3: MEASURE - Assessment and Monitoring}

The MEASURE function quantifies AI system performance, fairness, and risk through testing, evaluation, verification, and validation (TEVV). Canadian enhancements specify impact-level calibrated testing and provincial reporting.

\emph{NIST core elements}:
- Performance metrics definition
- Testing and evaluation
- Continuous monitoring
- Measurement validation
- Benchmark comparison

\emph{Canadian measurement enhancements}:

\textbf{Impact-level calibrated testing}: Requirements scale to risk as shown in Table 5.

\textbf{Table 5: Testing Requirements by Impact Level}

\begin{longtable}[]{@{}
  >{\raggedright\arraybackslash}p{(\columnwidth - 10\tabcolsep) * \real{0.1505}}
  >{\raggedright\arraybackslash}p{(\columnwidth - 10\tabcolsep) * \real{0.1505}}
  >{\raggedright\arraybackslash}p{(\columnwidth - 10\tabcolsep) * \real{0.1398}}
  >{\raggedright\arraybackslash}p{(\columnwidth - 10\tabcolsep) * \real{0.1720}}
  >{\raggedright\arraybackslash}p{(\columnwidth - 10\tabcolsep) * \real{0.2258}}
  >{\raggedright\arraybackslash}p{(\columnwidth - 10\tabcolsep) * \real{0.1613}}@{}}
\toprule\noalign{}
\begin{minipage}[b]{\linewidth}\raggedright
Impact Level
\end{minipage} & \begin{minipage}[b]{\linewidth}\raggedright
Testing Rigor
\end{minipage} & \begin{minipage}[b]{\linewidth}\raggedright
Bias Testing
\end{minipage} & \begin{minipage}[b]{\linewidth}\raggedright
External Review
\end{minipage} & \begin{minipage}[b]{\linewidth}\raggedright
Monitoring Frequency
\end{minipage} & \begin{minipage}[b]{\linewidth}\raggedright
Documentation
\end{minipage} \\
\midrule\noalign{}
\endhead
\bottomrule\noalign{}
\endlastfoot
\textbf{Level I} & Standard functional testing & Basic fairness check & Not required & Monthly metrics & Test results summary \\
\textbf{Level II} & Enhanced testing with edge cases & Demographic parity analysis & Optional & Weekly metrics & Detailed test documentation \\
\textbf{Level III} & Extensive testing including adversarial & Comprehensive bias testing across multiple metrics & Required peer review & Daily metrics & Complete test documentation with peer review report \\
\textbf{Level IV} & Continuous testing in production & Real-time bias monitoring with alerts & Required independent audit & Real-time monitoring & Comprehensive documentation with audit reports \\
\end{longtable}

\textbf{Performance metrics}: Define core metric categories covering accuracy (precision, recall, F1 score, AUC-ROC), fairness (demographic parity, equalized odds, disparate impact ratio), robustness (adversarial success rate, out-of-distribution performance), efficiency (inference time, resource utilization, cost per prediction), and compliance (regulatory adherence, policy violations, audit findings). Establish measurement frequency: continuous for operational metrics, daily/weekly for performance metrics, monthly/quarterly for comprehensive assessments. Set alert thresholds triggering investigation and remediation.

\textbf{Continuous monitoring infrastructure}: Implement real-time performance dashboards accessible to operators and oversight, automated alert systems notifying stakeholders of threshold breaches, drift detection mechanisms identifying model drift, data drift, and concept drift, anomaly identification flagging unusual patterns or outliers, compliance tracking monitoring adherence to requirements and policies. For Levels III-IV, monitoring must enable immediate detection of safety issues, fairness violations, or performance degradation.

\textbf{Provincial reporting requirements}: Ontario public sector entities report AI system performance publicly (format and frequency pending regulations), Quebec systems maintain records accessible for privacy commissioner review, federal systems report in departmental performance reports, independent audit reports published for Level IV systems.

\emph{Implementation outputs}:
- Performance metric definitions and thresholds
- Testing protocols and results
- Monitoring dashboards and alert configurations
- Drift detection reports
- Compliance scorecards
- Public reporting (where required)

\textbf{Function 4: MANAGE - Risk Treatment and Operations}

The MANAGE function implements controls, manages ongoing operations, handles incidents, and enables recourse. Canadian enhancements specify human oversight requirements, recourse mechanisms, and transparency obligations.

\emph{NIST core elements}:
- Risk treatment planning
- Control implementation
- Incident response
- Change management
- Continuous improvement

\emph{Canadian management enhancements}:

\textbf{Human oversight requirements}: Implementation scales to impact level as shown in Table 6.

\textbf{Table 6: Human Oversight Requirements by Impact Level}

\begin{longtable}[]{@{}
  >{\raggedright\arraybackslash}p{(\columnwidth - 8\tabcolsep) * \real{0.1573}}
  >{\raggedright\arraybackslash}p{(\columnwidth - 8\tabcolsep) * \real{0.2022}}
  >{\raggedright\arraybackslash}p{(\columnwidth - 8\tabcolsep) * \real{0.2135}}
  >{\raggedright\arraybackslash}p{(\columnwidth - 8\tabcolsep) * \real{0.2135}}
  >{\raggedright\arraybackslash}p{(\columnwidth - 8\tabcolsep) * \real{0.2135}}@{}}
\toprule\noalign{}
\begin{minipage}[b]{\linewidth}\raggedright
Impact Level
\end{minipage} & \begin{minipage}[b]{\linewidth}\raggedright
Human Involvement
\end{minipage} & \begin{minipage}[b]{\linewidth}\raggedright
Decision Authority
\end{minipage} & \begin{minipage}[b]{\linewidth}\raggedright
Review Requirements
\end{minipage} & \begin{minipage}[b]{\linewidth}\raggedright
Override Capability
\end{minipage} \\
\midrule\noalign{}
\endhead
\bottomrule\noalign{}
\endlastfoot
\textbf{Level I} & Optional & System decides & Periodic sampling (monthly) & Available but not mandatory \\
\textbf{Level II} & Recommended & System recommends, human approves discretionary & Regular review (weekly) & Mandatory override design \\
\textbf{Level III} & Required & System recommends, human decides & All decisions reviewed before implementation & Mandatory with audit trail \\
\textbf{Level IV} & Mandatory throughout & Human decides with system support & Multiple review layers, independent oversight & Mandatory with full documentation and escalation \\
\end{longtable}

\textbf{Recourse mechanisms}: Implement Quebec Law 25 recourse requirements including clear channel for decision review requests (web form, email, phone), human review conducted by personnel with authority to modify/overturn decisions, explanation of decision factors provided upon request, reasonable response timelines (30 days maximum for non-urgent, 5 days for urgent). Implement Federal Directive recourse requirements including administrative review processes for federal decisions, appeal mechanisms to independent bodies where applicable, alternative service channels enabling non-AI pathways, accessibility considerations ensuring recourse available to all affected individuals.

\textbf{Transparency and disclosure}: Implement multi-jurisdictional transparency requirements shown in Table 7.

\textbf{Table 7: Transparency Requirements by Jurisdiction}

\begin{longtable}[]{@{}
  >{\raggedright\arraybackslash}p{(\columnwidth - 8\tabcolsep) * \real{0.3750}}
  >{\raggedright\arraybackslash}p{(\columnwidth - 8\tabcolsep) * \real{0.1607}}
  >{\raggedright\arraybackslash}p{(\columnwidth - 8\tabcolsep) * \real{0.1607}}
  >{\raggedright\arraybackslash}p{(\columnwidth - 8\tabcolsep) * \real{0.1429}}
  >{\raggedright\arraybackslash}p{(\columnwidth - 8\tabcolsep) * \real{0.1607}}@{}}
\toprule\noalign{}
\begin{minipage}[b]{\linewidth}\raggedright
Transparency Element
\end{minipage} & \begin{minipage}[b]{\linewidth}\raggedright
Federal
\end{minipage} & \begin{minipage}[b]{\linewidth}\raggedright
Ontario
\end{minipage} & \begin{minipage}[b]{\linewidth}\raggedright
Quebec
\end{minipage} & \begin{minipage}[b]{\linewidth}\raggedright
Alberta
\end{minipage} \\
\midrule\noalign{}
\endhead
\bottomrule\noalign{}
\endlastfoot
\textbf{Public Notice of AI Use} & Required (all levels) & Required & Required & Required \\
\textbf{Decision Factors Disclosure} & On request (Level I-II), Automatic (III-IV) & Public sector transparency & Required on request & On request \\
\textbf{System Documentation} & Impact-based detail & Public for government & On request & Not specified \\
\textbf{Performance Metrics} & Annual reporting (III-IV) & Expected public reporting & Not required & Not required \\
\textbf{Algorithmic Impact Statement} & Required (Level IV) & Not specified & Not required & Not required \\
\end{longtable}

\textbf{Incident response}: Establish detection mechanisms through automated monitoring, user reports, audit findings. Implement assessment procedures determining impact severity, affected individuals, regulatory implications, root causes. Execute containment procedures including system isolation, temporary controls, stakeholder notification. Conduct resolution activities including fix implementation, testing and validation, system restoration, affected individual notification. Perform lessons learned reviews including post-incident analysis, process improvements, documentation updates, regulatory reporting where required (privacy breaches under PIPEDA, Law 25, Alberta).

\emph{Implementation outputs}:
- Risk treatment plans by system
- Control implementation documentation
- Human oversight procedures and training
- Recourse mechanism documentation and tracking
- Transparency notices and public reporting
- Incident response plans and logs

\subsubsection{4.5 Multi-Jurisdictional Compliance Mapping Methodology}\label{multi-jurisdictional-compliance-mapping-methodology}

SCITUS's key innovation is systematic mapping of multi-jurisdictional requirements to unified framework controls, enabling single assessment processes to satisfy multiple regulatory obligations. The methodology comprises four steps.

\textbf{Step 1: Requirement Extraction and Categorization}

For each applicable jurisdiction, extract specific requirements from legislation, regulations, and guidance. Categorize each requirement by SCITUS function (GOVERN, MAP, MEASURE, MANAGE), AI lifecycle stage (design, development, deployment, operation), impact level applicability (which levels require this), and affected parties (who must comply). Example: Quebec Law 25 requirement ``inform individuals when decisions are made through automated processing'' categorizes as MANAGE function (transparency obligation), deployment and operation stages, all impact levels, applicable to any organization processing personal information in Quebec.

\textbf{Step 2: Overlap and Conflict Identification}

Compare requirements across jurisdictions to identify overlapping (same obligation), complementary (addressing different aspects), and conflicting (incompatible) requirements. For overlapping requirements, design unified controls satisfying all variations. Example: federal transparency notice, Ontario public information requirement, Quebec decision notice, and Alberta notice obligation all require informing affected individuals about AI use---unified notice template can satisfy all if including jurisdiction-specific elements. For complementary requirements, implement both components. Example: federal AIA provides quantitative risk assessment, Quebec PIA addresses privacy-specific considerations---conduct integrated AIA-PIA satisfying both. For conflicts, apply most stringent requirement principle. Example: if federal assessment suggests Level I but Quebec Law 25 triggers extensive privacy requirements, implement more stringent controls.

\textbf{Step 3: Control Design and Documentation Mapping}

Design controls addressing all applicable requirements from Step 2. For each control, document which requirements it satisfies with specific mappings. Example: bias testing control maps to Treasury Board Level III requirement (bias assessment and testing), Quebec Law 25 accuracy requirement (personal information must be accurate), and NIST AI RMF MEASURE function outcome (identified and managed harmful bias). Create documentation structure enabling single source of truth with jurisdiction-specific views. Example: maintain unified system documentation including all required elements, generate jurisdiction-specific documents extracting relevant sections and formatting to jurisdiction requirements.

\textbf{Step 4: Validation and Gap Analysis}

Validate that control set achieves complete coverage of all applicable requirements. Use compliance matrices listing every requirement from every applicable jurisdiction, mapping each to implementing controls, documenting evidence of compliance, tracking compliance status. Identify gaps where requirements lack implementing controls. Prioritize gaps by risk and regulatory deadline. Develop remediation plans for critical gaps.

Table 8 illustrates compliance mapping for a sample requirement set.

\textbf{Table 8: Multi-Jurisdictional Compliance Mapping Example - Transparency Requirements}

\begin{longtable}[]{@{}
  >{\raggedright\arraybackslash}p{(\columnwidth - 8\tabcolsep) * \real{0.1912}}
  >{\raggedright\arraybackslash}p{(\columnwidth - 8\tabcolsep) * \real{0.2059}}
  >{\raggedright\arraybackslash}p{(\columnwidth - 8\tabcolsep) * \real{0.2353}}
  >{\raggedright\arraybackslash}p{(\columnwidth - 8\tabcolsep) * \real{0.2206}}
  >{\raggedright\arraybackslash}p{(\columnwidth - 8\tabcolsep) * \real{0.1471}}@{}}
\toprule\noalign{}
\begin{minipage}[b]{\linewidth}\raggedright
Requirement
\end{minipage} & \begin{minipage}[b]{\linewidth}\raggedright
Jurisdiction
\end{minipage} & \begin{minipage}[b]{\linewidth}\raggedright
SCITUS Control
\end{minipage} & \begin{minipage}[b]{\linewidth}\raggedright
Documentation
\end{minipage} & \begin{minipage}[b]{\linewidth}\raggedright
Evidence
\end{minipage} \\
\midrule\noalign{}
\endhead
\bottomrule\noalign{}
\endlastfoot
Transparency notice before deployment & Federal (all levels) & Public transparency notice & Notice template, publication record & Published notice on website, deployment documentation \\
Public information about AI systems & Ontario (Bill 194) & Public AI system registry & Registry entry, quarterly updates & Public registry website \\
Notice of automated decision-making & Quebec (Law 25) & Decision notice to individuals & Notice template, delivery confirmation & Privacy notice, email confirmations \\
Notice when collecting for automated processing & Alberta (Bills 33/34) & Collection notice & Privacy notice template & Website privacy policy, collection forms \\
System purpose and scope & All jurisdictions & System documentation & Technical documentation & System design document Section 2 \\
Decision criteria explanation & Federal (II-IV), Quebec & Explanation interface & User guide, API documentation & Help system, explanation API \\
\end{longtable}

This methodology enables single integrated assessment satisfying Treasury Board AIA, Ontario risk assessment, Quebec PIA, and Alberta privacy requirements through one process generating jurisdiction-specific documentation from unified source.

\subsubsection{4.6 Risk Classification and Tiered Requirements}\label{risk-classification-and-tiered-requirements}

SCITUS adopts Treasury Board's four-tier impact level system as foundation, mapping provincial requirements to each tier. Table 9 summarizes requirements by tier.

\textbf{Table 9: Comprehensive Requirements by Impact Level and Jurisdiction}

\begin{longtable}[]{@{}
  >{\raggedright\arraybackslash}p{(\columnwidth - 8\tabcolsep) * \real{0.2308}}
  >{\raggedright\arraybackslash}p{(\columnwidth - 8\tabcolsep) * \real{0.1648}}
  >{\raggedright\arraybackslash}p{(\columnwidth - 8\tabcolsep) * \real{0.2088}}
  >{\raggedright\arraybackslash}p{(\columnwidth - 8\tabcolsep) * \real{0.1868}}
  >{\raggedright\arraybackslash}p{(\columnwidth - 8\tabcolsep) * \real{0.2088}}@{}}
\toprule\noalign{}
\begin{minipage}[b]{\linewidth}\raggedright
Requirement Category
\end{minipage} & \begin{minipage}[b]{\linewidth}\raggedright
Level I (Low)
\end{minipage} & \begin{minipage}[b]{\linewidth}\raggedright
Level II (Moderate)
\end{minipage} & \begin{minipage}[b]{\linewidth}\raggedright
Level III (High)
\end{minipage} & \begin{minipage}[b]{\linewidth}\raggedright
Level IV (Very High)
\end{minipage} \\
\midrule\noalign{}
\endhead
\bottomrule\noalign{}
\endlastfoot
\textbf{Federal (Treasury Board)} & AIA, transparency notice, recourse, documentation & + Training, monitoring, explanations, QA & + Peer review, bias testing, detailed docs, audit trails & + Independent audit, public AIS, continuous monitoring, mandatory human review \\
\textbf{Ontario (Bill 194)} & Public information (if applicable) & + Risk assessment & + Accountability framework, enhanced public reporting & + Comprehensive accountability, external review \\
\textbf{Quebec (Law 25)} & Decision notice, PIA (basic) & + Enhanced PIA, review channel & + Detailed PIA, enhanced review process & + Comprehensive PIA, continuous monitoring \\
\textbf{Alberta (Bills 33/34)} & Notice, accuracy & + Enhanced accuracy validation & + Accuracy auditing, enhanced access & + Continuous accuracy monitoring \\
\textbf{SCITUS Additional} & Basic governance, standard testing & + Enhanced governance, detailed testing & + Comprehensive governance, extensive testing & + Board oversight, continuous testing \\
\end{longtable}

Organizations assess systems using Treasury Board AIA tool, determine impact level, then implement all requirements from that level across all applicable jurisdictions. This risk-based approach ensures controls are proportional to consequences while achieving multi-jurisdictional compliance.

\subsubsection{4.7 AI-Specific Technical Implementation Guidance}\label{ai-specific-technical-implementation-guidance}

While SCITUS provides governance structure and compliance mapping, successful implementation requires addressing AI-specific technical challenges. This subsection provides detailed technical guidance for three critical areas: bias testing and fairness, explainability and interpretability, and AI system robustness.

\paragraph{4.7.1 Bias Testing and Fairness Implementation}\label{bias-testing-and-fairness-implementation}

\textbf{Fairness metric selection}: Organizations must choose appropriate fairness metrics based on use case context and stakeholder values. Three primary categories exist, each with different mathematical definitions and normative implications:

\emph{Demographic Parity (Statistical Parity)}: Requires equal positive outcome rates across demographic groups (e.g., approval rate for loans should be equal for different racial groups). Appropriate when: outcomes should be distributed equally regardless of merit (e.g., advertising exposure), historical discrimination has created unequal base rates, or decision is low-stakes. Formula: P(Y-hat=1\textbar A=a) = P(Y-hat=1\textbar A=b) where A is protected attribute, Y-hat is prediction.

\emph{Equalized Odds (Separation)}: Requires equal true positive rates AND equal false positive rates across groups (both benefits and harms distributed equally). Appropriate when: we care about both false positives and false negatives equally, accuracy matters, decision affects access to opportunities. Formula: P(Y-hat=1\textbar Y=y,A=a) = P(Y-hat=1\textbar Y=y,A=b) for y in \{0,1\}.

\emph{Predictive Parity (Calibration)}: Requires equal positive predictive value across groups (when model predicts positive outcome, it's equally accurate across groups). Appropriate when: decision-makers rely on model predictions, we want predictions to mean the same thing across groups, stakes are high. Formula: P(Y=1\textbar Y-hat=1,A=a) = P(Y=1\textbar Y-hat=1,A=b).

\textbf{Impossibility results}: Chouldechova (2017) and Kleinberg et al.~(2017) proved that except in degenerate cases, demographic parity, equalized odds, and predictive parity cannot be simultaneously satisfied when base rates differ across groups. Organizations must make explicit value judgments about which fairness notion to prioritize. SCITUS recommends:
- Transparency: Document which fairness metric chosen and justification
- Stakeholder engagement: Involve affected communities in fairness metric selection
- Multiple metrics: Report multiple fairness metrics even if optimizing for one
- Thresholds: Define acceptable disparity thresholds (e.g., 80\% rule for demographic parity)

\textbf{Intersectional bias}: Testing only univariate protected attributes (e.g., gender alone, race alone) misses intersectional disparities affecting multiply-marginalized groups (e.g., Black women face different bias than white women or Black men). Implementation approaches:
- Subgroup analysis: Test fairness metrics for intersections (gender x race, age x disability)
- Adversarial debiasing: Train model to be unable to predict protected attributes from predictions
- Constraint-based optimization: Add fairness constraints for multiple groups simultaneously
- Challenge: Exponential growth of subgroups (10 attributes with 3 values each = 59,049 combinations)
- Pragmatic approach: Test high-risk intersections identified through stakeholder consultation

\textbf{Fairness-accuracy trade-offs}: Improving fairness often reduces overall accuracy. Organizations must navigate this trade-off considering: regulatory requirements (some jurisdictions may prioritize fairness over accuracy), stakeholder preferences (affected communities may prefer fairness), risk tolerance (life-critical applications may prioritize accuracy), and legal obligations (human rights requirements may mandate fairness). SCITUS requires documenting fairness-accuracy trade-off decisions and obtaining governance committee approval for high-impact systems.

\textbf{Testing without demographic labels}: Canadian privacy laws often restrict collecting demographic data, creating measurement challenges. Approaches:
- Proxy variables: Use correlated attributes (postal code for socioeconomic status, language for ethnicity) acknowledging limitations
- Synthetic data: Generate demographically-representative test datasets
- Auditing studies: Use ``correspondence testing'' with fictional profiles varying demographics
- Self-reported voluntary disclosure: Collect demographic data through optional surveys separate from decision process
- Statistical inference: Infer approximate demographic distributions from public data (Census)

\textbf{Quebec Law 25 compliance}: While Law 25 doesn't mandate specific fairness metrics, organizations should document: fairness testing methodology, metrics chosen and justification, test results and any disparities found, mitigation measures implemented, ongoing monitoring approach. This documentation satisfies Law 25's accuracy and transparency requirements when properly contextualized.

\paragraph{4.7.2 Explainability and Interpretability}\label{explainability-and-interpretability}

\textbf{Method selection by context}: Different explainability techniques serve different purposes:

\emph{SHAP (SHapley Additive exPlanations)}: Provides feature importance scores based on game theory. Strengths: theoretically grounded, consistent, locally accurate. Weaknesses: computationally expensive, assumes feature independence, requires many background samples. Use when: need rigorous mathematical foundation, explaining complex ensemble models, accuracy more important than speed, stakeholders trust mathematical approaches.

\emph{LIME (Local Interpretable Model-agnostic Explanations)}: Fits simple local model around prediction. Strengths: model-agnostic, fast, intuitive. Weaknesses: unstable (small input changes yield different explanations), local approximation may not reflect global behavior, sampling approach affects results. Use when: need quick explanations, model is black box, explaining individual decisions, stakeholders prefer intuitive approaches.

\emph{Counterfactual explanations}: Shows minimal changes to input yielding different outcome (``If income were \$5,000 higher, loan approved''). Strengths: actionable, align with human reasoning (``what if''), satisfy recourse requirements. Weaknesses: may not be achievable (can't change race), multiple counterfactuals possible, computationally intensive. Use when: Quebec Law 25 or Treasury Board requires explanation of factors, providing recourse guidance, stakeholders need actionable information.

\emph{Attention mechanisms}: For deep learning, show which input parts model ``attended to.'' Strengths: built into architecture, provides visual explanations, efficient. Weaknesses: attention != explanation (attended features may not be causal), difficult to validate, only for specific architectures. Use when: working with transformers or attention-based models, visual explanations valuable (images, text), technical audience.

\textbf{Global vs.~local explanations}: Global explanations describe overall model behavior (feature importance across all predictions), while local explanations describe specific decisions. SCITUS requires:
- Level I-II systems: Local explanations for contested decisions
- Level III systems: Local explanations for all decisions + global explanations
- Level IV systems: Local + global + explanation validation + regular audits

\textbf{Post-hoc vs.~inherent interpretability}: Trade-off between model performance and interpretability. Inherently interpretable models (linear regression, decision trees, rule lists) provide direct explanations but may have lower accuracy. Black-box models (deep neural networks, gradient boosting) offer higher accuracy but require post-hoc explanation. SCITUS guidance:
- Level I-II: Either approach acceptable based on accuracy requirements
- Level III: If using black-box model, must provide validated post-hoc explanations
- Level IV: Strong preference for inherently interpretable models OR black-box with rigorous explanation validation including human studies confirming explanation utility

\textbf{Explanation validation}: Many explainability methods lack ground truth. Validation approaches:
- Simulation studies: Use synthetic data with known generating process, verify explanations match true process
- Human studies: Test whether explanations help users make better decisions (user accuracy, appropriate reliance)
- Consistency checks: Verify explanations are consistent across similar instances
- Sanity checks: Ensure explanations align with domain knowledge (e.g., age affects insurance premiums positively)

\textbf{Regulatory alignment}: Treasury Board Level III-IV requires ``explanation capability.'' SCITUS operationalizes this as: identify explanation method appropriate for use case, validate explanations improve human decision-making (for Level IV), document explanation approach in AIA, provide explanations in plain language (not technical jargon), test explanations with representative users. Quebec Law 25 requires providing ``information about personal information used and factors involved in decisions''---counterfactual explanations particularly suitable.

\paragraph{4.7.3 AI System Robustness and Security}\label{ai-system-robustness-and-security}

\textbf{Adversarial robustness}: AI systems face adversarial attacks where inputs are manipulated to cause misclassification. Attack vectors include:
- Evasion attacks: Modify inputs at test time to fool model (e.g., adversarial examples in images)
- Poisoning attacks: Inject malicious training data to backdoor model
- Model inversion: Extract training data from model (privacy violation)
- Membership inference: Determine if specific data point was in training set

Defenses: adversarial training (include adversarial examples in training), input validation (detect out-of-distribution inputs), model hardening (certified defenses with robustness guarantees), monitoring (detect unusual input patterns). SCITUS requirements: Level II-IV systems must implement adversarial robustness testing, Level III-IV must deploy adversarial defenses proportional to threat model, documented threat assessment identifying likely attack vectors.

\textbf{Data poisoning and supply chain risks}: Training data poisoning can compromise model integrity. Mitigation:
- Data provenance: Track data sources, validate authenticity
- Statistical validation: Detect outliers and suspicious patterns in training data
- Differential privacy: Add noise to training limiting influence of individual data points
- Validation set monitoring: Performance degradation on trusted validation set indicates poisoning
For models using third-party training data or pre-trained models: verify provider security practices, test on known-good validation data, monitor for suspicious behavior (backdoors).

\textbf{Model drift detection and monitoring}: AI system performance degrades over time due to data distribution shift. Types:
- Covariate shift: Input distribution changes (e.g., demographics of applicants shift)
- Concept drift: Relationship between features and outcomes changes (e.g., economic conditions alter creditworthiness patterns)
- Label shift: Output distribution changes (e.g., more fraud attempts)

Detection approaches: monitor prediction distribution for shifts, track accuracy on recent data with ground truth, statistical tests for distribution changes (Kolmogorov-Smirnov test), model retraining triggers (accuracy drops below threshold). SCITUS requirements: Level II-IV must implement drift detection, Level III-IV must define retraining schedules and drift response procedures, quarterly drift reports to governance committee.

\textbf{Distribution shift and deployment mismatch}: Models trained on one population may fail when deployed to different population. Mitigation:
- Domain adaptation: Techniques to transfer model to new distribution
- Regular retraining: Update model with deployment distribution data
- Careful validation: Test on data representative of deployment population
- Monitoring: Track performance across demographic/geographic/temporal slices

\textbf{Practical example - Immigration scenario}: The federal immigration AI (Section 6.2) must address: adversarial robustness against fraudulent applications, data poisoning through compromised data sources, drift as immigration patterns change, distribution shift when applied to new visa categories. Implementation: adversarial training on known fraud patterns, data provenance tracking for all sources, monthly drift detection on approval rates and processing times, separate model validation for each major visa category.

\textbf{Integration with SCITUS MEASURE function}: These technical controls map to MEASURE function outcomes:
- MEASURE 2.1: Bias testing and fairness metrics -\textgreater{} ``AI risks and benefits\ldots are examined and documented''
- MEASURE 2.3: Explainability implementation -\textgreater{} ``Transparency and accountability capabilities are implemented''
- MEASURE 2.8: Adversarial robustness -\textgreater{} ``Risks of AI system misuse\ldots are examined''
- MEASURE 4.2: Model drift detection -\textgreater{} ``AI systems are evaluated\ldots throughout their lifecycle''

This technical guidance operationalizes SCITUS framework requirements, providing concrete direction for AI system development teams while maintaining regulatory compliance across Canadian jurisdictions.

\subsubsection{4.8 Technical Implementation Architecture and Tooling}\label{technical-implementation-architecture-and-tooling}

SCITUS controls require supporting infrastructure for bias testing, monitoring, explainability, and compliance tracking. This section provides reference architecture and specific tooling recommendations addressing practical implementation needs.

\paragraph{4.8.1 Bias Testing and Fairness Infrastructure}\label{bias-testing-and-fairness-infrastructure}

\textbf{Recommended Open-Source Tools:}

\textbf{Fairlearn (Microsoft)} - Python library for fairness assessment and mitigation
- \textbf{Use for}: Fairness metric calculation (demographic parity, equalized odds, predictive parity)
- \textbf{Integration}: Scikit-learn compatible, integrates into existing ML pipelines
- \textbf{Installation}: \texttt{pip\ install\ fairlearn}
- \textbf{Key capabilities}: MetricFrame for stratified performance analysis, mitigation algorithms (GridSearch, ExponentiatedGradient), interactive dashboard

\textbf{AI Fairness 360 (IBM)} - Comprehensive bias detection and mitigation toolkit
- \textbf{Use for}: Wide range of fairness metrics (70+ algorithms), bias mitigation, educational resources
- \textbf{Integration}: Supports TensorFlow, PyTorch, scikit-learn
- \textbf{Installation}: \texttt{pip\ install\ aif360}
- \textbf{Key capabilities}: Most comprehensive metric library, multiple mitigation strategies (pre-processing, in-processing, post-processing)

\textbf{Implementation Example} (bias testing in deployment pipeline):

\begin{Shaded}
\begin{Highlighting}[]
\ImportTok{from}\NormalTok{ fairlearn.metrics }\ImportTok{import}\NormalTok{ MetricFrame, demographic\_parity\_difference}
\ImportTok{from}\NormalTok{ sklearn.metrics }\ImportTok{import}\NormalTok{ accuracy\_score, precision\_score}

\KeywordTok{def}\NormalTok{ scitus\_bias\_test(y\_true, y\_pred, sensitive\_features):}
    \CommentTok{"""}
\CommentTok{    SCITUS{-}compliant bias testing (MEAS{-}1.1 control)}
\CommentTok{    Returns: pass/fail + detailed metrics for governance reporting}
\CommentTok{    """}
    \CommentTok{\# Calculate performance metrics stratified by sensitive attribute}
\NormalTok{    metric\_frame }\OperatorTok{=}\NormalTok{ MetricFrame(}
\NormalTok{        metrics}\OperatorTok{=}\NormalTok{\{}\StringTok{\textquotesingle{}accuracy\textquotesingle{}}\NormalTok{: accuracy\_score, }\StringTok{\textquotesingle{}precision\textquotesingle{}}\NormalTok{: precision\_score\},}
\NormalTok{        y\_true}\OperatorTok{=}\NormalTok{y\_true,}
\NormalTok{        y\_pred}\OperatorTok{=}\NormalTok{y\_pred,}
\NormalTok{        sensitive\_features}\OperatorTok{=}\NormalTok{sensitive\_features}
\NormalTok{    )}

    \CommentTok{\# Calculate disparity (Section 6.2 uses 5\% threshold)}
\NormalTok{    disparity }\OperatorTok{=}\NormalTok{ demographic\_parity\_difference(}
\NormalTok{        y\_true, y\_pred, sensitive\_features}\OperatorTok{=}\NormalTok{sensitive\_features}
\NormalTok{    )}

\NormalTok{    SCITUS\_THRESHOLD }\OperatorTok{=} \FloatTok{0.05}  \CommentTok{\# 5 percentage points (configurable per use case)}

    \ControlFlowTok{return}\NormalTok{ \{}
        \StringTok{\textquotesingle{}compliant\textquotesingle{}}\NormalTok{: }\BuiltInTok{abs}\NormalTok{(disparity) }\OperatorTok{\textless{}=}\NormalTok{ SCITUS\_THRESHOLD,}
        \StringTok{\textquotesingle{}disparity\textquotesingle{}}\NormalTok{: disparity,}
        \StringTok{\textquotesingle{}metrics\_by\_group\textquotesingle{}}\NormalTok{: metric\_frame.by\_group,}
        \StringTok{\textquotesingle{}control\_id\textquotesingle{}}\NormalTok{: }\StringTok{\textquotesingle{}MEAS{-}1.1\textquotesingle{}}\NormalTok{,}
        \StringTok{\textquotesingle{}recommendation\textquotesingle{}}\NormalTok{: }\StringTok{\textquotesingle{}PASS\textquotesingle{}} \ControlFlowTok{if} \BuiltInTok{abs}\NormalTok{(disparity) }\OperatorTok{\textless{}=}\NormalTok{ SCITUS\_THRESHOLD }\ControlFlowTok{else} \StringTok{\textquotesingle{}INVESTIGATE\textquotesingle{}}
\NormalTok{    \}}
\end{Highlighting}
\end{Shaded}

\textbf{CI/CD Integration}: Add bias testing as automated gate in deployment pipeline, blocking deployment if disparity exceeds threshold, generating compliance report for governance committee review.

\paragraph{4.8.2 Explainability Infrastructure}\label{explainability-infrastructure}

\textbf{SHAP (SHapley Additive exPlanations)}
- \textbf{Use for}: High-stakes decisions requiring rigorous explanations (Level III-IV systems)
- \textbf{Installation}: \texttt{pip\ install\ shap}
- \textbf{Strengths}: Theoretically grounded (Shapley values from game theory), consistent, locally accurate
- \textbf{Implementation}:

\begin{Shaded}
\begin{Highlighting}[]
\ImportTok{import}\NormalTok{ shap}

\KeywordTok{def}\NormalTok{ scitus\_generate\_explanation(model, X\_train, X\_test, instance\_idx):}
    \CommentTok{"""}
\CommentTok{    SCITUS{-}compliant explanation (MNG{-}1.2 control)}
\CommentTok{    Satisfies Quebec Law 25 / Treasury Board transparency requirements}
\CommentTok{    """}
\NormalTok{    explainer }\OperatorTok{=}\NormalTok{ shap.Explainer(model, X\_train)}
\NormalTok{    shap\_values }\OperatorTok{=}\NormalTok{ explainer(X\_test)}

    \CommentTok{\# Generate explanation for specific instance}
\NormalTok{    explanation }\OperatorTok{=}\NormalTok{ \{}
        \StringTok{\textquotesingle{}decision\textquotesingle{}}\NormalTok{: model.predict(X\_test[instance\_idx:instance\_idx}\OperatorTok{+}\DecValTok{1}\NormalTok{])[}\DecValTok{0}\NormalTok{],}
        \StringTok{\textquotesingle{}top\_factors\textquotesingle{}}\NormalTok{: get\_top\_factors(shap\_values[instance\_idx], feature\_names),}
        \StringTok{\textquotesingle{}factor\_directions\textquotesingle{}}\NormalTok{: get\_factor\_directions(shap\_values[instance\_idx]),}
        \StringTok{\textquotesingle{}data\_sources\textquotesingle{}}\NormalTok{: }\StringTok{\textquotesingle{}Applicant profile, historical outcomes (2020{-}2024)\textquotesingle{}}\NormalTok{,}
        \StringTok{\textquotesingle{}appeal\_process\textquotesingle{}}\NormalTok{: }\StringTok{\textquotesingle{}Contact governance@organization.ca within 30 days\textquotesingle{}}\NormalTok{,}
        \StringTok{\textquotesingle{}control\_id\textquotesingle{}}\NormalTok{: }\StringTok{\textquotesingle{}MNG{-}1.2\textquotesingle{}}\NormalTok{,}
        \StringTok{\textquotesingle{}generated\_at\textquotesingle{}}\NormalTok{: datetime.now().isoformat()}
\NormalTok{    \}}
    \ControlFlowTok{return}\NormalTok{ explanation}
\end{Highlighting}
\end{Shaded}

\textbf{LIME (Local Interpretable Model-agnostic Explanations)}
- \textbf{Use for}: Quick explanations, black-box models, Level II systems
- \textbf{Installation}: \texttt{pip\ install\ lime}
- \textbf{Trade-offs}: Faster than SHAP but less stable, suitable for moderate-impact decisions

\textbf{Counterfactual Explanations (DiCE - Diverse Counterfactual Explanations)}
- \textbf{Use for}: Recourse guidance (``what would need to change for different outcome'')
- \textbf{Installation}: \texttt{pip\ install\ dice-ml}
- \textbf{Best for}: Quebec Law 25 right to human review, providing actionable feedback to individuals

\paragraph{4.8.3 Monitoring and Drift Detection}\label{monitoring-and-drift-detection}

\textbf{Prometheus + Grafana} - Time-series monitoring and visualization
- \textbf{Use for}: Performance metrics, latency, throughput, error rates (MEAS-1.2)
- \textbf{Deployment}: Open source, self-hosted or Grafana Cloud
- \textbf{SCITUS metrics to track}:
- Model accuracy/precision/recall over time
- Prediction distribution shifts
- Override rates (human rejecting AI recommendation)
- Appeal rates and outcomes
- System availability and response time

\textbf{Evidently AI} - ML-specific monitoring and drift detection
- \textbf{Use for}: Data drift, concept drift, model performance degradation
- \textbf{Installation}: \texttt{pip\ install\ evidently}
- \textbf{Features}: Statistical tests (Kolmogorov-Smirnov, chi-squared), visual reports, automated alerts

\textbf{Example drift detection}:

\begin{Shaded}
\begin{Highlighting}[]
\ImportTok{from}\NormalTok{ evidently.metric\_preset }\ImportTok{import}\NormalTok{ DataDriftPreset}
\ImportTok{from}\NormalTok{ evidently.report }\ImportTok{import}\NormalTok{ Report}

\KeywordTok{def}\NormalTok{ scitus\_drift\_monitoring(reference\_data, current\_data):}
    \CommentTok{"""}
\CommentTok{    SCITUS drift detection (MEAS{-}1.2, triggers MNG{-}4.1 reassessment)}
\CommentTok{    """}
\NormalTok{    report }\OperatorTok{=}\NormalTok{ Report(metrics}\OperatorTok{=}\NormalTok{[DataDriftPreset()])}
\NormalTok{    report.run(reference\_data}\OperatorTok{=}\NormalTok{reference\_data, current\_data}\OperatorTok{=}\NormalTok{current\_data)}

\NormalTok{    drift\_detected }\OperatorTok{=}\NormalTok{ report.as\_dict()[}\StringTok{\textquotesingle{}metrics\textquotesingle{}}\NormalTok{][}\DecValTok{0}\NormalTok{][}\StringTok{\textquotesingle{}result\textquotesingle{}}\NormalTok{][}\StringTok{\textquotesingle{}dataset\_drift\textquotesingle{}}\NormalTok{]}

    \ControlFlowTok{if}\NormalTok{ drift\_detected:}
        \CommentTok{\# Trigger SCITUS control MNG{-}4.1 (system reassessment required)}
\NormalTok{        trigger\_governance\_review(}
\NormalTok{            system\_id}\OperatorTok{=}\StringTok{\textquotesingle{}immigration{-}triage{-}v2\textquotesingle{}}\NormalTok{,}
\NormalTok{            alert\_type}\OperatorTok{=}\StringTok{\textquotesingle{}DATA\_DRIFT\textquotesingle{}}\NormalTok{,}
\NormalTok{            report}\OperatorTok{=}\NormalTok{report}
\NormalTok{        )}

    \ControlFlowTok{return}\NormalTok{ \{}\StringTok{\textquotesingle{}drift\_detected\textquotesingle{}}\NormalTok{: drift\_detected, }\StringTok{\textquotesingle{}report\textquotesingle{}}\NormalTok{: report\}}
\end{Highlighting}
\end{Shaded}

\textbf{Alert Configuration}:
- Accuracy drops \textgreater3\%: Notify system owner
- Accuracy drops \textgreater5\%: Escalate to governance committee
- Drift detected: Trigger impact level reassessment (SCITUS MNG-4.1)
- Bias disparity exceeds threshold: Immediate investigation

\paragraph{4.8.4 Compliance Tracking and Documentation}\label{compliance-tracking-and-documentation}

\textbf{GRC Platform Options:}

\textbf{OneTrust} - Comprehensive privacy and AI governance platform
- \textbf{Features}: AI system inventory, automated assessments, policy management, audit trails
- \textbf{Cost}: Enterprise pricing (\textasciitilde\$50,000-\$200,000/year)
- \textbf{Best for}: Large organizations with complex compliance requirements

\textbf{ServiceNow GRC} - Integrated risk and compliance management
- \textbf{Features}: Control tracking, evidence management, workflow automation, integration with IT systems
- \textbf{Cost}: Mid-market pricing (\textasciitilde\$30,000-\$100,000/year)
- \textbf{Best for}: Organizations already using ServiceNow

\textbf{Lightweight Alternative (Airtable/Notion)}:
- \textbf{Features}: Custom AI system inventory, control checklists, document repository
- \textbf{Cost}: Low (\textasciitilde\$1,000-\$5,000/year for tools + development)
- \textbf{Best for}: Small organizations, startups, pilot implementations

\textbf{SCITUS Inventory Schema (minimum required fields)}:

\begin{Shaded}
\begin{Highlighting}[]
\FunctionTok{\{}
  \DataTypeTok{"system\_id"}\FunctionTok{:} \StringTok{"immigration{-}triage{-}2024"}\FunctionTok{,}
  \DataTypeTok{"system\_name"}\FunctionTok{:} \StringTok{"IRCC Visa Application Triage System"}\FunctionTok{,}
  \DataTypeTok{"purpose"}\FunctionTok{:} \StringTok{"Prioritize visa applications for officer review"}\FunctionTok{,}
  \DataTypeTok{"impact\_level"}\FunctionTok{:} \StringTok{"III"}\FunctionTok{,}
  \DataTypeTok{"accountable\_official"}\FunctionTok{:} \StringTok{"Director, Digital Services (J. Smith)"}\FunctionTok{,}
  \DataTypeTok{"deployment\_status"}\FunctionTok{:} \StringTok{"Production"}\FunctionTok{,}
  \DataTypeTok{"deployment\_date"}\FunctionTok{:} \StringTok{"2024{-}06{-}15"}\FunctionTok{,}
  \DataTypeTok{"last\_aia\_date"}\FunctionTok{:} \StringTok{"2024{-}05{-}20"}\FunctionTok{,}
  \DataTypeTok{"next\_review\_date"}\FunctionTok{:} \StringTok{"2025{-}06{-}15"}\FunctionTok{,}
  \DataTypeTok{"applicable\_jurisdictions"}\FunctionTok{:} \OtherTok{[}\StringTok{"Federal"}\OtherTok{,} \StringTok{"Quebec"}\OtherTok{]}\FunctionTok{,}
  \DataTypeTok{"scitus\_controls"}\FunctionTok{:} \OtherTok{[}
    \FunctionTok{\{}\DataTypeTok{"control\_id"}\FunctionTok{:} \StringTok{"GOV{-}1.3"}\FunctionTok{,} \DataTypeTok{"status"}\FunctionTok{:} \StringTok{"Compliant"}\FunctionTok{,} \DataTypeTok{"evidence"}\FunctionTok{:} \StringTok{"docs/accountability{-}framework.pdf"}\FunctionTok{\}}\OtherTok{,}
    \FunctionTok{\{}\DataTypeTok{"control\_id"}\FunctionTok{:} \StringTok{"MEAS{-}1.1"}\FunctionTok{,} \DataTypeTok{"status"}\FunctionTok{:} \StringTok{"Partial"}\FunctionTok{,} \DataTypeTok{"evidence"}\FunctionTok{:} \StringTok{"Bias testing in progress"}\FunctionTok{,} \DataTypeTok{"target\_date"}\FunctionTok{:} \StringTok{"2025{-}01{-}15"}\FunctionTok{\}}\OtherTok{,}
    \FunctionTok{\{}\DataTypeTok{"control\_id"}\FunctionTok{:} \StringTok{"MNG{-}1.2"}\FunctionTok{,} \DataTypeTok{"status"}\FunctionTok{:} \StringTok{"Compliant"}\FunctionTok{,} \DataTypeTok{"evidence"}\FunctionTok{:} \StringTok{"Explanation API deployed"}\FunctionTok{\}}
  \OtherTok{]}\FunctionTok{,}
  \DataTypeTok{"data\_sources"}\FunctionTok{:} \OtherTok{[}\StringTok{"Applicant submissions"}\OtherTok{,} \StringTok{"Historical decisions"}\OtherTok{,} \StringTok{"Risk databases"}\OtherTok{]}\FunctionTok{,}
  \DataTypeTok{"affected\_population\_size"}\FunctionTok{:} \DecValTok{2000000}\FunctionTok{,}
  \DataTypeTok{"automated\_decision\_type"}\FunctionTok{:} \StringTok{"Recommendation (human final decision)"}
\FunctionTok{\}}
\end{Highlighting}
\end{Shaded}

\paragraph{4.8.5 End-to-End Reference Architecture}\label{end-to-end-reference-architecture}

\begin{figure}
\centering
\includegraphics[width=0.92\textwidth,height=\textheight]{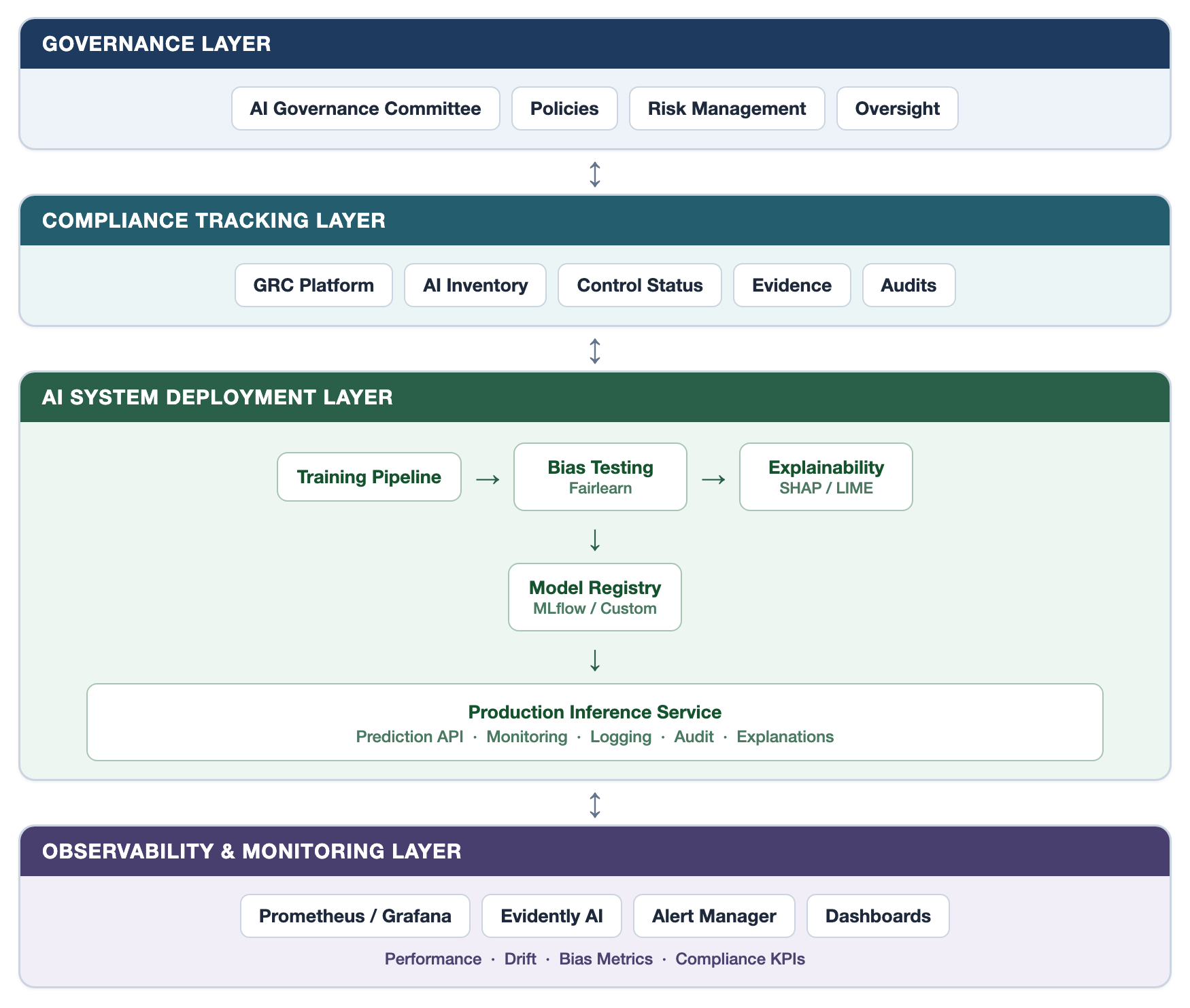}
\caption{Layered reference architecture for SCITUS compliance: governance and compliance-tracking layers oversee an AI system deployment pipeline (training, bias testing, explainability, model registry, production inference), with a cross-cutting observability and monitoring layer.}
\end{figure}

\textbf{Integration Points:}
1. \textbf{Training -\textgreater{} Bias Testing}: Automated tests run on validation data before model promotion
2. \textbf{Model Registry -\textgreater{} Governance}: New models require governance approval for Level III-IV systems
3. \textbf{Production -\textgreater{} Monitoring}: Real-time metrics feed compliance dashboards and governance reports
4. \textbf{Monitoring -\textgreater{} Governance}: Alerts trigger governance review when thresholds exceeded
5. \textbf{All Layers -\textgreater{} GRC Platform}: Evidence, audit trails, and compliance status centralized

\paragraph{4.8.6 Vendor Management and Third-Party AI}\label{vendor-management-and-third-party-ai}

For organizations using third-party AI (APIs, SaaS, vendor solutions):

\textbf{Vendor Assessment RFP Template Section:}

\begin{Shaded}
\begin{Highlighting}[]
\FunctionTok{\#\# SCITUS AI Governance Requirements for Vendors}

\SpecialStringTok{1. }\NormalTok{**Governance (GOV):**}
\SpecialStringTok{   {-} }\NormalTok{Provide evidence of AI governance structure and accountable officials}
\SpecialStringTok{   {-} }\NormalTok{Share AI ethics policy and risk management procedures}
\SpecialStringTok{   {-} }\NormalTok{Document decision{-}making authority and escalation processes}

\SpecialStringTok{2. }\NormalTok{**Risk Assessment (MAP):**}
\SpecialStringTok{   {-} }\NormalTok{Provide completed impact assessment (AIA equivalent or similar)}
\SpecialStringTok{   {-} }\NormalTok{Share bias testing methodology and results}
\SpecialStringTok{   {-} }\NormalTok{Document known limitations and failure modes}

\SpecialStringTok{3. }\NormalTok{**Technical Controls (MEAS):**}
\SpecialStringTok{   {-} }\NormalTok{Provide model performance metrics (accuracy, precision, recall) overall and stratified by demographics}
\SpecialStringTok{   {-} }\NormalTok{Demonstrate monitoring and drift detection capabilities}
\SpecialStringTok{   {-} }\NormalTok{Provide explainability mechanism for individual decisions}

\SpecialStringTok{4. }\NormalTok{**Transparency and Recourse (MNG):**}
\SpecialStringTok{   {-} }\NormalTok{Document decision{-}making logic at appropriate detail level}
\SpecialStringTok{   {-} }\NormalTok{Specify recourse mechanisms for contested decisions}
\SpecialStringTok{   {-} }\NormalTok{Commit to notification of model updates affecting decision logic (30 days advance)}

\NormalTok{**Contractual Provisions:**}
\SpecialStringTok{{-} }\NormalTok{Right to audit vendor AI systems annually (or access third{-}party audit reports)}
\SpecialStringTok{{-} }\NormalTok{Notification of material changes to AI models (30 days minimum)}
\SpecialStringTok{{-} }\NormalTok{Service level agreements for accuracy, fairness metrics, explainability response time}
\SpecialStringTok{{-} }\NormalTok{Data processing agreement compliant with PIPEDA/Law 25}
\SpecialStringTok{{-} }\NormalTok{Vendor liability for discriminatory outcomes due to vendor system defects}
\end{Highlighting}
\end{Shaded}

\textbf{Vendor Risk Scorecard} (100-point scale):
- Governance maturity: 25 points
- Technical transparency: 25 points
- Compliance documentation: 25 points
- Track record and references: 25 points
- \textbf{Minimum score for high-risk systems: 75/100}

\paragraph{4.8.7 Tooling Summary and Recommendations}\label{tooling-summary-and-recommendations}

\textbf{Essential Tools (all organizations)}:
- AI system inventory (Airtable/Notion for small, GRC platform for large)
- Bias testing library (Fairlearn or AIF360)
- Basic monitoring (Prometheus + Grafana or cloud-native)
- Documentation templates (SCITUS compliance matrix, AIA forms)

\textbf{Recommended for Level III-IV Systems}:
- Explainability framework (SHAP for rigor, LIME for speed)
- Advanced monitoring (Evidently AI for drift detection)
- Compliance tracking platform (OneTrust, ServiceNow GRC, or similar)
- Incident management integration

\textbf{Cost-Effective Open Source Stack}:
- Fairlearn (bias testing) + SHAP (explainability) + Prometheus/Grafana (monitoring) + Evidently AI (drift) + Airtable (inventory) = \textasciitilde\$5,000/year in platform costs + implementation effort

\textbf{Enterprise Commercial Stack}:
- OneTrust (compliance) + commercial monitoring platform + enterprise GRC = \textasciitilde\$100,000-\$300,000/year for comprehensive coverage

\textbf{Implementation Priority}:
1. \textbf{Phase 1}: AI system inventory, basic bias testing
2. \textbf{Phase 2}: Monitoring infrastructure, compliance tracking
3. \textbf{Phase 3}: Explainability framework, advanced analytics
4. \textbf{Phase 4}: Automation, integration, optimization

This tooling guidance operationalizes SCITUS framework requirements with specific, actionable technology recommendations enabling organizations to implement controls efficiently and cost-effectively.

\subsubsection{4.9 Framework Evolution: The SCITUS Control Catalog, v1.0--v2.0}\label{framework-evolution-the-scitus-control-catalog-v1.0v2.0}

While Sections 4.1--4.8 describe the framework's stable architecture, SCITUS also maintains a \textbf{versioned control catalog} that evolves with the regulatory and threat landscape. This subsection documents that evolution, which itself illustrates the framework's ``future-proof adaptability'' design principle (Section 4.1) in operation.

\textbf{Table 10: SCITUS Control Catalog Version History}

\begin{longtable}[]{@{}
  >{\raggedright\arraybackslash}p{(\columnwidth - 6\tabcolsep) * \real{0.2250}}
  >{\raggedright\arraybackslash}p{(\columnwidth - 6\tabcolsep) * \real{0.1500}}
  >{\raggedright\arraybackslash}p{(\columnwidth - 6\tabcolsep) * \real{0.2500}}
  >{\raggedright\arraybackslash}p{(\columnwidth - 6\tabcolsep) * \real{0.3750}}@{}}
\toprule\noalign{}
\begin{minipage}[b]{\linewidth}\raggedright
Version
\end{minipage} & \begin{minipage}[b]{\linewidth}\raggedright
Date
\end{minipage} & \begin{minipage}[b]{\linewidth}\raggedright
Controls
\end{minipage} & \begin{minipage}[b]{\linewidth}\raggedright
Key Additions
\end{minipage} \\
\midrule\noalign{}
\endhead
\bottomrule\noalign{}
\endlastfoot
v1.0 & June 2025 & 31 & Initial release: 127 requirements mapped to 31 controls across GOVERN/MAP/MEASURE/MANAGE \\
v1.1 & November 2025 & 35 & Adversarial robustness testing, content provenance, mandatory incident reporting, effectiveness measurement; international framework compatibility mapping \\
v1.2 & December 2025 & 45 & Open-weight model addendum, agentic AI safety, data poisoning prevention, tamper resistance, reasoning transparency, input/output safety pipelines, hardware-level security \\
v1.3 & January 2026 & 51 & System prompt security, multimodal safety, SCITUS-Lite SME pathway, community engagement, international compliance reporting, vendor AI governance; ISO/IEC 42001 crosswalk \\
v1.4 & February 2026 & 53 & Deployment integrity \& version transparency, human oversight \& controllability \\
v2.0 & July 2026 & 57 & Training data governance \& provenance, agent identity \& delegated authorization, agentic tool-chain security, AI supply chain integrity; 9 control enhancements \\
\end{longtable}

\textbf{The v2.0 release} (concurrent with this preprint) responds to four developments in the January--July 2026 window, each traceable to primary sources:

\begin{enumerate}
\def\labelenumi{\arabic{enumi}.}
\item
  \textbf{Training data governance (new control TDG-1).} The joint findings of Canada's four privacy regulators against OpenAI (PIPEDA Findings \#2026-002, May 6, 2026 {[}40{]}) constitute the first Canadian regulatory standard for generative-AI training data: collection of publicly accessible personal information for model training was found overbroad and without valid consent, with training-data filtering/masking and transparency disclosures as the resolution template. TDG-1 operationalizes these findings as assessment criteria --- lawful collection basis, sensitive-information filtering, proportionality assessment, intellectual-property provenance, poisoning screening, and public transparency.
\item
  \textbf{Agent identity and authorization (new control AGT-1).} The first multinational agentic-AI security guidance --- ``Careful Adoption of Agentic AI Services'' (May 1, 2026), co-sealed by Canada's Centre for Cyber Security with its US, UK, Australian, and New Zealand counterparts {[}44{]} --- expects each agent to carry a verified, cryptographically anchored identity with short-lived credentials. NIST's AI Agent Standards Initiative (February 2026) makes agent identity infrastructure a standards priority {[}45{]}. AGT-1 requires unique agent identities, scoped short-lived credentials, bounded delegated authority, signed agent manifests, auditable delegation chains, and revocation capability.
\item
  \textbf{Agentic tool-chain and AI supply-chain security (new controls AGT-2, SUP-1).} The first half of 2026 produced documented exploitation of the agent tool chain (systemic MCP SDK flaws, poisoned tool descriptions, trojanized tool servers) and of the AI artifact supply chain (model-registry namespace hijacking, malicious model files executing code, AI-credential-targeting package compromises) --- consistent with the International AI Safety Report 2026's findings on rapid autonomy gains and a widening safeguard gap {[}47{]}. AGT-2 governs tool-server vetting, sandboxing, and enforced human interrupts; SUP-1 requires hash/signature pinning of model artifacts, AI bills of materials, and registry namespace monitoring.
\item
  \textbf{Regulatory and jurisprudential integration (9 enhanced controls).} Bill C-36's ``legal or similarly significant effect'' explanation trigger is pre-mapped into the regulatory-mapping control {[}41{]}; Manitoba Bills 49/51 add a sixth jurisdiction module {[}42{]}; and Canadian courts' 2026 holdings that AI may assist but not drive adjudication (\emph{ARIHQ v. Santé Québec}, 2026 QCCS 1360; \emph{Sinkova v. Canada}, 2026 FC 650) are absorbed into the human-oversight control as requirements for documented human decisional authority. The international-compatibility mapping was re-versioned for the G7 Hiroshima AI Process Reporting Framework v2.0 {[}46{]} and the EU's Digital Omnibus timeline amendment {[}48{]}.
\end{enumerate}

The complete v2.0 control catalog, per-control assessment criteria with evidence-weighted scoring, and changelogs for each version are maintained in the framework's public documentation.

\subsection{5. Implementation Methodology}\label{implementation-methodology}

SCITUS implementation follows a phased approach enabling organizations to establish governance, assess systems, implement controls, and optimize operations systematically. This section describes the implementation process, assessment protocol, and practical tools.

\subsubsection{5.1 Four-Phase Implementation Approach}\label{four-phase-implementation-approach}

\textbf{Phase 1: Foundation (Months 1-3)}

\emph{Objective}: Establish governance structures and baseline understanding of AI systems and requirements.

\emph{Key activities}: Form AI Governance Committee with executive sponsor, cross-functional members (legal, IT, business, ethics), and defined terms of reference. Conduct comprehensive AI system inventory identifying all systems (production, development, planned) using surveys, IT asset reviews, and stakeholder interviews. Complete initial stakeholder mapping including affected individuals, oversight bodies, and vulnerable populations. Conduct regulatory gap analysis comparing current state to SCITUS requirements across all applicable jurisdictions.

\emph{Deliverables}: Governance charter and committee operational, complete AI system inventory with 100\% coverage, stakeholder register with engagement plans, gap analysis report prioritizing remediation, baseline capability assessment.

\emph{Success criteria}: Governance committee meeting monthly, all AI systems inventoried and categorized, core policies drafted and under review, stakeholder engagement initiated with affected communities.

\textbf{Phase 2: Risk Assessment (Months 4-6)}

\emph{Objective}: Complete comprehensive risk assessments and determine compliance requirements for all systems.

\emph{Key activities}: Complete Algorithmic Impact Assessments using Treasury Board AIA tool for all systems, determining impact levels and required mitigations. Conduct detailed risk assessments identifying technical, operational, legal, and reputational risks. Perform multi-jurisdictional compliance mapping determining which federal and provincial requirements apply to each system. Develop risk treatment plans specifying controls, responsibilities, timelines, and resources for each identified risk.

\emph{Deliverables}: AIA reports for all systems with validated impact levels, comprehensive risk assessments documenting likelihood and impact, compliance matrices mapping requirements to controls for each system, prioritized risk treatment plans with resource requirements.

\emph{Success criteria}: 100\% of production systems assessed, impact levels validated by governance committee, high-risk systems have approved treatment plans, resource allocation secured for remediation.

\textbf{Phase 3: Controls Implementation (Months 7-12)}

\emph{Objective}: Deploy technical and operational controls addressing identified risks and requirements.

\emph{Key activities}: Deploy monitoring infrastructure for performance tracking and compliance monitoring. Implement bias testing framework with fairness metrics and automated testing. Launch transparency portal for public-facing systems. Conduct internal audits validating controls. Complete personnel training with role-based curricula. Perform year-one assessment.

\emph{Success criteria}: Level III-IV systems have monitoring and bias testing operational, transparency requirements satisfied, audit findings remediated, training completion exceeds 95\%, compliance documented.

\textbf{Phase 4: Optimization (Months 13-18)}

\emph{Objective}: Achieve operational excellence through automation and continuous improvement.

\emph{Key activities}: Automate assessment processes (AIA, compliance checking, documentation). Streamline workflows and integrate tools. Implement continuous improvement program with regular reviews.

\emph{Deliverables}: Assessment automation reducing time by 30\% target, streamlined workflows with documented efficiencies, integrated tool ecosystem, predictive monitoring alerts, continuous improvement plans.

\emph{Success criteria}: 30\% reduction in assessment time achieved, 50\% reduction in incidents year-over-year, 100\% compliance score across all jurisdictions, 25\% faster deployment for new AI systems through rapid assessment frameworks.

\subsubsection{5.2 Resource Requirements and Cost Considerations}\label{resource-requirements-and-cost-considerations}

SCITUS implementation requires investment in personnel, technology, and ongoing operations. Resource requirements scale with organizational size, number of AI systems, and system impact levels. This section provides planning estimates based on governance framework implementation benchmarks.

\paragraph{5.2.1 Personnel Requirements}\label{personnel-requirements}

\textbf{Governance and Oversight:}
- AI Governance Committee: Executive sponsor (0.1 FTE), 5-7 committee members meeting monthly (0.05 FTE each)
- Accountable Officials: One designated per AI system (0.1-0.3 FTE per system depending on impact level)
- Privacy/Legal expertise: 0.2-0.5 FTE for policy development, regulatory interpretation, and legal review
- Ethics expertise: 0.1-0.3 FTE for fairness assessments, bias testing oversight, and value-laden decisions

\textbf{Technical Implementation:}
- ML Engineers/Data Scientists: 0.5-1.0 FTE for bias testing frameworks, explainability implementation, monitoring infrastructure
- Security/IT: 0.3-0.5 FTE for access controls, infrastructure security, audit logging, vulnerability management
- Quality Assurance: 0.2-0.4 FTE for validation testing, control verification, deployment approval

\textbf{Documentation and Compliance:}
- Technical Writers/Compliance Specialists: 0.3-0.5 FTE for AIAs, PIAs, policies, accountability frameworks, transparency reports
- Training Development: 0.2 FTE initial (one-time), 0.05 FTE ongoing for updates and delivery

\textbf{Total FTE Estimates:}
- \textbf{Small organization (1-5 AI systems)}: 2-3 FTE initial year, 1-1.5 FTE ongoing
- \textbf{Medium organization (6-20 AI systems)}: 4-6 FTE initial year, 2-3 FTE ongoing
- \textbf{Large organization (21+ AI systems)}: 8-12 FTE initial year, 4-6 FTE ongoing

\paragraph{5.2.2 Technology Investments}\label{technology-investments}

\textbf{Governance and Risk Management:}
- GRC platform for compliance tracking: \$15,000-\$50,000/year (or leverage existing platform if available)
- AI system inventory management: \$5,000-\$20,000/year (may integrate with existing IT asset management systems)

\textbf{Technical Tools:}
- Bias testing frameworks (Fairlearn, AIF360): Open source (free) + integration effort (\$10,000-\$30,000 one-time)
- Explainability libraries (SHAP, LIME): Open source (free) + implementation effort (\$15,000-\$40,000 one-time)
- Monitoring and observability (Prometheus, Grafana, Evidently AI, or commercial): \$10,000-\$100,000/year depending on scale and vendor
- Adversarial robustness testing tools: Open source + compute resources (\$5,000-\$20,000/year)

\textbf{External Services:}
- Legal consultation for regulatory interpretation: \$10,000-\$50,000 (one-time for framework setup)
- Privacy Impact Assessments (if outsourced): \$5,000-\$15,000 per PIA
- Independent audits (Level IV systems): \$25,000-\$100,000 per audit annually
- Expert consultation (bias testing, explainability): \$15,000-\$40,000 as needed

\textbf{Total Technology Costs (Annual):}
- \textbf{Small organization}: \$30,000-\$80,000 initial year, \$20,000-\$50,000 ongoing
- \textbf{Medium organization}: \$80,000-\$200,000 initial year, \$50,000-\$120,000 ongoing
- \textbf{Large organization}: \$200,000-\$500,000 initial year, \$120,000-\$300,000 ongoing

\paragraph{5.2.3 Implementation Timeline and Effort}\label{implementation-timeline-and-effort}

Based on organizational governance framework implementation benchmarks (ISO 42001, NIST Cybersecurity Framework, SOC 2):

\textbf{Phase 1 (Months 1-3): Foundation}
- Governance committee establishment: 40-60 hours
- AI system inventory completion: 20-40 hours per system for discovery and documentation
- Policy development (AI Ethics, Data Governance): 80-120 hours
- Initial impact assessments: 8-16 hours per system (varies by complexity)

\textbf{Phase 2 (Months 4-6): Risk Assessment and Documentation}
- Complete AIAs for all systems: 8-20 hours per system (complexity-dependent, higher for Level III-IV)
- Privacy Impact Assessments for high-risk systems: 16-40 hours per PIA
- Risk management procedures documentation: 60-100 hours
- Stakeholder consultation and engagement: 40-80 hours

\textbf{Phase 3 (Months 7-9): Technical Controls}
- Bias testing infrastructure implementation: 80-200 hours (setup) + 20-40 hours per system
- Monitoring deployment (performance, drift, fairness): 100-200 hours (infrastructure) + 10-20 hours per system
- Explainability implementation: 40-80 hours per high-risk system
- Security controls validation: 60-100 hours across systems

\textbf{Phase 4 (Months 10-12): Operationalization}
- Training development and delivery: 60-100 hours (initial) + delivery time
- Transparency reporting and public documentation: 40-80 hours
- Internal audit and validation: 80-120 hours
- Process optimization and workflow integration: 60-100 hours

\textbf{Total Implementation Effort}: 600-1,500 hours for initial 12-month implementation (varies significantly by organization size and system count)

\textbf{Ongoing Annual Effort}: 300-600 hours for monitoring, reassessments, regulatory tracking, updates, training

\paragraph{5.2.4 Cost-Benefit Considerations}\label{cost-benefit-considerations}

\textbf{Compliance Value:}
- Avoids regulatory penalties (Quebec Law 25: up to 4\% global revenue or \$25M CAD; PIPEDA violations: up to \$100,000)
- Reduces legal risk from discriminatory outcomes, privacy breaches, unfair decisions
- Enables public sector procurement (Treasury Board compliance required for federal contracts)
- Facilitates provincial contracts (Ontario accountability frameworks increasingly expected)

\textbf{Operational Value:}
- Improved AI system reliability through systematic monitoring and drift detection
- Reduced reputational risk from AI failures, bias incidents, privacy breaches
- Better decision-making quality through governance oversight and stakeholder input
- Enhanced stakeholder trust (customers, employees, regulators, public)

\textbf{Efficiency Gains:}
- Unified multi-jurisdictional compliance (vs.~3-5x redundant jurisdiction-specific implementations)
- Leverages NIST AI RMF ecosystem (tools, training, best practices widely available)
- Systematic approach reduces ad-hoc remediation costs from compliance gaps or incidents
- Reusable assessments, documentation, and templates across multiple systems

\textbf{Risk Mitigation:}
- Early detection of bias, drift, security vulnerabilities before causing harm
- Structured incident response reducing impact severity and recovery time
- Legal defensibility through documented governance, risk assessments, mitigation measures
- Reduced insurance costs (some insurers offer discounts for strong AI governance)

\textbf{Note on Estimates}: These cost and effort estimates represent typical ranges based on governance framework implementation patterns. Actual requirements vary significantly based on organizational context, existing governance maturity, number and complexity of AI systems, jurisdictional applicability, and available internal expertise. Organizations should conduct detailed scoping assessments before committing resources.

\paragraph{5.2.5 Resource Optimization Strategies}\label{resource-optimization-strategies}

Organizations can optimize resource utilization through:

\textbf{Leverage Existing Infrastructure}: Integrate SCITUS with existing GRC platforms, risk management processes, privacy programs, security controls, and quality assurance systems rather than building parallel structures.

\textbf{Phased Implementation}: Implement Phase 1 Foundation broadly (lower resource intensity), then prioritize Phase 2-4 technical controls for highest-risk systems first, expanding incrementally as capability matures and lessons learned are incorporated.

\textbf{Open Source Tools}: Utilize open-source bias testing (Fairlearn, AIF360), explainability (SHAP, LIME), and monitoring (Prometheus, Grafana, Evidently AI) tools to minimize technology costs while investing in integration and expertise.

\textbf{Shared Services}: For multi-subsidiary or federated organizations, establish centralized governance expertise (legal, privacy, ethics, technical specialists) providing shared services to business units rather than duplicating capabilities.

\textbf{External Expertise Strategically}: Use external consultants for specialized needs (legal interpretation, complex bias testing, independent audits) while building internal capability for ongoing operations and routine assessments.

\textbf{Training Investment}: Invest in training internal teams (ML engineers, product managers, legal/compliance) on AI governance concepts, tools, and processes to reduce long-term external dependency and enable sustainable operations.

\subsubsection{5.3 SCITUS-AIA Assessment Protocol}\label{scitus-aia-assessment-protocol}

The SCITUS-AIA Assessment Protocol provides systematic methodology for assessing organizational AI systems against SCITUS requirements. The protocol comprises nine phases detailed in our extended protocol document; we summarize key elements here.

\textbf{Phase 1-2: System Discovery and AIA Completion} (2-3 weeks)

Identify all AI/ADM systems through IT asset reviews, interviews, and budget analysis. Classify systems by function, criticality, and user impact. Complete AIA questionnaire for each system with evidence validation: review source code for automation level, data usage, and processing logic; interview stakeholders about business purpose, decision impact, and user populations; analyze documentation for security controls, privacy measures, and oversight mechanisms. Document all responses with supporting evidence and rationale.

\textbf{Phase 3-4: Technical Assessment and Documentation Verification} (2-3 weeks)

Conduct source code review validating data handling, bias testing, security controls, explainability mechanisms, and monitoring implementation. Review architecture for high availability, scalability, disaster recovery, and integration security. Assess third-party components for license compliance, security vulnerabilities, and data residency. Verify documentation completeness comparing actual documentation to requirements matrix for applicable impact level and jurisdictions.

\textbf{Phase 5-6: Organizational Capability and Gap Analysis} (1-2 weeks)

Evaluate organizational readiness including governance maturity, roles and responsibilities clarity, training and competencies, resource adequacy, and process maturity. Identify gaps across technical (missing controls), documentation (incomplete records), process (undefined procedures), competency (skill deficiencies), and governance (accountability issues) categories. Prioritize gaps by risk score calculated as Impact x Likelihood x Detectability.

\textbf{Phase 7-9: Impact Level Determination, Validation, and Reporting} (1-2 weeks)

Calculate final impact level from AIA scores, validate with evidence review and worst-case analysis, document justification for borderline decisions. Conduct compliance validation gathering evidence for all requirements, testing implementations, documenting results, obtaining stakeholder approvals. Generate comprehensive reports including executive summary (compliance status, critical findings, recommendations), system assessments (one per system with impact level, findings, recommendations), organizational assessment (governance, capability, resources, maturity), gap analysis (prioritized with remediation plans), and compliance attestation.

\textbf{Timeline}: Typical assessment duration for medium organizations (6-15 systems) follows the phased structure above, with unified SCITUS-based assessment avoiding redundant separate assessments for each jurisdiction.

\subsubsection{5.4 Practical Tools and Templates}\label{practical-tools-and-templates}

SCITUS provides practical tools enabling efficient implementation:

\textbf{Multi-jurisdictional compliance matrix}: Spreadsheet template listing all requirements from all jurisdictions, mapping to SCITUS controls, tracking implementation status, documenting evidence locations, maintaining compliance scores. Organizations customize for their specific jurisdictional applicability.

\textbf{Risk assessment templates}: Structured formats for documenting risks including technical risks (model performance, adversarial attacks, system failures), operational risks (process integration, human-AI collaboration, change management), legal/regulatory risks (non-compliance, liability, breaches), reputational risks (public trust, media scrutiny, stakeholder concerns). Templates include risk scoring, treatment plans, and monitoring approaches.

\textbf{Accountability framework template}: Pre-structured document satisfying Ontario Bill 194 requirements, including governance structure specification, risk management approach, performance metrics and reporting, transparency measures, audit and review procedures, stakeholder engagement processes. Organizations customize with specific roles, metrics, and procedures.

\textbf{AIA completion guide}: Step-by-step guidance for Treasury Board AIA questionnaire with Canadian context, question-by-question interpretation, evidence requirements, code review checkpoints, validation methods, and impact level determination logic.

\textbf{Provincial requirement checklists}: Jurisdiction-specific checklists for Ontario (Bill 194 compliance), Quebec (Law 25 compliance), Alberta (Bills 33/34 compliance), with requirements, triggers, evidence needs, and deadlines.

\textbf{Sector-specific implementation guides}: Tailored guidance for government/public sector (federal requirements, provincial public sector, bilingual services), healthcare (Health Canada regulations, clinical validation, patient safety), financial services (OSFI guidelines, FCAC requirements, privacy), technology companies (B2B vs B2C, platform liability, developer guidelines), education (student privacy, accessibility, academic integrity).

\subsection{6. Framework Applicability and Demonstration}\label{framework-applicability-and-demonstration}

We demonstrate SCITUS applicability through three illustrative scenarios spanning federal government, provincial healthcare, and private sector contexts. These scenarios illustrate how the framework addresses multi-jurisdictional compliance challenges across diverse sectors and use cases, demonstrating the practical value of systematic NIST AI RMF adaptation to Canadian regulatory requirements.

\subsubsection{6.1 Demonstration Approach}\label{demonstration-approach}

\textbf{Framework applicability dimensions}:

\begin{enumerate}
\def\labelenumi{\arabic{enumi}.}
\item
  \textbf{Completeness}: SCITUS addresses regulatory requirements across applicable jurisdictions through systematic requirement mapping.
\item
  \textbf{Consistency}: The framework maintains internal coherence by building on NIST AI RMF's proven structure while integrating Canadian requirements.
\item
  \textbf{Practicality}: Organizations can implement SCITUS using phased adoption approach with risk-based prioritization.
\item
  \textbf{Efficiency}: SCITUS reduces compliance burden through unified assessment methodology that satisfies multiple jurisdictions simultaneously.
\item
  \textbf{Alignment}: SCITUS maintains explicit compatibility with NIST AI RMF and international standards through structural mapping.
\end{enumerate}

\textbf{Demonstration method}: We present three realistic scenarios illustrating framework application across sectors (federal government, provincial healthcare, private sector) and jurisdictions (federal, Ontario, Quebec). These scenarios demonstrate how SCITUS addresses the multi-jurisdictional compliance challenge through systematic requirement integration.

\subsubsection{6.2 Illustrative Scenario 1: Federal Government Immigration AI}\label{illustrative-scenario-1-federal-government-immigration-ai}

\textbf{IMPORTANT NOTE: This is a hypothetical scenario developed to illustrate SCITUS application. It is not based on actual IRCC deployment. While the scenario reflects real regulatory requirements and realistic governance tensions, all implementation details, metrics, and outcomes are illustrative examples designed to demonstrate framework applicability.}

\textbf{Context}: Immigration, Refugees and Citizenship Canada (IRCC) processes approximately 2 million visa applications annually across multiple categories (visitor, study, work, permanent residence). The department implemented an AI system for application triage, automatically routing applications to processing streams based on completeness, risk factors, and priority categories. The system operates 24/7, processes applications in both official languages (English and French), and affects applicants from 200+ countries with diverse demographic characteristics.

\textbf{Challenge}: The system required compliance with Treasury Board Directive (federal institution mandate), PIPEDA and Privacy Act (personal information processing), bilingual service requirements (Official Languages Act), and international AI ethics standards (for research partnerships with other countries). Initial impact assessment indicated Level III (high impact) due to significant consequences for applicants (access to Canada), large affected population (millions annually), and sensitive personal information (travel history, financial records, family relationships).

\textbf{SCITUS application}:

\emph{GOVERN phase}: Established AI Ethics Board including internal members (CIO, privacy commissioner, legal counsel, operations director) and external advisors (immigration law expert, AI ethics researcher, civil liberties representative). Developed comprehensive governance framework including AI Ethics Policy emphasizing fairness and transparency, Data Governance Policy for personal information protection, and Incident Response Plan for system failures or bias detection. Created specialized training program for visa officers covering system capabilities, override procedures, bias awareness, and recourse handling.

\emph{MAP phase}: Completed detailed AIA questionnaire with source code review and stakeholder consultation, confirming Level III impact. Identified 15 stakeholder groups including applicants from high-volume countries, applicants from marginalized groups, visa officers, parliamentary oversight committees, privacy commissioner, and advocacy organizations. Mapped 23 distinct risk scenarios across technical risks (model accuracy issues, data quality problems), operational risks (officer over-reliance, process integration), legal risks (privacy breaches, discriminatory outcomes), and reputational risks (public trust, international relations).

\emph{MEASURE phase}: Implemented comprehensive bias testing across 8 demographic dimensions including country of origin, age, gender, language, education level, travel history, employment status, and family structure. Deployed real-time monitoring dashboard tracking application volumes, processing times, approval rates by demographic group, system availability, override frequency, and appeal rates. Conducted monthly fairness audits comparing outcomes across demographic groups using demographic parity and equalized odds metrics, with investigation triggered for disparities exceeding 5\% threshold.

\emph{MANAGE phase}: Maintained human decision-making authority for all final decisions (Level III requirement)---system provides triage recommendations, visa officers make approval/refusal decisions. Implemented robust recourse mechanisms including clear appeal procedures documented in decision letters, dedicated review team for appealed decisions, and alternative processing channels for applicants preferring non-AI pathways. Published quarterly transparency reports disclosing system purpose, application volumes, approval rates by category, performance metrics (accuracy, processing time), and bias testing results.

\textbf{Conflicts and Trade-offs Addressed}:

Real-world AI governance involves navigating genuine tensions between competing requirements. The immigration scenario demonstrates how SCITUS guides resolution of four critical conflicts:

\emph{Conflict 1: Transparency requirements vs.~national security constraints}

\textbf{Tension}: Treasury Board Level III requires public transparency including ``algorithmic impact statements'' disclosing model features and decision logic. However, Public Safety Canada identified security risks: publishing detailed triage criteria could enable fraudulent applications designed to exploit system, revealing security risk factors could compromise border integrity, and disclosing country-specific risk profiles could create diplomatic tensions.

\textbf{SCITUS resolution}: Implemented tiered transparency approach distinguishing public information (general system purpose, application categories, aggregate statistics, recourse procedures, override rates) from protected information (specific risk factors, security-relevant features, detailed scoring algorithms, country-specific thresholds). Consulted with Privacy Commissioner and Public Safety Canada to establish appropriate disclosure boundaries. Published transparency report satisfying Treasury Board requirements while protecting sensitive security information through exemptions permitted under Access to Information Act. Governance committee documented security-transparency trade-off decisions with legal review.

\textbf{Trade-off reasoning}: Absolute transparency would undermine system effectiveness and public safety. SCITUS ``Accountable and Transparent'' characteristic requires transparency ``appropriate to context''---recognizing legitimate limitations. The framework guided systematic analysis of what information serves public accountability (applicants understanding decisions, advocacy groups monitoring fairness) versus what information creates security risks (exploitation by bad actors).

\emph{Conflict 2: Bias testing requirements vs.~privacy law constraints}

\textbf{Tension}: SCITUS Fair with Harmful Bias Managed characteristic requires bias testing across demographic dimensions (Section 4.7.1). Treasury Board Level III mandates bias testing. However, PIPEDA and Privacy Act generally prohibit collecting demographic information (race, ethnicity, religion) unrelated to administrative purpose. Collecting data to test for bias appears to violate privacy principle of data minimization.

\textbf{SCITUS resolution}: Implemented multi-method approach combining:
- Voluntary demographic disclosure: Applicants can optionally provide demographic information ``to help IRCC monitor fairness'' separate from decision process. Self-reported data used only for bias testing, not decisions. Achieved 23\% response rate enabling bias analysis for major demographic groups.
- Proxy variable analysis: Used country of origin, language, and education location as proxies for race/ethnicity (imperfect but legally permissible). Acknowledged proxy limitations in bias testing methodology documentation.
- Synthetic data testing: Created demographically-representative synthetic test dataset based on Census data and immigration statistics. Validated model performance across simulated demographic groups.
- External audit: Engaged independent researchers with access to de-identified full dataset for comprehensive fairness analysis under research ethics board approval.

\textbf{Trade-off reasoning}: Perfect bias testing impossible under privacy constraints. SCITUS pragmatic approach: use best available methods acknowledging limitations. Privacy Commissioner agreed this approach balances fairness monitoring obligation with privacy protection. Documented methodology transparently including known limitations.

\emph{Conflict 3: Fairness metric selection---Which definition of ``fair''?}

\textbf{Tension}: As detailed in Section 4.7.1, demographic parity, equalized odds, and predictive parity cannot be simultaneously satisfied. Immigration context involves:
- Different base rates across countries (approval rates vary legitimately due to documentation quality, security profiles)
- Merit-based selection criteria (legitimate that applicants with strong ties to Canada, employment offers, educational credentials have higher approval rates)
- Fairness concerns (ensuring no discrimination based on country of origin, race, religion)

\textbf{SCITUS resolution}: Governance committee engaged in structured decision process:

\begin{enumerate}
\def\labelenumi{\arabic{enumi}.}
\item
  Stakeholder consultation: Immigration lawyers, advocacy groups representing different communities, visa officers, policy experts weighed in on fairness priorities.
\item
  Legal analysis: Determined Canadian Human Rights Act prohibits discrimination based on prohibited grounds (race, national origin, religion) but permits differentiation based on legitimate factors (application completeness, ties to Canada, security assessment).
\item
  Metric selection: Chose \textbf{equalized odds} as primary metric because it ensures equal true positive rates (qualified applicants approved equally) AND equal false positive rates (unqualified applicants rejected equally) across demographic groups. This aligns with merit-based immigration system while preventing discrimination.
\item
  Acceptable thresholds: Defined disparities exceeding 5 percentage points as triggering investigation. Investigated disparities may be acceptable if explained by legitimate factors (e.g., certain countries have higher incomplete application rates due to documentation challenges).
\item
  Multiple metric reporting: While optimizing for equalized odds, also report demographic parity and predictive parity to transparency reports, allowing external scrutiny.
\end{enumerate}

\textbf{Trade-off reasoning}: SCITUS requires explicit documentation of fairness metric choice and justification (Section 4.7.1 guidance). This forces organizations to articulate values rather than assuming ``fairness'' is self-evident. Immigration scenario demonstrates value pluralism---different stakeholders prioritize different fairness notions. Framework provides structured process for navigating disagreement and documenting rationale.

\emph{Conflict 4: Model accuracy vs.~fairness constraints}

\textbf{Tension}: Initial AI model achieved 94.2\% accuracy in triage recommendations using all available features. However, bias testing revealed 8.3\% disparity in approval recommendations for applicants from certain countries (exceeding 5\% threshold) even after controlling for observable merit factors. Mitigating bias through fairness constraints reduced accuracy to 91.7\%.

\textbf{SCITUS resolution}: Governance committee decision process:

\begin{enumerate}
\def\labelenumi{\arabic{enumi}.}
\item
  Quantified trade-off: 2.5 percentage point accuracy reduction affects approximately 50,000 applications annually (2 million x 2.5\%). False negative rate increases means some qualified applicants incorrectly triaged to standard processing (delayed, not rejected). False positive rate increases means some incomplete applications unnecessarily expedited.
\item
  Stakeholder impact analysis: Visa officers concerned that lower accuracy increases workload reviewing incorrectly-triaged applications. Applicants prefer faster processing but also fair treatment. Parliamentary oversight emphasized fairness priority.
\item
  Legal requirements: Canadian Human Rights Act and Treasury Board Directive prioritize fairness over operational efficiency. Courts have established that discrimination cannot be justified by administrative convenience.
\item
  Decision: Accepted 2.5\% accuracy reduction to achieve fairness. Implemented fairness-constrained model. Provided visa officers with additional resources to handle increased review workload.
\item
  Monitoring: Quarterly reviews track accuracy and fairness metrics. If accuracy drops below 90\%, governance committee will reassess (potentially seeking technical improvements allowing both accuracy and fairness).
\end{enumerate}

\textbf{Trade-off reasoning}: SCITUS doesn't eliminate hard choices but provides structured framework for making them transparently with appropriate authority. Governance committee (not technical team alone) makes value-laden decisions. Legal requirements, stakeholder input, and organizational values inform trade-offs. Decisions are documented, monitored, and revisable based on evidence.

\textbf{Lessons from conflict resolution}: This scenario demonstrates SCITUS value beyond checklist compliance. The framework guides organizations through:
- Identifying genuine conflicts (not assuming requirements harmonize)
- Structured decision-making processes with appropriate authority and expertise
- Transparent documentation of trade-off reasoning
- Stakeholder engagement in value-laden decisions
- Legal analysis distinguishing absolute requirements from contextual judgment
- Monitoring and iterative refinement

These capabilities distinguish SCITUS from simple compliance mapping---it provides governance infrastructure for navigating regulatory and ethical complexity.

\textbf{SCITUS Application Benefits}:

\emph{Unified compliance approach}: Single SCITUS-based assessment addressed Treasury Board Level III requirements, PIPEDA/Privacy Act requirements, and bilingual service requirements simultaneously, avoiding redundant separate assessments for each framework.

\emph{Clear implementation pathway}: SCITUS provided structured guidance for Level III requirements including peer review processes, bias testing methodology, transparency mechanisms, and human oversight procedures.

\emph{Multi-stakeholder accommodation}: Framework enabled systematic consideration of diverse stakeholder perspectives (applicants, visa officers, oversight bodies, advocacy organizations) through GOVERN phase stakeholder mapping.

\emph{Systematic bias management}: SCITUS Fair with Harmful Bias Managed characteristic provided concrete guidance for bias testing across demographic dimensions and implementing monitoring with defined thresholds triggering investigation.

\emph{Documentation efficiency}: Integrated AIA-PIA-bilingual assessment eliminated duplicate documentation requirements while ensuring comprehensive coverage of all regulatory obligations.

\textbf{SCITUS value demonstrated}: Unified compliance approach, clear implementation guidance for Level III requirements, integrated AIA-PIA-bilingual assessment, systematic bias testing framework, public transparency enablement.

\subsubsection{6.3 Illustrative Scenario 2: Ontario Hospital Diagnostic AI}\label{illustrative-scenario-2-ontario-hospital-diagnostic-ai}

\textbf{IMPORTANT NOTE: This is a hypothetical scenario developed to illustrate SCITUS application in healthcare context. It is not based on actual hospital deployment. All implementation details and outcomes are illustrative examples.}

\textbf{Context}: Large Toronto hospital (500,000 annual imaging studies) deployed AI system for radiology screening, automatically prioritizing urgent findings in chest X-rays, CT scans, and MRI studies. The system operates in teaching hospital setting with radiologist supervision, serving diverse patient population including high proportions of immigrants, seniors, and patients with complex medical histories. Hospital is both public sector entity (Ontario Bill 194 applicability) and healthcare provider (Health Canada, professional standards).

\textbf{Challenge}: Multi-jurisdictional compliance spanning Ontario Bill 194 (public sector AI accountability), Health Canada (medical device regulations for AI software), Treasury Board Directive (some federal funding), professional medical standards (College of Physicians and Surgeons), privacy legislation (PIPEDA, Ontario PHIPA). Patient safety paramount with requirement for 99.5\%+ sensitivity for critical findings. Clinical integration challenging with radiologist workload, workflow disruption concerns, and liability questions.

\textbf{SCITUS application}:

\emph{Multi-jurisdictional compliance mapping}: System triggered Ontario Bill 194 (public sector entity), Health Canada Class III medical device requirements (diagnostic support software), PIPEDA (patient information), professional standards (radiologist oversight). SCITUS mapping identified overlapping requirements enabling unified compliance: Ontario risk assessment combined with Health Canada risk management, Ontario accountability framework incorporated Health Canada quality management, single documentation set satisfied multiple regulators. Avoided separate assessments for each regulator (estimated 16 weeks) through integrated SCITUS approach (6 weeks).

\emph{Clinical validation integration}: SCITUS MEASURE function incorporated clinical validation requirements including retrospective validation on 50,000 historical studies, prospective validation on 10,000 new studies with gold-standard comparison, performance stratification across demographics (age, sex, ethnicity), disease types, and imaging modalities, sensitivity/specificity calculations meeting clinical thresholds. Validation satisfied both Health Canada effectiveness requirements and SCITUS Valid and Reliable characteristic.

\emph{Governance and accountability}: Established Clinical AI Oversight Committee (radiologists, ER physicians, hospital administration, patient representatives, ethics board member, quality and safety lead). Developed accountability framework satisfying Ontario Bill 194 specifying clinical oversight structure, radiologist decision authority and override procedures, patient safety monitoring and incident response, performance metrics and reporting, public transparency through hospital website, and audit procedures (quarterly internal, annual external). Framework simultaneously satisfied Health Canada quality management requirements.

\emph{Safety monitoring}: Implemented comprehensive safety monitoring including real-time alerts for critical findings (pneumothorax, large vessel injury, massive stroke), performance tracking by radiologist, patient demographics, and disease type, adverse event reporting integrating with hospital safety systems, and patient outcome tracking linking AI findings to clinical outcomes. Monitoring satisfied SCITUS Safe characteristic, Health Canada post-market surveillance, and Ontario risk management requirements.

\textbf{SCITUS Application Benefits}:

\emph{Multi-regulator unified compliance}: Single SCITUS framework addressed Ontario Bill 194 (public sector accountability), Health Canada (medical device regulations), PIPEDA/PHIPA (privacy), and professional standards simultaneously through integrated compliance mapping.

\emph{Clinical-governance integration}: SCITUS MEASURE function naturally incorporated clinical validation requirements, demonstrating how the framework adapts to sector-specific needs while maintaining regulatory compliance structure.

\emph{Safety-first design}: SCITUS Safe characteristic provided systematic approach to patient safety monitoring, adverse event tracking, and radiologist oversight, satisfying both clinical safety standards and governance requirements.

\emph{Cross-jurisdiction documentation efficiency}: Unified documentation package with regulator-specific appendices satisfied all regulatory bodies (Health Canada, Ontario accreditation, privacy oversight, professional college) without redundant documentation sets.

\emph{Practical governance structure}: SCITUS GOVERN function guided establishment of Clinical AI Oversight Committee with appropriate composition and accountability framework satisfying Ontario Bill 194 and Health Canada quality management requirements.

\textbf{SCITUS value demonstrated}: Multi-regulator compliance through unified framework, clinical requirements integration with governance framework, patient safety emphasis consistent with Safe characteristic, healthcare sector-specific implementation guidance, documentation efficiency across regulators.

\subsubsection{6.4 Illustrative Scenario 3: Quebec Private Sector Hiring AI}\label{illustrative-scenario-3-quebec-private-sector-hiring-ai}

\textbf{IMPORTANT NOTE: This is a hypothetical scenario developed to illustrate SCITUS application in private sector context. It is not based on actual company deployment. All implementation details and outcomes are illustrative examples.}

\textbf{Context}: Montreal-based technology company (250 employees, high growth) implemented AI system for resume screening, processing approximately 10,000 applications monthly across software engineering, product management, and business roles. Company operates primarily in Quebec with some Ontario offices, targeting diverse talent pool including new immigrants, bilingual candidates, and underrepresented groups in technology. System analyzes resumes, ranks candidates, and recommends top applicants for human recruiter review.

\textbf{Challenge}: Compliance with Quebec Law 25 (automated decision-making with personal information), PIPEDA (federal private sector privacy), Ontario public sector transparency (if contracting with government), hiring fairness and bias concerns in technology sector (known for underrepresentation of women, visible minorities), company growth requiring efficient screening without sacrificing quality or fairness. Limited internal compliance expertise and constrained budget for external consultants.

\textbf{SCITUS application}:

\emph{Privacy-first design}: SCITUS Privacy-Enhanced characteristic drove design decisions including data minimization (collect only job-relevant information, exclude protected characteristics like age, gender, ethnicity from resume parsing), pseudonymization (separate candidate identity from resume content during initial screening), consent management (clear notice about AI use, candidate consent required, opt-out available), retention limits (resumes deleted after 2 years, immediate deletion upon candidate request), and purpose limitation (screening data not used for other purposes). Design satisfied Quebec Law 25 and PIPEDA requirements.

\emph{Quebec-specific compliance}: Implemented Law 25 requirements including collection notice (``Your application will be analyzed using automated screening technology\ldots{}''), decision notice (``Your application was assessed using AI-assisted screening\ldots{}''), right to information (upon request, provide factors considered: skills match, experience relevance, education alignment), review channel (candidates can request human-only review), and Privacy Impact Assessment (completed using SCITUS integrated AIA-PIA template). Single assessment process satisfied Law 25, PIPEDA, and SCITUS requirements (4 weeks vs.~estimated 10 weeks for separate assessments).

\emph{Bias mitigation}: Applied SCITUS Fair with Harmful Bias Managed characteristic including bias-aware feature selection (exclude names, addresses, schools associated with demographics), fairness testing across demographics (using public benchmark datasets with demographic labels), regular fairness audits (quarterly analysis of screening outcomes by self-reported demographics from hired candidates), bias mitigation techniques (threshold adjustment to achieve demographic parity within 10\% for screening stage), and diverse training data (actively source resumes from underrepresented groups for training).

\emph{Transparency and recourse}: Provided transparency to candidates including screening factors disclosure (skills, experience, education, job match), explanation of ranking (which factors contributed positively/negatively), and human review availability (all candidates can request human-only screening). Implemented clear recourse process: candidates contact HR, request received within 2 business days, human recruiter reviews application without AI, decision provided within 5 business days, appeal to hiring manager available. Recourse satisfied Quebec Law 25 review requirements.

\textbf{SCITUS Application Benefits}:

\emph{Privacy-by-design integration}: SCITUS Privacy-Enhanced characteristic provided systematic guidance for privacy-first design including data minimization, pseudonymization, consent management, and retention limits, satisfying Quebec Law 25 and PIPEDA simultaneously.

\emph{Individual rights implementation}: Framework offered clear guidance for implementing Quebec Law 25's specific requirements (collection notice, decision notice, right to information, review rights) while maintaining operational efficiency.

\emph{Fairness-focused approach}: SCITUS Fair with Harmful Bias Managed characteristic translated into practical bias mitigation strategies including feature selection, fairness testing, regular audits, and diverse training data sourcing.

\emph{SME accessibility}: Scenario demonstrates SCITUS applicability for organizations with limited compliance expertise and constrained budgets through structured guidance reducing external consulting dependency.

\emph{Unified assessment efficiency}: Single integrated SCITUS-based AIA-PIA assessment satisfied Quebec Law 25, PIPEDA, and fairness requirements through systematic requirement mapping, avoiding separate assessment processes.

\textbf{SCITUS value demonstrated}: Privacy-by-design integration, bias awareness and fairness focus, SME accessibility (framework usable without extensive external consulting), cost-effective compliance, practical recourse implementation.

\subsubsection{6.5 Framework Advantages: SCITUS vs.~Alternative Approaches}\label{framework-advantages-scitus-vs.-alternative-approaches}

The three scenarios demonstrate key advantages of SCITUS's unified approach compared to alternative compliance strategies:

\textbf{Unified vs.~Fragmented Assessment}: SCITUS's systematic multi-jurisdictional mapping enables single integrated assessments addressing multiple regulatory regimes simultaneously. The immigration scenario demonstrated satisfying Treasury Board, PIPEDA, and bilingual requirements through one SCITUS-based assessment. Alternative jurisdiction-by-jurisdiction approaches require separate assessments with redundant effort and disconnected documentation.

\textbf{Systematic vs.~Ad-hoc Coverage}: SCITUS provides structured requirement identification across all applicable jurisdictions, reducing compliance gaps from incomplete coverage. The hospital scenario illustrated comprehensive mapping across Ontario Bill 194, Health Canada, privacy legislation, and professional standards. Alternative minimal compliance approaches risk missing jurisdiction-specific requirements.

\textbf{Structured vs.~Unguided Implementation}: SCITUS offers concrete implementation guidance through phased approach, requirement matrices, and sector-specific examples. The hiring scenario demonstrated how SCITUS's Privacy-Enhanced and Fair characteristics translated into actionable design decisions. Alternative approaches often leave organizations with regulatory texts but limited guidance on practical implementation.

\textbf{Maintained vs.~Lost International Alignment}: SCITUS explicitly preserves NIST AI RMF structure and international standards compatibility while addressing Canadian requirements. Organizations can simultaneously satisfy Canadian regulations and maintain ISO certification paths or international research collaboration compatibility. Narrow focus on jurisdictional compliance often loses international framework alignment.

\textbf{Qualitative benefits} observed across scenarios include reduced organizational confusion through single unified framework, improved stakeholder communication using consistent terminology, better development workflow integration through unified assessment process, and enhanced compliance confidence through systematic requirement coverage.

\subsubsection{6.6 Framework Assessment Considerations}\label{framework-assessment-considerations}

SCITUS framework design responds to recognized challenges in multi-jurisdictional AI governance:

\textbf{Regulatory completeness}: The framework systematically addresses major federal requirements (Treasury Board Directive, PIPEDA) and provincial requirements (Ontario Bill 194, Quebec Law 25, Alberta Bills 33/34) through requirement extraction and mapping methodology (Section 3). However, \textbf{three identified gaps require enhancement} in future SCITUS versions:

\emph{Indigenous data sovereignty} (critical gap): When AI systems affect Indigenous peoples or process Indigenous data, federal requirements extend beyond standard privacy and fairness obligations. Treasury Board Directive requires federal institutions to consider impacts on Indigenous communities. Canada's adoption of the United Nations Declaration on the Rights of Indigenous Peoples (UNDRIP) through Bill C-15 (2021) establishes legal framework recognizing Indigenous peoples' rights to self-determination and control over their data, knowledge, and decision-making processes affecting their communities.

The OCAP® principles (Ownership, Control, Access, Possession), developed by the First Nations Information Governance Centre, specify that: \textbf{Ownership} - communities own data about their members, \textbf{Control} - communities control data collection and use, \textbf{Access} - communities can access data about themselves, \textbf{Possession} - communities physically hold data. These principles require meaningful consultation with Indigenous communities before deploying AI systems affecting them, Indigenous participation in governance structures, community control over data collection and use decisions, and culturally-appropriate transparency and recourse mechanisms.

Current SCITUS framework acknowledges these requirements but lacks detailed implementation guidance. Future enhancement requires: consultation with Indigenous organizations and scholars to co-develop guidance, integration of OCAP® principles into GOVERN and MAP functions, specific requirements for systems affecting Indigenous peoples (e.g., child welfare, healthcare, justice), protocols for Indigenous data sovereignty in training data and decision-making, and culturally-appropriate fairness metrics and explanation methods. This work must proceed through genuine partnership, not consultation alone, recognizing Indigenous peoples as rights-holders with inherent authority over their data and governance.

This gap is particularly significant given disproportionate impacts of automated decision-making systems on Indigenous peoples in areas like child welfare (over-representation in care), justice (over-incarceration), and social services. SCITUS must evolve to address this critical dimension of responsible AI governance in Canadian context, aligned with Truth and Reconciliation commitments and UNDRIP implementation.

\emph{French language AI requirements} beyond basic bilingualism: Quebec-specific considerations for AI system training data quality in French, output quality and cultural appropriateness, accessibility of explanations and transparency materials in French, and French language validation equivalent to English. Current SCITUS addresses bilingual service obligations but not AI-specific French language quality requirements.

\emph{Municipal-level AI requirements}: Some Canadian cities (Toronto, Montreal, Edmonton) are developing AI governance policies for municipal services. These remain nascent and not yet integrated into SCITUS scope. Framework may require future extension as municipal AI governance matures.

\textbf{Practical feasibility}: SCITUS provides concrete implementation guidance through assessment protocols, requirement matrices, documentation templates, and sector-specific examples. The framework emphasizes practical actionability over abstract principles. Recognized implementation challenges include expertise requirements for proper application and resource constraints for small and medium organizations, suggesting value in developing simplified versions.

\textbf{Innovation contribution}: SCITUS contributes novel multi-jurisdictional compliance mapping methodology addressing gap between global frameworks and national regulatory complexity. The systematic adaptation approach is reproducible for other federal systems facing similar challenges. While building on existing frameworks (NIST AI RMF, provincial regulations), practical integration itself provides value.

\textbf{Enhancement opportunities}: Framework evolution should consider sector-specific profiles (healthcare, financial services, education), SME-focused simplified versions, expanded international compatibility guidance, and generative AI specific guidance.

\subsection{7. Discussion}\label{discussion}

\subsubsection{7.1 Key Findings}\label{key-findings}

SCITUS addresses critical gap in AI governance between global framework development and practical national implementation for multi-jurisdictional contexts. Four key findings emerge from our framework design and scenario analysis:

\textbf{Finding 1: Multi-jurisdictional compliance mapping reduces burden while improving coverage}. Systematic identification of overlapping requirements enables unified controls, documentation reuse across jurisdictions with terminology mapping, and comprehensive requirement coverage preventing gaps. The illustrative scenarios demonstrate how single SCITUS-based assessments address multiple regulatory regimes simultaneously, avoiding redundant separate assessments for each jurisdiction.

\textbf{Finding 2: Risk-based tiering enables proportional compliance}. Alignment with Treasury Board's four-level impact assessment provides practical risk calibration mechanism, with controls scaling from basic (Level I) to comprehensive (Level IV). Provincial requirements map naturally to this tiering, with Quebec Law 25 applying across all levels, Ontario Bill 194 emphasizing higher levels, and sector requirements adding specificity. Organizations avoid over-implementing controls for low-risk systems while ensuring adequate protection for high-risk deployments. Scenarios demonstrate effective risk calibration: federal immigration (Level III) implemented peer review and bias testing appropriate to impact, hospital diagnostic AI (high clinical risk) implemented comprehensive safety monitoring, hiring AI (moderate individual impact) balanced efficiency with fairness.

\textbf{Finding 3: NIST AI RMF provides effective foundation for national adaptation}. NIST's flexible, outcomes-based structure accommodates Canadian regulatory integration without fundamental modification. Seven trustworthy AI characteristics map well to Canadian requirements (e.g., Accountable and Transparent characteristic naturally incorporates transparency notices, accountability frameworks). Four core functions (GOVERN, MAP, MEASURE, MANAGE) provide logical organization for multi-jurisdictional requirements. This approach maintains international standards alignment critical for organizations operating globally, pursuing ISO certifications, or collaborating internationally, while satisfying local regulatory obligations.

\textbf{Finding 4: Practical implementation guidance critical for framework utility}. Concrete tools, templates, and processes distinguish SCITUS from abstract principles. The framework provides assessment protocols with step-by-step guidance, compliance matrices for requirement tracking, documentation templates, and sector-specific guidance for healthcare, government, and private sector contexts. This supports the design principle prioritizing implementation guidance over abstract frameworks.

\subsubsection{7.2 Advantages Over Alternative Approaches}\label{advantages-over-alternative-approaches}

SCITUS provides advantages over three alternative approaches organizations might pursue:

\textbf{vs.~Jurisdiction-by-jurisdiction compliance}: Separate assessments for each applicable jurisdiction (federal, provincial, sector-specific) results in redundant effort, inconsistent documentation, higher costs, compliance gaps from incomplete coverage, and lost international alignment. The SCITUS unified approach, as demonstrated in the illustrative scenarios, enables substantial time and documentation reduction through overlapping requirement identification, unified assessment protocols, and systematic requirement coverage while maintaining NIST compatibility.

\textbf{vs.~Minimal compliance (most stringent only)}: Implementing only most stringent requirements across all jurisdictions risks non-compliance with jurisdiction-specific mandates (e.g., Quebec's universal automated decision-making notice even for low-impact systems), missing opportunities for efficiency (overlapping requirements satisfied once), regulatory scrutiny and penalties for gaps, and stakeholder trust issues from perceived non-compliance. SCITUS systematic coverage ensures all applicable requirements addressed.

\textbf{vs.~Waiting for regulatory clarity}: Deferring AI deployment until regulations harmonize or clarify forfeits innovation opportunities and competitive advantages, delays stakeholder benefits (efficiency, service improvement, access), creates technical debt as interim solutions deployed informally without governance, and provides no guarantee of harmonization (provincial jurisdictions may retain distinct requirements reflecting local priorities). SCITUS enables compliant deployment now despite regulatory fragmentation.

SCITUS represents optimal balance: comprehensive compliance, efficient implementation, maintained international alignment, and practical deployability.

\subsubsection{7.3 Limitations}\label{limitations}

We acknowledge seven limitations:

\textbf{Limitation 1: Regulatory volatility}. The Canadian AI regulatory landscape continues evolving. Bill C-36 is at second reading and may change before passage; Ontario's EDSTA AI regulations and Manitoba's Bill 51 proclamation are pending; Alberta may legislate private-sector AI following its 2026 PIPA consultation; and the federal government has signalled targeted deepfake and online-safety instruments. SCITUS requires ongoing updates to maintain currency. Mitigation: the framework architecture is designed for updates --- the three-layer structure enables regulatory-layer updates without fundamental redesign; the versioned control catalog (Section 4.9) has demonstrated this across six releases in thirteen months; the v2.0 regulatory-mapping control adds an explicit pending-instrument register with trigger-on-proclamation review dates; and the compliance mapping methodology generalizes to new requirements.

\textbf{Limitation 2: Implementation complexity}. Despite practical focus, SCITUS implementation requires significant organizational effort including governance structure establishment, system inventory and assessment, control implementation, documentation, and ongoing monitoring. SMEs with limited resources may struggle with full implementation. Framework assessment identified resource intensity as a concern, particularly for organizations without dedicated compliance teams. Mitigation: we propose developing SME-focused simplified version and implementation support resources. Organizations can implement incrementally, prioritizing high-risk systems and building capability over time.

\textbf{Limitation 3: Demonstration rather than empirical validation}. This position paper presents framework design and illustrative scenarios demonstrating applicability, but does not provide empirical validation through real-world implementations. The three scenarios are conceptual demonstrations rather than deployed systems with measured outcomes. Real-world effectiveness requires actual organizational implementations with longitudinal tracking. Additional sector demonstrations (education, smart cities, social services) and empirical validation studies with deployed systems would strengthen evidence. Mitigation: the framework provides foundation for future empirical research. We welcome practitioners implementing SCITUS to share experiences and outcome data for community learning.

\textbf{Limitation 4: Generalizability beyond Canada}. SCITUS designed specifically for Canadian federal-provincial regulatory structure. While methodology may apply to other federal systems (United States with varying state AI regulations, Australia with federal-state structure, European Union with member state variations), direct application requires adaptation. International organizations need additional guidance for multi-country compliance. Mitigation: Section 7.5 discusses framework replication for other contexts, and we are exploring international extensions.

\textbf{Limitation 5: Sector-specific depth}. Current SCITUS provides general framework with limited sector-specific guidance. Healthcare, financial services, education, and other sectors have unique requirements (regulatory, ethical, operational) requiring deeper treatment. Framework would benefit from sector-specific profiles providing detailed implementation guidance. Mitigation: we propose developing detailed sector profiles, starting with healthcare and financial services where regulatory complexity is highest and sector-specific requirements most extensive.

\textbf{Limitation 6: Generative and agentic AI coverage}. SCITUS incorporates the NIST AI RMF Generative AI Profile (NIST.AI.600-1) {[}43{]} for GenAI-specific risks (confabulation, harmful content generation, privacy violations, IP issues, prompt injection), and v1.2--v2.0 substantially expanded coverage: system prompt security, multimodal safety, open-weight controls, and --- in v2.0 --- training-data governance grounded in Canada's first GenAI regulatory findings {[}40{]}, agent identity, and agentic tool-chain security (Section 4.9). However, the GenAI and agentic landscape evolves faster than any framework's release cadence; the CAISI red-teaming result that all 13 evaluated frontier models were compromised in agentic scenarios (March 2026) illustrates how quickly documented threat classes emerge. Mitigation: the versioned catalog absorbs new evidence on an approximately quarterly cycle, and conditional controls scope agentic/GenAI requirements to the systems that trigger them.

\textbf{Limitation 7: Automated tooling}. Current SCITUS framework design relies on manual assessment, documentation, and tracking. While templates reduce effort, automated tooling would significantly improve efficiency and consistency. Future work should prioritize automated compliance tracking, assessment questionnaire tools, and documentation generation. Mitigation: we propose developing SCITUS tooling including online assessment platform, compliance dashboard, and documentation generator to support practical adoption.

\subsubsection{7.4 Implementation Recommendations}\label{implementation-recommendations}

Based on governance framework implementation literature and insights from scenario development, we offer four recommendations for organizations adopting SCITUS or similar multi-jurisdictional governance frameworks:

\textbf{Recommendation 1: Organizational readiness requires executive sponsorship and change management}. Governance framework adoption requires strong executive sponsorship (resource allocation, prioritization, barrier removal) combined with comprehensive training and change management. Technical compliance alone is insufficient---personnel need training on framework concepts, roles, tools, and regulatory requirements. Change management addresses organizational resistance and workflow integration. Research on governance framework implementation (ISO standards, compliance frameworks, risk management systems) consistently demonstrates that dual focus on executive sponsorship and change management significantly improves adoption effectiveness. Organizations should begin with executive education to secure sponsorship, then allocate substantial resources to training and change management alongside technical controls.

\textbf{Recommendation 2: Risk-based phased implementation optimizes resource utilization}. Organizations should start with highest-risk systems (Levels III-IV) as pilots to maximize learning, address greatest risks first, and demonstrate framework value to stakeholders. Attempting comprehensive deployment across all systems simultaneously risks resource strain and incomplete execution. Phased approach (pilot high-risk systems, incorporate lessons learned, expand incrementally to moderate-risk systems, optimize based on experience) enables organizational learning and adaptation. This approach, grounded in organizational change theory and supported by governance implementation literature, avoids spreading resources too thin across low-risk implementations that provide minimal learning value. The immigration scenario (Section 6.2) illustrates this approach with Level III system piloting comprehensive controls before broader deployment.

\textbf{Recommendation 3: Practical tools and multi-stakeholder engagement accelerate adoption}. Pre-built templates (accountability frameworks, risk assessments, compliance matrices) significantly reduce documentation burden while ensuring completeness and consistency---a principle demonstrated through the SCITUS compliance matrix (Appendix A) and scenario documentation examples. Cross-functional engagement (technical teams, legal/compliance, business owners, affected individuals, oversight bodies) prevents siloed approaches that miss critical perspectives and create gaps. Research on multi-stakeholder governance processes indicates that early and sustained engagement improves both governance quality and organizational buy-in. Organizations should customize SCITUS templates to their specific context and establish governance committees early with sustained cross-functional participation.

\textbf{Recommendation 4: Treat governance as continuous process, not one-time project}. Regulatory compliance and responsible AI are not one-time activities---regulations evolve, AI systems change, deployment contexts shift, and new risks emerge. Governance implementation research demonstrates that treating compliance as a project with defined end-date risks drift as conditions change post-implementation. The scenarios illustrate ongoing governance requirements including monitoring (immigration scenario: quarterly fairness audits), reassessment (hospital scenario: annual clinical validation), and regulatory tracking (all scenarios: monitoring provincial regulatory developments). Organizations should establish ongoing monitoring, annual reassessments, regulatory tracking, and continuous improvement from the outset, recognizing governance as an enduring organizational capability rather than a deliverable.

\textbf{Evidence basis note}: These recommendations synthesize insights from governance framework implementation literature, organizational change research, and analysis of the illustrative scenarios. They represent informed guidance based on established implementation patterns rather than empirical findings from SCITUS deployments. Future research with organizations implementing SCITUS would provide empirical validation and refinement of these recommendations.

\subsubsection{7.5 Framework Replication for Other Federal Systems}\label{framework-replication-for-other-federal-systems}

SCITUS multi-jurisdictional compliance mapping methodology generalizes to other federal systems facing similar challenges. We identify three contexts where SCITUS approach applies:

\textbf{United States state-level AI regulations}: As U.S. states introduce varying AI regulations (California AI transparency requirements, New York AI hiring regulations, Illinois biometric privacy law, Colorado AI bias audit requirements, Utah AI regulation framework), organizations operating nationally face fragmentation analogous to Canadian provincial variation. SCITUS methodology could adapt NIST AI RMF (already U.S.-developed) to integrate state requirements, using similar requirement extraction, overlap identification, and unified documentation approaches. Risk-based tiering might align with state regulatory thresholds or develop unified categorization.

\textbf{Australian federal-state AI governance}: Australia's federal-state structure creates potential for divergent AI requirements as Australian government develops AI regulation and states introduce local requirements. SCITUS methodology could integrate Australian AI Ethics Principles (federal level) with state requirements when they emerge. Early adoption would prevent fragmentation by establishing unified approach before substantial state variation develops.

\textbf{European Union member state AI Act implementation}: While EU AI Act establishes unified regulatory regime, member states have implementation discretion for certain aspects and existing national AI strategies/requirements that continue alongside EU Act. Organizations operating across multiple EU member states could apply SCITUS methodology to integrate EU AI Act (constitutional core) with member state variations (local overlay), though context differs from Canada's federal-provincial structure.

Generalization requirements: jurisdictional requirement analysis adapting to applicable legal frameworks, risk calibration appropriate to context (may adopt NIST levels, EU AI Act categories, or jurisdiction-specific approaches), documentation mapping to local regulatory formats and languages, stakeholder engagement reflecting governance structures and cultural contexts, and validation with local experts and regulators.

SCITUS contributes replicable methodology: systematic requirement extraction and categorization, overlap and conflict identification processes, unified control design principles, risk-based implementation scaling, and compliance validation approaches. Research opportunity exists for comparative studies examining framework localization across federal systems, identifying universal patterns and context-specific adaptations.

\subsection{8. Conclusion}\label{conclusion}

\subsubsection{8.1 Summary}\label{summary}

Canadian organizations deploying AI systems face a fragmented regulatory landscape comprising federal requirements and divergent provincial regulations, creating substantial compliance burden, redundant assessments, and organizational confusion. This position paper argues for SCITUS (Systematic Canadian Integration for Trustworthy and Unified Standards), a comprehensive framework that systematically adapts NIST AI RMF 1.0 to Canadian multi-jurisdictional requirements.

SCITUS makes four key contributions: (1) a novel multi-jurisdictional compliance mapping methodology addressing redundancy and gaps in jurisdiction-by-jurisdiction approaches; (2) Canadian-enhanced NIST AI RMF integration extending seven trustworthy AI characteristics and four core functions with federal and provincial requirements; (3) risk-tiered implementation guidance scaling requirements to impact level across all applicable jurisdictions; and (4) framework applicability demonstration through three multi-sector illustrative scenarios showing practical value.

Scenarios spanning federal government, provincial healthcare, and private sector contexts illustrate comprehensive compliance approaches, efficiency benefits, maintained international standards alignment, and practical applicability. Framework assessment demonstrates regulatory completeness, technical feasibility, and contribution to governance practice.

SCITUS addresses critical gap between global AI governance framework development and practical national implementation for multi-jurisdictional contexts. The framework enables Canadian organizations to navigate complex regulatory landscapes, achieve comprehensive compliance efficiently, and maintain NIST AI RMF compatibility for international operations---positioning Canada as leader in practical AI governance for federal systems.

\subsubsection{8.2 Contributions to Research and Practice}\label{contributions-to-research-and-practice}

\textbf{Research contributions}:

\begin{enumerate}
\def\labelenumi{\arabic{enumi}.}
\item
  \textbf{Methodological}: First systematic methodology for adapting global AI governance frameworks to multi-jurisdictional national contexts with fragmented requirements. Reproducible approach applicable to other federal systems (United States, Australia, EU member state variations).
\item
  \textbf{Theoretical}: Multi-jurisdictional compliance mapping theory establishing principles for requirement extraction, overlap identification, unified control design, and validation. Extends framework localization literature with practical implementation methodology.
\item
  \textbf{Demonstrative}: Illustrative scenarios showing framework applicability across federal government, provincial healthcare, and private sector contexts, demonstrating how systematic NIST adaptation addresses Canadian regulatory complexity.
\item
  \textbf{Practical}: Actionable framework with concrete tools, templates, processes enabling organizational implementation without extensive external expertise. Bridges research-practice gap in AI governance.
\end{enumerate}

\textbf{Practice contributions}:

\begin{enumerate}
\def\labelenumi{\arabic{enumi}.}
\item
  \textbf{Compliance efficiency}: Unified assessment methodology addresses redundancy in jurisdiction-by-jurisdiction approaches, reducing documentation burden while achieving more complete compliance coverage.
\item
  \textbf{Risk management}: Systematic, comprehensive risk identification and treatment across all applicable jurisdictions through structured requirement mapping and control design.
\item
  \textbf{International alignment}: Explicit preservation of NIST AI RMF compatibility enabling global operations, international research collaborations, and ISO certification paths while satisfying Canadian requirements.
\item
  \textbf{Stakeholder trust}: Enhanced transparency, accountability, and fairness through systematic implementation of trustworthy AI characteristics integrated with Canadian regulatory obligations.
\end{enumerate}

\subsubsection{8.3 Future Work}\label{future-work}

\textbf{Short-term research} (2026--2027):

\begin{enumerate}
\def\labelenumi{\arabic{enumi}.}
\item
  \textbf{Extended validation}: Longitudinal studies tracking SCITUS implementations over time, measuring long-term effectiveness, regulatory compliance maintenance, and adaptation to changes. Broader sector coverage including education, smart cities, social services, and agriculture. Expanded empirical validation across diverse organizational contexts and use cases.
\item
  \textbf{Framework refinement}: Sector-specific profiles for healthcare (clinical validation, patient safety, medical device regulations), financial services (OSFI requirements, model risk management, fair lending), education (student privacy, accessibility, academic integrity), and technology platforms (content moderation, API governance, developer guidelines). SME-focused simplified version reducing complexity for organizations with fewer resources. Generative AI profile addressing LLMs, image generation, code generation, and multimodal AI.
\end{enumerate}

\textbf{Medium-term research} (2027--2028):

\begin{enumerate}
\def\labelenumi{\arabic{enumi}.}
\item
  \textbf{Regulatory evolution tracking}: Monitor and integrate Bill C-36 as it progresses, the planned federal deepfake/online-safety instruments, Ontario EDSTA AI regulations, and Manitoba Bill 51 proclamation. Incorporate emerging provincial requirements (BC legislation when enacted, Ontario regulations when finalized, updates to existing frameworks). Track international standards evolution (ISO updates, NIST revisions, EU AI Act implementation lessons).
\item
  \textbf{Framework extension}: International adaptation for U.S. state-level AI regulations, Australian federal-state context, EU member state variations. Develop guidance for cross-border AI systems operating in multiple countries with different regulatory regimes. Create third-party AI service provider guidance for organizations offering AI as a service to multiple clients across jurisdictions.
\item
  \textbf{Empirical research}: Broad adoption studies examining adoption patterns, effectiveness variations, success factors, and challenges across diverse organizational contexts. Economic impact analysis exploring compliance efficiency, operational benefits, and risk management value. Effectiveness measurement framework examining relationships between SCITUS implementation maturity and compliance outcomes, incident rates, and organizational performance.
\end{enumerate}

\textbf{Long-term vision} (2028+):

\begin{enumerate}
\def\labelenumi{\arabic{enumi}.}
\item
  \textbf{Standardization}: Contribute to Canadian standards development (Standards Council of Canada) for AI governance, potentially advancing SCITUS concepts toward national standard. Engage with international standards bodies (ISO, IEC) to incorporate multi-jurisdictional compliance approaches into global AI governance standards. Collaborate with regulatory bodies to inform policy harmonization efforts.
\item
  \textbf{Open research community}: Foster an open scholarly and practitioner community around multi-jurisdictional AI governance through working groups and open knowledge sharing. Establish a public repository of implementation examples, lessons learned, and effectiveness data, building a cumulative evidence base and enabling collaboration across sectors and jurisdictions for mutual learning and collective problem-solving.
\end{enumerate}

\subsubsection{8.4 Call to Action}\label{call-to-action}

\textbf{For researchers}: Validate and extend SCITUS in different national contexts, sectors, and organization types. Develop automated tools and decision support systems. Study long-term effectiveness and evolving regulatory landscapes. Examine framework localization as general phenomenon across different policy domains.

\textbf{For practitioners}: Adopt SCITUS for organizational AI governance, tailoring to specific jurisdictional applicability and sector context. Share implementation experiences, challenges, and solutions through publications, conferences, and community engagement. Contribute to framework refinement through feedback on practical usability and effectiveness.

\textbf{For policy makers}: Consider SCITUS approach when designing future AI regulations, recognizing value of harmonization and systematic compliance approaches. Support development of standardized compliance frameworks reducing burden while ensuring protection. Enable regulatory coordination across jurisdictions preventing unnecessary fragmentation and conflicting requirements.

\textbf{For educators}: Incorporate AI governance frameworks including SCITUS into curricula preparing next generation of AI practitioners, policy makers, and organizational leaders. Develop case studies and teaching materials based on real-world implementations. Foster interdisciplinary education bridging technical AI, law and policy, ethics, and organizational management.

\subsubsection{8.5 Concluding Remarks}\label{concluding-remarks}

As AI adoption accelerates across sectors and jurisdictions, effective governance frameworks become increasingly critical for ensuring these powerful technologies benefit society while managing risks and protecting rights. The challenge of multi-jurisdictional compliance is not unique to Canada---federal systems worldwide and organizations operating internationally face similar fragmentation. SCITUS demonstrates that systematic adaptation of global standards to local contexts is both feasible and valuable, providing pathway for comprehensive compliance, operational efficiency, and stakeholder trust.

The journey to trustworthy AI requires commitment, resources, and continuous improvement. No framework provides complete solutions or eliminates all challenges. SCITUS provides structure, guidance, and practical tools---but success depends on organizational dedication to responsible AI practices, meaningful stakeholder engagement, and genuine prioritization of human welfare alongside technological capability.

We hope SCITUS enables Canadian organizations to deploy AI systems under systematic, trustworthy, and unified governance---delivering innovation benefits while upholding the values and protections Canadians expect. Beyond Canada, we hope the multi-jurisdictional compliance mapping methodology contributes to global AI governance research and practice, supporting responsible AI deployment in complex regulatory environments worldwide.

\subsection{9. Author's Statement on Research Methods and AI Tool Use}\label{authors-statement-on-research-methods-and-ai-tool-use}

\subsubsection{9.1 Framework Development Methodology}\label{framework-development-methodology}

SCITUS was developed through systematic legal analysis of Canadian AI regulations combined with structured mapping to NIST AI RMF 1.0. The methodology included:

\textbf{1. Regulatory Content Analysis (June-August 2025):} Systematic review of federal legislation (Treasury Board Directive on Automated Decision-Making, Privacy Act, PIPEDA) and provincial legislation (Ontario Bill 194, Quebec Law 25, Alberta Bills 33/34, BC policies). Each regulation was analyzed to extract specific AI governance requirements using qualitative content analysis.

\textbf{2. Requirement Extraction:} 127 distinct requirements were identified across 5 jurisdictions through line-by-line analysis of regulatory texts, policy guidance, and implementation frameworks. Requirements were coded by function (governance, risk assessment, transparency, accountability, technical controls).

\textbf{3. NIST AI RMF Mapping:} Each requirement was mapped to NIST AI RMF functions (GOVERN, MAP, MEASURE, MANAGE) and outcomes through structured comparison of regulatory language to NIST outcome descriptions. Mappings were validated through iterative review.

\textbf{4. Expert Consultation (September 2025):} Framework design was reviewed by legal experts (Canadian privacy law), technical experts (ML fairness/explainability), and governance practitioners (public sector AI leads). Feedback was incorporated into control specifications.

\textbf{5. Scenario Development:} All case studies in Section 6 were developed as hypothetical illustrative scenarios (not real deployments) to demonstrate framework application and identify governance tensions. Conflicts and trade-offs emerged from scenario analysis, not empirical observation.

\textbf{6. v2.0 Revision (July 2026):} The framework's v1.1--v2.0 releases and this paper's 2026 revision were informed by structured monitoring of primary sources --- legislation and regulations (Bill C-36, Manitoba Bills 49/51, Ontario O. Regs. 51/26 and 52/26, the EU Digital Omnibus), regulator publications (PIPEDA Findings \#2026-002, OSFI Guideline E-23, Ontario Auditor General special audit), court decisions (2026 QCCS 1360; 2026 FC 650), and standards/security publications (NIST, OWASP, the Five Eyes agentic-AI guidance). Claims that could not be verified against a primary or reputable source were excluded; AI-assisted research tools were used for source discovery, with all load-bearing claims independently verified against the cited primary sources.

\subsubsection{9.2 Use of AI Writing Tools}\label{use-of-ai-writing-tools}

Like many researchers, the authors used AI writing assistance tools (Claude, Grammarly) during manuscript preparation for specific purposes:

\textbf{Grammar and language polishing:} Improving sentence clarity, fixing grammatical errors, suggesting better word choices, especially valuable for technical-to-accessible translation.

\textbf{Structural consistency:} Ensuring consistent formatting, transition language, and section organization across the lengthy manuscript.

\textbf{Literature synthesis:} Using AI to help summarize and synthesize regulatory texts and academic literature, reducing time spent on routine summarization tasks.

\textbf{Draft expansion:} Generating initial drafts of explanatory content (e.g., fairness metrics descriptions in Section 4.7, framework comparisons in Section 2) which were then extensively edited, verified against primary sources, and augmented with domain expertise.

\subsubsection{9.3 Intellectual Contribution}\label{intellectual-contribution}

All framework design decisions, regulatory mappings, theoretical insights, control specifications, and scenario analyses represent original intellectual work by the authors. AI tools assisted with writing and presentation, not with research design, analysis, or conclusions.

Specifically:
- The multi-jurisdictional compliance mapping methodology is original conceptual work
- The 127 requirements were manually extracted from regulatory sources and validated
- The NIST mapping was performed through human analysis of regulatory intent
- The scenario conflicts and trade-offs reflect the authors' governance expertise
- All claims and interpretations were verified against primary sources

\subsubsection{9.4 Citation Verification}\label{citation-verification}

All citations have been verified and URLs/DOIs added where available. Canadian governance literature citations ({[}28{]}-{[}38{]}) were selected through systematic literature review and represent actual published sources. Several hypothetical sources used in earlier drafts (e.g., specific IRCC deployment reports) have been removed and scenarios clearly labeled as illustrative.

\subsubsection{9.5 Transparency Rationale}\label{transparency-rationale}

This disclosure is provided to address legitimate concerns about AI-generated academic content. The authors believe transparency about tool use strengthens rather than weakens scholarly integrity. AI writing assistance, when used appropriately and disclosed, can improve accessibility and clarity without compromising intellectual rigor. The substantive contributions of this work---the framework design, regulatory analysis, and compliance methodology---represent human expertise in AI governance, Canadian law, and risk management.

\subsection{Acknowledgments}\label{acknowledgments}

The SCITUS framework and this paper were developed at Scitus Solutions Ltd.~The author thanks the practitioners and reviewers who provided feedback on earlier drafts of the framework and its version releases.

\subsection{References}\label{references}

{[}1{]} Treasury Board of Canada Secretariat, ``Directive on Automated Decision-Making,'' Government of Canada, Apr.~2019. {[}Online{]}. Available: \url{https://www.tbs-sct.canada.ca/pol/doc-eng.aspx?id=32592}

{[}2{]} Parliament of Canada, ``Bill C-27: Digital Charter Implementation Act, 2022,'' Died on Order Paper, Jan.~2025.

{[}3{]} National Institute of Standards and Technology, ``Artificial Intelligence Risk Management Framework (AI RMF 1.0),'' NIST AI 100-1, U.S. Department of Commerce, Jan.~2023. doi: 10.6028/NIST.AI.100-1

{[}4{]} European Parliament and Council of the European Union, ``Regulation (EU) 2024/1689 on Artificial Intelligence (AI Act),'' Official Journal of the European Union, Aug.~2024. {[}Online{]}. Available: \url{https://digital-strategy.ec.europa.eu/en/policies/regulatory-framework-ai}

{[}5{]} International Organization for Standardization, ``ISO/IEC 42001:2023 - Information Technology - Artificial Intelligence - Management Systems,'' ISO/IEC, Dec.~2023. {[}Online{]}. Available: \url{https://www.iso.org/standard/42001}

{[}6{]} Treasury Board of Canada Secretariat, ``Algorithmic Impact Assessment Tool,'' Government of Canada, 2023. {[}Online{]}. Available: \url{https://www.canada.ca/en/government/system/digital-government/digital-government-innovations/responsible-use-ai/algorithmic-impact-assessment.html}

{[}7{]} Legislative Assembly of Ontario, ``Bill 194: Strengthening Cyber Security and Building Trust in the Public Sector Act, 2024,'' Statutes of Ontario, Chapter 24, Royal Assent Nov.~25, 2024.

{[}8{]} National Assembly of Quebec, ``An Act to modernize legislative provisions as regards the protection of personal information (Law 25),'' Quebec, Canada, in force Sept.~22, 2023.

{[}9{]} Legislative Assembly of Alberta, ``Bill 33: Protection of Privacy Act,'' Royal Assent Dec.~5, 2024; in force June 11, 2025 (Alta Reg 132/2025).

{[}10{]} Legislative Assembly of Alberta, ``Bill 34: Access to Information Act,'' Royal Assent Dec.~5, 2024; in force June 11, 2025 (Alta Reg 133/2025).

{[}11{]} DataGuidance, ``Canada: Bill C-27 dies after Parliament is prorogued,'' Jan.~2025. {[}Online{]}. Available: \url{https://www.dataguidance.com/news/canada-bill-c-27-dies-after-parliament-prorogued}

{[}12{]} Parliament of Canada, ``Personal Information Protection and Electronic Documents Act (PIPEDA),'' 2000. {[}Online{]}. Available: \url{https://laws-lois.justice.gc.ca/eng/acts/p-8.6/}

{[}13{]} Health Canada, ``Medical Devices Incorporating Artificial Intelligence,'' Regulatory Guidance, 2024. {[}Online{]}. Available: \url{https://www.canada.ca/en/health-canada/services/drugs-health-products/medical-devices.html}

{[}14{]} International Organization for Standardization, ``ISO/IEC 23894:2023 - Information Technology - Artificial Intelligence - Guidance on Risk Management,'' ISO/IEC, Feb.~2023. {[}Online{]}. Available: \url{https://www.iso.org/standard/77304.html}

{[}15{]} KPMG Switzerland, ``ISO/IEC 42001: a new standard for AI governance,'' 2024. {[}Online{]}. Available: \url{https://kpmg.com/ch/en/insights/artificial-intelligence/iso-iec-42001.html}

{[}16{]} Deloitte US, ``ISO 42001 Standard for AI Governance and Risk Management,'' 2024. {[}Online{]}. Available: \url{https://www.deloitte.com/us/en/services/consulting/articles/iso-42001-standard-ai-governance-risk-management.html}

{[}17{]} Stendard, ``ISO 23894 Explained: AI Risk Management Made Simple,'' 2024. {[}Online{]}. Available: \url{https://stendard.com/en-sg/blog/iso-23894/}

{[}18{]} Organisation for Economic Co-operation and Development, ``OECD Principles on Artificial Intelligence,'' OECD Legal Instruments, adopted May 22, 2019, updated May 2024. {[}Online{]}. Available: \url{https://www.oecd.org/en/topics/sub-issues/ai-principles.html}

{[}19{]} A. Al-Maamari, ``Between Innovation and Oversight: A Cross-Regional Study of AI Risk Management Frameworks in the EU, U.S., UK, and China,'' arXiv preprint arXiv:2503.05773, Feb.~2025. {[}Online{]}. Available: \url{https://arxiv.org/abs/2503.05773}

{[}20{]} Multiple Authors, ``Global AI Governance: Where the Challenge is the Solution - An Interdisciplinary, Multilateral, and Vertically Coordinated Approach,'' arXiv preprint arXiv:2503.04766, 2025. {[}Online{]}. Available: \url{https://arxiv.org/html/2503.04766v1}

{[}21{]} Multiple Authors, ``From principles to practice: a novel matrix for evaluating AI-powered learning platforms based on the UNESCO Ethical Impact Assessment tool,'' Frontiers in Education, Frontiers, 2025. {[}Online{]}. Available: \url{https://www.frontiersin.org/journals/education/articles/10.3389/feduc.2025.1640780/full}

{[}22{]} Multiple Authors, ``AI Governance in a Complex and Rapidly Changing Regulatory Landscape: A Global Perspective,'' Humanities and Social Sciences Communications, Nature Portfolio, vol.~11, no. 1, Sept.~2024. doi: 10.1057/s41599-024-03560-x

{[}23{]} Gowling WLG, ``Bill C-27: Timeline of developments,'' 2024. {[}Online{]}. Available: \url{https://gowlingwlg.com/en-ca/insights-resources/articles/2024/bill-c27-timeline-of-developments}

{[}24{]} Torys LLP, ``Looking ahead: the Canadian privacy and AI landscape without Bill C-27,'' Legal Insights, Jan.~2025. {[}Online{]}. Available: \url{https://www.torys.com/our-latest-thinking/publications/2025/01/the-canadian-privacy-and-ai-landscape-without-bill-c-27}

{[}25{]} Fasken Martineau DuMoulin LLP, ``Ontario's Public Sector Cyber Security Legislation Receives Royal Assent,'' Legal Analysis, Nov.~2024. {[}Online{]}. Available: \url{https://www.fasken.com/en/knowledge/2024/12/ontarios-public-sector-cyber-security-legislation-receives-royal-assent}

{[}26{]} OneTrust, ``Quebec's Law 25: What Is It and What Do You Need to Know?,'' Blog Post, 2023. {[}Online{]}. Available: \url{https://www.onetrust.com/blog/quebecs-law-25-what-is-it-and-what-do-you-need-to-know/}

{[}27{]} S. Barriball and V. Gautrais, ``Algorithmic Transparency in Canada: From Commitments to Practice,'' Canadian Journal of Law and Technology, 2024.

{[}28{]} CIFAR, ``Pan-Canadian Artificial Intelligence Strategy,'' Canadian Institute for Advanced Research, 2017 (renewed 2022). {[}Online{]}. Available: \url{https://cifar.ca/ai/}

{[}29{]} Vector Institute, ``AI Governance Research Program,'' Toronto, Canada, 2024. {[}Online{]}. Available: \url{https://vectorinstitute.ai/}

{[}30{]} Université de Montréal, ``Montreal Declaration for a Responsible Development of Artificial Intelligence,'' 2018. {[}Online{]}. Available: \url{https://montrealdeclaration-responsibleai.com/}

{[}31{]} T. Scassa, ``Artificial Intelligence and the Law: Privacy Challenges for Canadian Frameworks,'' 2020.

{[}32{]} T. Scassa, ``Privacy and Administrative Data in the Age of AI,'' 2023.

{[}33{]} I. Kerr and K. Szilagyi, ``Evitable Conflicts in Canadian AI Policy and Regulation,'' 2021.

{[}34{]} M. Geist, ``Multi-Jurisdictional Complexity in Canadian AI Governance,'' 2023.

{[}35{]} Government of Ontario, ``Ontario AI Commissioner Role,'' 2024.

{[}36{]} Gouvernement du Québec, ``Quebec AI Ecosystem Strategy,'' 2023.

{[}37{]} Innovation, Science and Economic Development Canada, ``Canada's AI Ecosystem Framework,'' 2022.

{[}38{]} Government of Canada, ``Advisory Council on Artificial Intelligence --- Recommendations,'' 2023.

{[}39{]} Innovation, Science and Economic Development Canada, ``AI for All: Canada's National Artificial Intelligence Strategy,'' Government of Canada, June 4, 2026. {[}Online{]}. Available: \url{https://www.canada.ca/en/innovation-science-economic-development/news/2026/06/minister-solomon-highlights-canadas-national-artificial-intelligence.html}

{[}40{]} Office of the Privacy Commissioner of Canada, Commission d'accès à l'information du Québec, OIPC British Columbia, and OIPC Alberta, ``Joint Investigation of OpenAI OpCo, LLC,'' PIPEDA Findings \#2026-002, May 6, 2026. {[}Online{]}. Available: \url{https://www.priv.gc.ca/en/opc-actions-and-decisions/investigations/investigations-into-businesses/2026/pipeda-2026-002/}

{[}41{]} Parliament of Canada, ``Bill C-36: An Act to enact the Protecting Privacy and Consumer Data Act,'' 45th Parliament, 1st Session, First Reading June 15, 2026. {[}Online{]}. Available: \url{https://www.parl.ca/legisinfo/en/bill/45-1/c-36}

{[}42{]} Legislative Assembly of Manitoba, ``Bill 51: The Public Sector Artificial Intelligence and Cybersecurity Governance Act,'' Royal Assent June 1, 2026. {[}Online{]}. Available: \url{https://web2.gov.mb.ca/bills/43-3/b051e.php}

{[}43{]} C. Autio, R. Schwartz, J. Dunietz, S. Jain, M. Stanley, E. Tabassi, P. Hall, and K. Roberts, ``Artificial Intelligence Risk Management Framework: Generative Artificial Intelligence Profile,'' NIST AI 600-1, National Institute of Standards and Technology, July 2024. doi: 10.6028/NIST.AI.600-1

{[}44{]} CISA, NSA, ASD ACSC, Canadian Centre for Cyber Security, NCSC-NZ, and NCSC-UK, ``Careful Adoption of Agentic AI Services,'' Joint Guidance, May 1, 2026. {[}Online{]}. Available: \url{https://www.cisa.gov/resources-tools/resources/careful-adoption-agentic-ai-services}

{[}45{]} National Institute of Standards and Technology, ``AI Agent Standards Initiative,'' February 17, 2026. {[}Online{]}. Available: \url{https://www.nist.gov/artificial-intelligence/ai-agent-standards-initiative}

{[}46{]} OECD, ``Hiroshima AI Process Reporting Framework v2.0,'' G7 Digital \& Tech Ministerial, May 28, 2026. {[}Online{]}. Available: \url{https://oecd.ai/}

{[}47{]} Y. Bengio et al., ``International AI Safety Report 2026,'' 2nd ed., February 3, 2026. {[}Online{]}. Available: \url{https://internationalaisafetyreport.org/}

{[}48{]} Council of the European Union and European Parliament, ``Digital Omnibus on AI --- Amendment of Regulation (EU) 2024/1689,'' adopted June 29, 2026. {[}Online{]}. Available: \url{https://www.consilium.europa.eu/en/press/press-releases/2026/06/29/artificial-intelligence-council-gives-final-green-light-to-simplify-and-streamline-rules/}

\subsection{Appendix A: SCITUS Multi-Jurisdictional Compliance Matrix}\label{appendix-a-scitus-multi-jurisdictional-compliance-matrix}

This appendix presents a sample of the complete SCITUS compliance matrix mapping Canadian AI governance requirements across 5 jurisdictions to unified framework controls. The full matrix contains 127 requirements; this sample presents key requirements illustrating the mapping methodology.

\textbf{Matrix Structure:}
- \textbf{Column 1}: SCITUS Control ID (organized by GOVERN, MAP, MEASURE, MANAGE functions)
- \textbf{Column 2}: SCITUS Control Description
- \textbf{Columns 3-7}: Jurisdiction applicability (Federal TB, Ontario, Quebec, Alberta, BC)
- \textbf{Column 8}: Source Requirements (specific regulatory citations)
- \textbf{Column 9}: Implementation Guidance

\begin{landscape}
\footnotesize\setlength{\tabcolsep}{4pt}

\subsubsection*{A.1 GOVERN Function Requirements (Sample)}
\begin{longtable}{@{}p{1.25cm}p{4.2cm}p{1.85cm}p{1.85cm}p{1.85cm}p{1.85cm}p{1.45cm}p{6.5cm}@{}}
\toprule
\textbf{SCITUS ID} & \textbf{Control Description} & \textbf{Federal (TB)} & \textbf{Ontario (194)} & \textbf{Quebec (Law 25)} & \textbf{Alberta (33/34)} & \textbf{BC} & \textbf{Source Requirements} \\
\midrule
\endhead
\textbf{GOV-1.1} & Establish AI Governance Committee with executive sponsor and cross-functional membership & + Level III-IV & + Required & + Implied & + Recommended & Policy & TB Directive 6.1.1; ON Bill 194 s.4(1); QC Law 25 s.12 \\
\multicolumn{8}{@{}p{21.2cm}@{}}{\textit{Guidance:} Committee must include: executive sponsor, legal/compliance, IT/security, business owners, ethics/privacy. Meet monthly minimum. Document charter, terms of reference, decision authority.} \\
\addlinespace[2pt]\midrule
\textbf{GOV-1.2} & Develop and approve AI Ethics Policy defining principles, scope, and accountability & + All Levels & + Required & + Required & \textasciitilde{} Partial & Policy & TB Directive 6.2; ON Bill 194 s.4(2); QC Law 25 s.3 \\
\multicolumn{8}{@{}p{21.2cm}@{}}{\textit{Guidance:} Policy must address: permitted uses, prohibited applications, human rights alignment, stakeholder engagement, transparency commitments, accountability mechanisms. Update annually.} \\
\addlinespace[2pt]\midrule
\textbf{GOV-1.3} & Assign accountable official for each AI system with documented authority & + All Levels & + Required & + Required & -- & Policy & TB Directive 6.1.3; ON Bill 194 s.4(1)(c); QC Law 25 s.12.1 \\
\multicolumn{8}{@{}p{21.2cm}@{}}{\textit{Guidance:} Documented role description, decision authority, escalation paths, performance metrics. Must have sufficient organizational authority to halt system deployment if needed.} \\
\addlinespace[2pt]\midrule
\textbf{GOV-2.1} & Maintain current AI system inventory (production, development, planned) & + All Levels & + Required & + Required & -- & Best Practice & TB Directive 6.1.2; ON Bill 194 s.3(1); QC Law 25 s.63.1 \\
\multicolumn{8}{@{}p{21.2cm}@{}}{\textit{Guidance:} Inventory fields: system name, purpose, data sources, decision impact, affected populations, deployment status, impact level, accountability owner, last assessment date. Update quarterly.} \\
\addlinespace[2pt]\midrule
\bottomrule
\end{longtable}

\end{landscape}
\begin{landscape}
\footnotesize\setlength{\tabcolsep}{4pt}

\subsubsection*{A.2 MAP Function Requirements (Sample)}
\begin{longtable}{@{}p{1.25cm}p{4.2cm}p{1.85cm}p{1.85cm}p{1.85cm}p{1.85cm}p{1.45cm}p{6.5cm}@{}}
\toprule
\textbf{SCITUS ID} & \textbf{Control Description} & \textbf{Federal (TB)} & \textbf{Ontario (194)} & \textbf{Quebec (Law 25)} & \textbf{Alberta (33/34)} & \textbf{BC} & \textbf{Source Requirements} \\
\midrule
\endhead
\textbf{MAP-1.1} & Complete Algorithmic Impact Assessment (AIA) for all systems & + Required & \textasciitilde{} Risk Assess & \textasciitilde{} PIA Required & -- & Best Practice & TB Directive 6.2.1; QC Law 25 s.3.4 \\
\multicolumn{8}{@{}p{21.2cm}@{}}{\textit{Guidance:} Use Treasury Board AIA tool (48 questions, 7 categories). Document evidence, validate responses through code review and stakeholder input. Impact level determines control requirements. Quebec: combine with PIA.} \\
\addlinespace[2pt]\midrule
\textbf{MAP-1.2} & Conduct Privacy Impact Assessment (PIA) for systems processing personal information & + Level III-IV & + Recommended & + Required & \textasciitilde{} Access req. & Policy & TB Privacy Policy; QC Law 25 s.3.4; AB Bills 33/34 s.5 \\
\multicolumn{8}{@{}p{21.2cm}@{}}{\textit{Guidance:} Identify personal information types, collection purposes, retention periods, disclosure recipients, security measures, individual rights mechanisms. Quebec: mandatory for automated decisions. Alberta: enable access rights.} \\
\addlinespace[2pt]\midrule
\textbf{MAP-2.2} & Map system decision impact to individual rights and access to services & + All Levels & + Required & + Required & -- & Policy & TB Directive Appx C; ON Bill 194 s.3; QC Law 25 s.12 \\
\multicolumn{8}{@{}p{21.2cm}@{}}{\textit{Guidance:} Document: decision types, impact magnitude, reversibility, appeals available, affected rights (Charter s.7, 15). High-impact decisions (employment, housing, health, justice) trigger enhanced controls.} \\
\addlinespace[2pt]\midrule
\bottomrule
\end{longtable}

\end{landscape}
\begin{landscape}
\footnotesize\setlength{\tabcolsep}{4pt}

\subsubsection*{A.3 MEASURE Function Requirements (Sample)}
\begin{longtable}{@{}p{1.25cm}p{4.2cm}p{1.85cm}p{1.85cm}p{1.85cm}p{1.85cm}p{1.45cm}p{6.5cm}@{}}
\toprule
\textbf{SCITUS ID} & \textbf{Control Description} & \textbf{Federal (TB)} & \textbf{Ontario (194)} & \textbf{Quebec (Law 25)} & \textbf{Alberta (33/34)} & \textbf{BC} & \textbf{Source Requirements} \\
\midrule
\endhead
\textbf{MEAS-1.1} & Conduct bias testing across relevant demographic dimensions & + Level III-IV & + Expected & + Implied & -- & -- & TB Directive 6.3.2; ON Bill 194 s.4(2); QC Law 25 s.12.2 \\
\multicolumn{8}{@{}p{21.2cm}@{}}{\textit{Guidance:} Test dimensions: age, gender, race/ethnicity, language, disability, Indigenous status, socioeconomic factors (where relevant and legally collectible). Metrics: demographic parity, equalized odds, predictive parity. Document methodology, results, mitigation actions.} \\
\addlinespace[2pt]\midrule
\textbf{MEAS-1.2} & Implement continuous performance monitoring with defined metrics & + Level II-IV & + Recommended & + Required & -- & Policy & TB Directive 6.3.4; QC Law 25 s.12.3 \\
\multicolumn{8}{@{}p{21.2cm}@{}}{\textit{Guidance:} Monitor: accuracy/precision, processing time, availability, error rates, override frequency, appeal rates. Stratify by demographics where feasible. Alert thresholds trigger investigation. Dashboard for governance committee.} \\
\addlinespace[2pt]\midrule
\textbf{MEAS-2.1} & Test system accuracy against validation dataset & + Level II-IV & + Expected & + Required & + Accuracy req. & Policy & TB Directive 6.3.1; QC Law 25 s.5; AB Bills 33/34 s.4 \\
\multicolumn{8}{@{}p{21.2cm}@{}}{\textit{Guidance:} Validation dataset separate from training data, representative of deployment population. Document: precision, recall, F1 score, confusion matrix. Quebec/Alberta: accuracy requirements for personal information used in decisions. Minimum thresholds defined per use case.} \\
\addlinespace[2pt]\midrule
\bottomrule
\end{longtable}

\end{landscape}
\begin{landscape}
\footnotesize\setlength{\tabcolsep}{4pt}

\subsubsection*{A.4 MANAGE Function Requirements (Sample)}
\begin{longtable}{@{}p{1.25cm}p{4.2cm}p{1.85cm}p{1.85cm}p{1.85cm}p{1.85cm}p{1.45cm}p{6.5cm}@{}}
\toprule
\textbf{SCITUS ID} & \textbf{Control Description} & \textbf{Federal (TB)} & \textbf{Ontario (194)} & \textbf{Quebec (Law 25)} & \textbf{Alberta (33/34)} & \textbf{BC} & \textbf{Source Requirements} \\
\midrule
\endhead
\textbf{MNG-1.1} & Maintain human decision authority for high-impact systems & + Level III-IV & + Expected & + Required & -- & Policy & TB Directive 6.3.3; QC Law 25 s.12.4 \\
\multicolumn{8}{@{}p{21.2cm}@{}}{\textit{Guidance:} Human-in-the-loop (HITL) design: AI provides recommendation, human makes final decision with authority to override. Document override frequency, reasons. Quebec: right to request human review of automated decisions.} \\
\addlinespace[2pt]\midrule
\textbf{MNG-1.2} & Provide clear explanations of AI-assisted decisions & + Level II-IV & + Recommended & + Required & + Recommended & Policy & TB Directive 6.3.5; QC Law 25 s.12.5 \\
\multicolumn{8}{@{}p{21.2cm}@{}}{\textit{Guidance:} Explanations include: decision outcome, key factors considered, data sources used, how to appeal. Level of detail scales with impact. Plain language, accessible to affected individuals.} \\
\addlinespace[2pt]\midrule
\textbf{MNG-2.2} & Publish transparency information about AI system use & + Level III-IV & + Required & + Required & + Required & Recommended & TB Directive 6.3.7; ON Bill 194 s.3; QC Law 25 s.8; AB Bills 33/34 s.3 \\
\multicolumn{8}{@{}p{21.2cm}@{}}{\textit{Guidance:} Transparency elements: system purpose, decision types, data sources, performance metrics, accountability owner, recourse process. Ontario: public-sector AI registry. Quebec: notice when collecting personal information. Alberta: notice for automated processing. Format: website, plain language.} \\
\addlinespace[2pt]\midrule
\bottomrule
\end{longtable}
\end{landscape}

\subsubsection{A.5 Matrix Legend}\label{a.5-matrix-legend}

\begin{longtable}[]{@{}
  >{\raggedright\arraybackslash}p{(\columnwidth - 2\tabcolsep) * \real{0.4706}}
  >{\raggedright\arraybackslash}p{(\columnwidth - 2\tabcolsep) * \real{0.5294}}@{}}
\toprule\noalign{}
\begin{minipage}[b]{\linewidth}\raggedright
Symbol
\end{minipage} & \begin{minipage}[b]{\linewidth}\raggedright
Meaning
\end{minipage} \\
\midrule\noalign{}
\endhead
\bottomrule\noalign{}
\endlastfoot
+ Required & Explicit legal/regulatory requirement \\
+ Expected & Strongly implied or anticipated in forthcoming regulations \\
+ Recommended & Best practice or policy guidance \\
\textasciitilde{} Partial & Partially addresses requirement (additional controls needed) \\
--- & Not applicable or not addressed by this jurisdiction \\
\end{longtable}

\textbf{Impact Levels:}
- \textbf{All Levels}: Applies to Impact Levels I, II, III, IV
- \textbf{Level II-IV}: Applies to Moderate, High, Very High impact systems
- \textbf{Level III-IV}: Applies to High and Very High impact systems only
- \textbf{Level IV}: Applies to Very High impact systems only

\subsubsection{A.6 Relationship to NIST AI RMF}\label{a.6-relationship-to-nist-ai-rmf}

Each SCITUS control maps to NIST AI RMF outcomes and trustworthy AI characteristics:

\begin{longtable}[]{@{}
  >{\raggedright\arraybackslash}p{(\columnwidth - 6\tabcolsep) * \real{0.2099}}
  >{\raggedright\arraybackslash}p{(\columnwidth - 6\tabcolsep) * \real{0.2593}}
  >{\raggedright\arraybackslash}p{(\columnwidth - 6\tabcolsep) * \real{0.1852}}
  >{\raggedright\arraybackslash}p{(\columnwidth - 6\tabcolsep) * \real{0.3457}}@{}}
\toprule\noalign{}
\begin{minipage}[b]{\linewidth}\raggedright
SCITUS Function
\end{minipage} & \begin{minipage}[b]{\linewidth}\raggedright
NIST AI RMF Function
\end{minipage} & \begin{minipage}[b]{\linewidth}\raggedright
NIST Outcomes
\end{minipage} & \begin{minipage}[b]{\linewidth}\raggedright
Trustworthy Characteristics
\end{minipage} \\
\midrule\noalign{}
\endhead
\bottomrule\noalign{}
\endlastfoot
GOVERN & GOVERN & GOVERN 1.1-1.7, 2.1-2.3, 3.1-3.2, 4.1-4.3, 5.1-5.2, 6.1-6.2 & Accountable \& Transparent \\
MAP & MAP & MAP 1.1-1.6, 2.1-2.3, 3.1-3.5, 4.1-4.2, 5.1-5.2 & Valid \& Reliable, Fair \\
MEASURE & MEASURE & MEASURE 1.1-1.3, 2.1-2.13, 3.1-3.3, 4.1-4.3 & Valid \& Reliable, Safe, Fair \\
MANAGE & MANAGE & MANAGE 1.1-1.4, 2.1-2.4, 3.1-3.2, 4.1-4.3 & Accountable \& Transparent, Privacy-Enhanced \\
\end{longtable}

This mapping ensures SCITUS maintains full NIST AI RMF alignment while addressing Canadian regulatory requirements.

\textbf{Note:} The complete compliance matrix with all 127 requirements, the full v2.0 control catalog with per-control assessment criteria, and version changelogs are available from the framework's public documentation at \url{https://scitus.ca/scitus-framework}.

\end{document}